\documentclass[journal]{IEEEtran}
\usepackage{cite}
\usepackage{bm}
\usepackage{amsmath}
\usepackage{extarrows}
\usepackage{amssymb}
\usepackage{graphicx}
\usepackage{xcolor}
\usepackage{enumerate}
\usepackage{bookmark} 
\graphicspath{{figure}}
\usepackage{stfloats}
\usepackage{setspace}
\usepackage{multirow}
\usepackage{amsmath} 
\usepackage{framed}

\newcommand{\mb}{\bm}
\newcommand{\mr}{\mathrm}
\newcommand{\ms}{\mathrm}
\newcommand{\BE}{\begin{equation}}
\newcommand{\EE}{\end{equation}}
\newcommand{\BS}{\begin{subequations}}
\newcommand{\ES}{\end{subequations}}
\renewcommand{\bf}{\bm}
\newcommand{\bb}{\mathbb}
\renewcommand{\cal}{\mathcal}

\newtheorem{theorem}{Theorem}
\newtheorem{proposition}[theorem]{Proposition}

\newtheorem{definition}[theorem]{Definition}

\newtheorem{property}[theorem]{Property}

\newtheorem{lemma}[theorem]{Lemma}

\newtheorem{conjecture}[theorem]{Conjecture}

\ifCLASSINFOpdf
\graphicspath{{figure/}}
 
\else
 
\fi

\usepackage{ulem}

\newcommand{\LL}[1]{\textcolor{black}{#1}}

\newcommand{\LLC}[1]{\textcolor{black}{#1}}%violet

\begin{document}
%\begin{spacing}{1.3}
\title{On Orthogonal Approximate Message Passing} %On OAMP: Impact of Orthogonal Principle

\author{{Lei~Liu, \textit{Member, IEEE}, Yiyao~Cheng, Shansuo~Liang, \\Jonathan~H.~Manton, \textit{Fellow, IEEE}, and~Li~Ping, \textit{Fellow, IEEE}}

\thanks{Lei~Liu was with the Department of Electronic Engineering, City University of Hong Kong (CityU), Hong Kong, SAR, China, and is currently with the School of Information Science, Japan Institute of Science and Technology (JAIST), Ishikawa 923-1292, Japan (e-mail: leiliu@jaist.ac.jp).} %The work was conducted at his stay with CityU.}
\thanks{Yiyao~Cheng, Shansuo~Liang and~Li~Ping are with the Department of Electronic Engineering, CityU, Hong Kong, SAR, China (e-mail:   \{yiycheng2-c, ssliang3-c\}@my.cityu.edu.hk, eeliping@cityu.edu.hk).}
\thanks{Jonathan~H.~Manton is with the Department of Electrical and Electronic Engineering, The University of Melbourne, VIC 3010, Australia (e-mail: j.manton@ieee.org).}
}

\maketitle

\begin{abstract}
Approximate Message Passing (AMP) is an efficient iterative parameter-estimation technique for certain high-dimensional linear systems with non-Gaussian distributions, such as sparse systems. In AMP, a so-called Onsager term is added to keep estimation errors approximately Gaussian. Orthogonal AMP (OAMP) does not require this Onsager term, relying instead on an orthogonalization procedure to keep the current errors uncorrelated with (i.e., orthogonal to) past errors. \LL{In this paper, we show the generality and significance of the orthogonality in ensuring that errors are ``asymptotically independently and identically distributed Gaussian'' (AIIDG).} This AIIDG property, which is essential for the attractive performance of OAMP, holds for separable functions. \LL{We present a simple and versatile procedure to establish the orthogonality through  Gram-Schmidt (GS) orthogonalization, which is applicable to any prototype. We show that different AMP-type algorithms, such as expectation propagation (EP), turbo, AMP and OAMP, can be unified under the orthogonal principle.} The simplicity and generality of OAMP provide efficient solutions for estimation problems beyond the classical linear models. \LL{As an example, we study the optimization of OAMP via the GS model and GS orthogonalization.} More related applications will be discussed in a companion paper where new  algorithms are developed for problems with multiple constraints and multiple measurement variables.

%Approximate Message Passing (AMP) is an efficient iterative parameter-estimation technique for certain high-dimensional linear systems with non-Gaussian distributions, such as sparse systems. In AMP, a so-called Onsager term is added to keep the errors of the current estimate approximately Gaussian. The analysis of AMP is somewhat involved.

%Orthogonal AMP (OAMP) is a modified version of AMP that is simpler to understand and analyse: it does not require an Onsager term, relying instead on an orthogonalisation procedure to keep the current errors uncorrelated with past errors. Moreover, OAMP is shown to be more general than AMP and other AMP-type algorithms such as Expectation Propagation (EP) and Vector AMP (VAMP). Specifically, given an AMP algorithm, an OAMP algorithm can be derived having the same performance: there is a mapping from the sequence of estimates produced by OAMP to the sequence of estimates produced by AMP. The converse is not always possible: OAMP can handle certain classes of systems that AMP cannot.

%Being simpler to understand makes it easier to apply OAMP in novel situations. This is elaborated on in a companion paper which applies OAMP to a wide range of applications.

\end{abstract}

\begin{IEEEkeywords}
Expectation propagation (EP), turbo, belief propagation (BP), approximate message passing (AMP), vector AMP (VAMP), unified framework, state evolution, Haar  matrices.
\end{IEEEkeywords}

\IEEEpeerreviewmaketitle
\vspace{0.5cm}

\newcommand{\bx}{\bf{x}}
\newcommand{\bX}{\bf{\mathcal{X}}}
\newcommand{\hx}{\hat{\bf{x}}}
\newcommand{\hX}{\hat{\bf{\mathcal{X}}}}
\newcommand{\bV}{\bf{V}}

\section{Introduction}

\subsection{Motivation and Insight}
\begin{figure}[htb]
  \centering 
  \includegraphics[width=7.5cm]{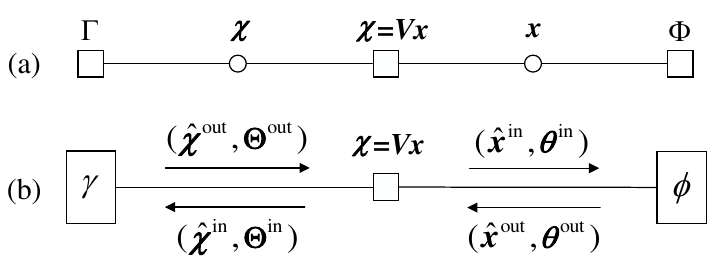}\\ 
  \caption{Graphic illustrations for (a) a system model with two pieces of statistical information (called constraints) $\Phi$ and $\Gamma$, and (b) an iterative process involving two local processors $\gamma$ and $\phi$. The parameter vectors $\bf{\theta}$ and $\bf{\Theta}$ carry small amounts of information about the estimates they accompany, such as the error variance. Normally they can be computed offline.
  }
  \label{Fig:model}
\end{figure}

Fig.~\ref{Fig:model}(a) illustrates a message passing scheme for estimating the random vector $\bf{x}$ in a Bayesian setting. Here, $\Gamma$ and $\Phi$ represent statistical information about $\bf{x}$ such as observations or known distributions. They are called \textbf{constraints} for brevity. Message passing repeatedly applies each constraint in turn, aiming to converge to an estimate that optimally combines both pieces of statistical information. This is attractive when applying both pieces of information at once is computationally infeasible. This paper provides new insight into when and how message passing can be made to work.

The following example explains some of the insights. Let the elements of $\bf{x}$ be source symbols chosen uniformly at random from the finite alphabet $\Omega_x \subset \mathbb{R}$. Additional statistical information comes from observing the received signal
\BE\label{Eqn:linear_system}
\bf{y}=\bf{A}\bf{x} + \bf{n},
\EE					
where $\bf{A} \in \mathbb{R}^{M \times N}$ is a fat or square sensing matrix of size $M \times N$ with $M\le N$, and $\bf{n}\sim\cal{N}(\bf{0},\sigma_n^2\bf{I})$ represents additive white Gaussian noise (AWGN). Finding the optimal maximum \textit{a posteriori} (MAP) estimate of $\bf{x}$ is computationally prohibitive when $N$ is large.

Directly mapping this example to Fig.~\ref{Fig:model} leads to $\Phi$ being the constraint $x_i \sim \textrm{uniform}(\Omega_x)$ ($1 \leq i \leq N$), the matrix $\bf{V}$ being the sensing matrix $\bf{A}$, and $\Gamma$ being the constraint $\bf{y} = \bf{\mathcal{X}} + \bf{n}$. Naively, given an estimate $\hat{\bf{x}}$ of $\bf{x}$, the constraint $\Phi$ can improve on it by bringing it closer to an element of $\Omega_x$ (e.g., by soft thresholding), while an estimate $\hat{\bf{\mathcal{X}}}$ of $\bf{\mathcal{X}}$ can be improved on by bringing it closer to the observation $\bf{y}$. This naive approach will not work due to data incest; the incoming estimate at a local processor in Fig.~\ref{Fig:model}(b) is treated as statistically independent of the statistical constraint imposed by the processor, but in general it is not.

The original way of message passing was made to work for this example, under the strong assumption that $\bf{A}$ ``looked like'' it was generated at random with independently and identically distributed (IID) Gaussian entries, was by including an Onsager correction term to mitigate the data incest problem by filtering out the dependent information~\cite{Donoho2009}. The idea of using an Onsager correction term comes from simulating spin glasses in physics and does not provide an elementary understanding of how to make message-passing schemes work in general.

\subsection{Iterative Detection}
For the estimation of $\bf{x}$ in Fig.~\ref{Fig:model}(a), jointly processing $\Gamma$ and $\Phi$ generally requires excessively high complexity. Fig.~\ref{Fig:model}(b) illustrates a low-cost iterative detector. Two local processors, denoted by $\gamma$ and $\phi$, process the two constraints $\Gamma$ and $\Phi$ separately. In each iteration, the estimates for $\bf{\mathcal{X}}$ and $\bf{x}$, denoted by $\bf{\mathcal{\hat{X}}}^{\mr{out}}$ and $\hat{\bf{x}}^{\mr{out}}$, are generated by $\gamma$ and $\phi$ respectively. They are referred to as output messages and, after transforming them by $\bf{V}^{\rm T}$ or $\bf{V}$ respectively, they form the input (i.e., \textit{a priori}) messages $\hat{\bf{x}}^{\mr{in}}$ and $\bf{\mathcal{\hat{X}}}^{\mr{in}}$ for the next iteration. In this way, $\bf{\mathcal{\hat{X}}}^{\mr{out}}$ and $\hat{\bf{x}}^{\mr{out}}$ are refined iteratively. Although omitted to reduce notational clutter, the distributions of the estimates are tracked. Each local processor treats the distribution of its input as a prior distribution. The computed (approximate) posterior distribution of the outgoing $\bf{\mathcal{\hat{X}}}^{\mr{out}}$ or $\hat{\bf{x}}^{\mr{out}}$ serves as the incoming prior for the next processor.

The local estimators $\gamma$ and $\phi$ can be constructed using minimum mean square error (MMSE) or MAP principles, which are broadly referred to as direct methods.  Implementations of such methods include expectation maximization (EM)-based techniques and MAP filtering \cite{Moon1996EM, Lodge1993, Loeliger2009EM}. Directly coupling two local estimators as in Fig.~\ref{Fig:model}(b) may suffer from an error correlation problem. Specifically,  errors at the input of a local estimator may become correlated with its previous output errors during iterative processing. Such correlated errors are difficult to track and may have detrimental effects.  

Various solutions have been proposed before. In the message passing decoding techniques for turbo and LDPC codes~\cite{Richardson2001,TurboCode}, an output of an estimator is ``extrinsic'', meaning it is independent of the related input message. This extrinsic principle can be asymptotically met by forward error control (FEC) codes designed on large sparse graphs. %In such codes, $\bf{A}$ is typically a permutation matrix, such as that in \eqref{Eqn:example_2}. 

%This extrinsic principle can be asymptotically met when $\Gamma$ and $\Phi$ are representable by large sparse graphs and $\bf{V}$ is a random permutation. In such cases, an LDPC decoder can offer capacity-approaching performance by matching the EXIT functions of the two local estimators \cite{Brink2001, Brink2004}.

The turbo principle has been extended to the linear system in  \eqref{Eqn:linear_system} involving FEC codes \cite{Wang1999}. The latter is not sparse when $\bf{A}$ is full \cite{Tuchler2002, Douillard1995, Loeliger2007,  Yuan2014, LiuLei2019TSP}. Expectation propagation (EP) is a closely related technique based on the Gaussian approximation of messages \cite{Cakmak2018, Minka2001}. The work on EP was initially heuristic but analysis techniques have been derived recently \cite{Takeuchi2017}. Approximate message passing (AMP) treats the correlation problem using a so-called Onsager term \cite{Donoho2009}. AMP assumes IID Gaussian (IIDG) entries for the sensing matrix $\bf{A}$ in \eqref{Eqn:linear_system}. 

\LL{State-evolution (SE), derived heuristically in \cite{Donoho2009} and proved rigorously in \cite{Bayati2011}, is an analysis tool for AMP as well as other related algorithms \cite{Rangan2016, Ma2016, Takeuchi2017, Dudeja2022, LiuAMP2019, MaLiu2018}. The discussions on SE in \cite{Rangan2016, Ma2016, Takeuchi2017, Dudeja2022, LiuAMP2019,Bayati2011, MaLiu2018} focus on mathematical rigor but are less intuitive. It is not straightforward to see a common core among different approaches. Recently, \cite{Takeuchi2019} shows an interesting direction toward a unified framework. However, \cite{Takeuchi2019} heavily relies on \cite{Takeuchi2017}. The latter involves detailed algorithmic operations of orthogonal AMP (OAMP) \cite{Ma2016} (categorized as a special case of EP), which cannot be easily generalized.}

\LL{AMP has been successfully applied to various communication systems, including grant-free machine-type communications (MTC) \cite{Senel2018}, massive random access \cite{ChenAMP2019}, compressed coding \cite{LiangAMP2020}, synchronous/asynchronous massive connectivity \cite{SunAMP2019,ZhuAMP2021}, and reconfigurable intelligent surface (RIS) aided multi-user multi-input multi-output (MIMO) systems \cite{Ruan2022}.}
\LL{\subsection{Orthogonal Approximate Message Passing (OAMP)}}
The assumption of IID sensing matrices in AMP is relaxed in OAMP \cite{Ma2016}, which offers a solution to systems with non-IID sensing matrices. SE for OAMP is conjectured in \cite{Ma2016} and proved in \cite{Takeuchi2017}. The key to OAMP is to maintain error orthogonality during iterative processing \cite{Ma2016}.  Based on SE, the optimality of OAMP is derived in \cite{Ma2016} under certain conditions. Vector AMP (VAMP) \cite{Rangan2016} is algorithmically similar to OAMP. An elegant proof of SE is given in \cite{Rangan2016} for VAMP. It is shown that AMP and OAMP-based decoding can outperform the turbo algorithm in coded linear systems \cite{LiuAMP2019, MaLiu2018}.

OAMP has been extensively investigated for communication and signal processing applications with non-IID (such as ill-conditioned and correlated) channel/pilot matrices, including massive MIMO \cite{KhaniANSD2020}, clipped sparse regression codes \cite{IC-SRC2021}, coded linear systems \cite{MaLiu2018}, orthogonal-frequency-division-multiplexing (OFDM) \cite{Yiyao_ofdm, HWang2019, XZhou2022, ZhangOAMP2017}, grant-free MTC \cite{SLiu2022}, massive access \cite{Yiyao_mmv} and orthogonal time frequency space (OTFS) modulation \cite{LiOTFS2022}. The connection between unfolded AMP/OAMP and artificial neural networks is studied in \cite{He2018AI, Zhang2019AI,  Takabe2019}. It has been demonstrated that the orthogonalization parameters can be acquired via deep learning, resulting in deep unfolded AMP/OAMP algorithms with improved performance \cite{He2018AI, Zhang2019AI, Takabe2019}.\vspace{3mm}

\subsection{The Haar Distribution of \texorpdfstring{$\bf{V}$}{TEXT}}

Throughout this paper, we assume that $\bf{V}$ in Fig.~\ref{Fig:model} is randomly selected and its size approaches infinity. Consequently, we may expect by some generalized law of large numbers that the errors $\bf{\mathcal{\hat{X}}}^{\rm in}= \bf{V}\hat{\bf{x}}^{\rm in}$ and  $\hat{\bf{x}}^{\rm in}= \bf{V}^{\rm T}\bf{\mathcal{\hat{X}}}^{\rm out}$ are close to being IID Gaussian. We however have to scrutinize this issue carefully as the errors may become correlated with $\bf{V}$ during iterative processing. To make the problem tractable, we will focus on the average behavior of OAMP over all possible $\bf{V}$ in a Haar ensemble. This means, although a fixed $\bf{V}$ is being used to generate the sequence of estimates, the fact that it was originally chosen at random and will be exploited at each step to simplify the analysis. The hope is that the behavior with a fixed $\bf{V}$, the one of interest, will be similar to that predicted by SE using the ``ensemble". Clearly, this is not always the case, e.g. $\bf{V}=\bf{I}$ is a bad choice. It is left to numerical studies to confirm this estimate. 

``Average performance" is also implied in other works on AMP family of algorithms \cite{Donoho2009, Takeuchi2019, Bayati2011, Rangan2016, Takeuchi2017}. This paper uses this argument explicitly.

{\color{black}\subsection{Contributions of This Paper}
This paper aims at a comprehensive understanding on the impact of orthogonality in OAMP and other AMP-type algorithms. Following the basic concept in \cite{Ma2016} and inspired by the works in \cite{Rangan2016, Takeuchi2019}, we take a high-level approach. We separate the overall problem into two inter-coupled sub-ones: 
\begin{enumerate}[(i)]
    \item the impacts of orthogonality, and
    \item how to establish orthogonality.
\end{enumerate} 
For (i), the orthogonal principle leads to a unified framework for different AMP-type algorithms \cite{ Donoho2009, Ma2016, Rangan2016, Takeuchi2017, Bayati2011, Takeuchi2019}. This is implied in \cite{Takeuchi2019}. We make it explicit in this paper. For (ii), we outline a Gram-Schmidt orthogonalization (GSO) procedure to establish orthogonality, which leads to a general realization technique for OAMP. 

 The separated approach above reveals new insights and new treatments/applications, as the contributions listed below. 
\begin{itemize}
    \item OAMP discussed in this paper is based on a class of orthogonal local estimators that are more general than those used in the standard EP/AMP/OAMP/VAMP \cite{ Donoho2009, Ma2016, Rangan2016, Takeuchi2017, Bayati2011, Takeuchi2019}. Hence, new realization methods are revealed. For example, we can optimize performance by selecting the best among the class of orthogonal local estimators. (See Section \ref{Sec:Optimization_OAMP}.) Another example is the integral approach discussed in the next bullet.
\item GSO does not require the differentiability required by the standard EP/AMP/OAMP/VAMP. In sub-section \ref{Sec:B_methods}, we outline an integral approach to OAMP via GSO, which works well when, e.g., a local estimator is realized by a software package in a black-box manner without guaranteed differentiability. Empirical advantages of integral-based OAMP can be found in \cite{Yiyao_integral}.
\item  The derivation of SE for OAMP in this paper is inspired by the techniques used \cite{Rangan2016, Takeuchi2017, Bayati2011, Takeuchi2019}. The new approach is more concise for the following reasons. 
\begin{itemize}
    \item Under the orthogonal principle, the behavior of SE can be directly characterized as the Bolthausen’s conditioning problem, which avoids the lengthy step-by-step tracking as in \cite{Rangan2016, Takeuchi2017, Bayati2011}. %Under the orthogonal principle, the behavior of SE can be directly characterized as the  Bolthausen’s conditioning problem. The derivations in \cite{Rangan2016, Takeuchi2017, Bayati2011, Takeuchi2019} arrive at the same problem by step-by-step tracking algorithm details, which is more lengthy and strenuous.    
\item  We analyze the Bolthausen’s conditioning problem using a few conjectures that can be bridged by more rigorous treatments in  \cite{Rangan2016, Takeuchi2017,  Takeuchi2019}. Our aim is conciseness, which can provide useful insights.
\item The algorithm discussed in this paper is symmetric, i.e., it consists of two local estimators under the same orthogonal principle. Hence we only need to analyze one local estimator during induction, since the result is applicable to the other one. %This is simpler than the works in \cite{Rangan2016, Takeuchi2017, Bayati2011} involving two asymmetric local estimators, one linear and one non-linear that have to be analyzed separately. 
\end{itemize}
 \end{itemize}
 
 The following are more implications of the findings in this paper. Interested readers are referred to the references listed.
\begin{itemize}
    \item Most discussions on AMP-family algorithms are for simple single measurement vector (SMV) systems \cite{Donoho2009, Cakmak2018, Minka2001, Rangan2016, Bayati2011, Ma2016, Takeuchi2017, Dudeja2022, LiuAMP2019, MaLiu2018}. GSO-based OAMP can be generalized to more complex multiple measurement vector (MMV) systems, as reported in \cite{Yiyao_mmv} for correlated massive-access channels. GSO also allows the extension of OAMP to systems with multiple Haar matrices. We are currently working on this issue \cite{Lei_TSP_2_2019}.
\item Based on GSO, we can prove that OAMP together with a properly designed decoder is capacity approaching. Such information-theoretic optimality is reported in \cite{OAMP_ISIT22, OAMP_TCOM}.
 \end{itemize}
 
Overall, we expect that the findings in this paper provide justifications for the applications of OAMP in \cite{He2018AI, Zhang2019AI,  Takabe2019, OAMP_ISIT22, OAMP_TCOM, Yiyao_integral, Lei_TSP_2_2019, KhaniANSD2020, IC-SRC2021, Yiyao_ofdm, HWang2019, XZhou2022, ZhangOAMP2017, SLiu2022, LiOTFS2022,Yiyao_mmv, Fletcher2016},   as well as guidelines for new treatments in communications and signal processing.}

\subsection{Notation}
For convenience, a vector is said IID (resp. IIDG) if its entries are IID (resp. IIDG). A vector is said joint-Gaussian if its entries are jointly Gaussian. Boldface lowercase letters represent column vectors and boldface uppercase symbols denote matrices. $\mb{I}$ denotes the identity matrix of appropriate size, $\bf{\mathcal{U}}^N$ the set of all $N\times N$ orthogonal matrices, $\mathbb{R}^N$ the set of all length-$N$ vectors, $\mb{0}$  the zero matrix, $\bm{a}^{\rm T}$ the transpose of $\bm{a}$, $\|\bm{a}\|$  the $\ell_2$-norm of $\bm{a}$, $\cal{N}(\bm{\mu},\bm{\Sigma})$ the Gaussian distribution with mean $\bm{\mu}$ and covariance $\bm{\Sigma}$, and $\mr{Diag}[{\bf{A}}_1,{\bf{A}}_2,\cdots {\bf{A}}_M]$ the block-diagonal matrix with diagonal blocks $\{{\bf{A}}_m\}$. $\ms{E}\{\cdot\}$ is the expectation operation over all random variables involved in the brackets, except when otherwise specified. $\mr{E}\{a|b\}$ is the expectation of $a$ conditional on $b$, $\ms{Var}\{{a}\}=\ms{E}\left\{ ({a} - \ms{E}\{{a}\})^2 \right\}$, $\ms{mmse}\{{a}|{b}\}=\ms{E}\left\{ ({a} - \ms{E}\{{a}|{b}\})^2\,\big|{b} \right\}$, and $\eta'(r)=\frac{\partial}{\partial r} \eta(r)$.  $\mr{Var}(\bf{x})$ is the common variance of the entries in $\bf{x}$, where $\bf{x}$ is a vector of IID entries. $\bf{a} \overset{\rm d}{\to} \bf{b}$ denotes that the distribution of $\bf{a}$ converges to that of $\bf{b}$ as the length goes to infinity. %\LC{For two random vectors $\bf{a}$ and $\bf{b}$, we write $\bf{a}\overset{\rm P}{\simeq} \bf{b}$ when their difference converges in probability to $\bf{0}$, i.e., $\bf{a}-\bf{b}\overset{\rm P}{\simeq} \bf{0}$.}

To aid the reader in finding a referenced Definition, Property, Theorem and so forth, a single counter is used for labelling (e.g., Definition 1, Property 2, \dots).

\section{Preliminaries}\label{Sec:Prelim}
%This section outlines the concept of orthogonality, based on which we will develop OAMP in Sections \ref{Sec:OAMP_Principle} and \ref{Sec:GS_orth}. 

\subsection{Polar Coordinates}
Let the Cartesian coordinate of $\bf{a}\in\mathbb{R}^N$ be $\bf{a}=\{a_n, n=1,\dots, N\}$. Define
 \BE
     \rho(\bf{a}) \equiv \|\bf{a}\| = \sqrt{\textstyle\sum_{n=1}^Na_n^2} \quad {\rm and} \quad \zeta(\bf{a})=\bf{a}/\|\bf{a}\|. 
 \EE
We call $\rho(\bf{a})$ and $\zeta(\bf{a})$ the  amplitude and angle of $\bf{a}$ respectively, and call $\big(\rho(\bf{a}), \zeta(\bf{a})\big)$ jointly as the polar coordinate of $\bf{a}$. (Note that $\zeta(\bf{a})$ has $N-1$ degrees of freedom since $\|\zeta(\bf{a})\|=1$.) Denote by $S^N(\rho)$ a sphere of amplitude $\rho$ over $\mathbb{R}^N$ in an $N$-dimensional Euclidean space. Any $\bf{a}\in S^N(\rho)$ has a fixed norm $\|\bf{a}\|=\rho$ \cite{Arfken1985_polar}.  Clearly, $\zeta(\bf{a})\in S^N(1)$, i.e., it is a point on the unit sphere. 
 
There are other ways to define the polar coordinates \cite{Bronshtein2004}. Our definition above serves the purpose in this paper.

 \subsection{Haar Distribution and Haar Transform}

 \begin{definition}[Haar Distribution]\label{Def:Haar}
An $N\times N$ matrix\footnote{Although we only consider a square $\bf{V}$, the results of this paper are still valid for non-square $\bf{V}$. For example, $\bf{V}$ is $M\times N$ with $M<N$, we can expand $\bf{V}$ to a square one \LL{without destroying the Haar} property and add constraints $\{{\mathcal{X}}_i=0, i=M+1,\cdots,N\}$ at $\Gamma$. The same method applies for $M>N$. Then the problem is the same as the square case.} $\bf{V}$ is said Haar distributed, denoted by $\bf{V}\sim \bf{\cal{H}}^N$, if $\bf{Vc}$ \LL{is uniformly distributed} over $S^N(\rho)$ for any fixed $\bf{c}\in S^N(\rho)$.  %, denoted as $\bf{Vc}\sim \mathbb{S}^N(\rho)$,
\end{definition} 
% \begin{definition}[Haar \cite{Tulino2004}]\label{Def:Haar}
% The set of $N \times N$ orthogonal matrices, $\bf{\mathcal{U}}^N$, is a compact Lie group and hence can be endowed with a unique bi-invariant probability measure called Haar measure. Saying $\bf{V}$ is Haar distributed means $\bf{V}$ is a randomly chosen orthogonal matrix with respect to this Haar probability measure. 
% \end{definition}

For brevity, we say $\bf{V}$ is Haar to mean $\bf{V}$ is Haar distributed. (This should not be confused with the concept of deterministic Haar matrices used in Fourier analysis.) 

It can be verified that Definition \ref{Def:Haar} is consistent with the conventional definition of the Haar distribution \cite{Tulino2004}. Some common linear transforms, such as discrete cosine transform (DCT) \cite{Ahmed1974}  and Hadamard transform \cite{Yarlagadda1997}, can be regarded as samples of a Haar transform. The following follows from Definition \ref{Def:Haar}  directly. 

\begin{lemma}
    Let $\bf{\Omega}\in \bf{\cal{U}}^N$, $\bf{V}\sim \bf{\cal{H}}^N$ and $\bf{c}\in \bb{R}^N$. Then,  $\bf{\Omega V}\sim \bf{\cal{H}}^N$ and $\bf{V\Omega}\sim \bf{\cal{H}}^N$. Furthermore, $\zeta(\bf{Vc})$ and $\zeta(\bf{c})$ are mutually independent.
\end{lemma}

\subsection{Pseudo-IID (PIID) Variables}\label{Sec:PIIDV}
We will prove in \ref{Sec:IID_error} that the distortions in OAMP are pseudo IID Gaussian (PIIDG) defined below. 
\begin{definition}
  A sequence $\bf{a}=\{a_n\}$ of size $N$ is said to be pseudo IID (PIID) if any subset of size $M$ in $\bf{a}$ is asymptotically IID when $N\to\infty$ and $M$ remains fixed. 
\end{definition} 
 
\begin{lemma}\label{Lem:PIIDG}
Let $\bf{a} = \bf{Va}'$ where $\bf{V}$ is Haar of size $N\times N$ and $\bf{a}'$ is an arbitrary fixed vector. From Theorem 2.8 in \cite{Meckes2014}, the entries of $\bf{a}$ are PIIDG. The marginal symbol distribution of $a_n, \forall n$, converges to ${\cal{N}}(0, v)$  with $v\!=\!\tfrac{1}{N}\mr{E}\{\|\bf{a}'\|^2\}$ as $N\!\to \!\infty$. 
\end{lemma}   

% Similarly, we define a jointly PIIDG as follows.
% \begin{definition}
%   Sequences $\{\bf{a}_1,\dots,\bf{a}_n\}$ with each of size $N$ are said to be jointly PIIDG if any subset of size $M$ in $\{\bf{a}_1,\dots,\bf{a}_n\}$ are asymptotically jointly IIDG when $N\to\infty$ and $M$ remains fixed. 
% \end{definition}

Some properties/conjectures of PIID/PIIDG (corresponding to those of IID/IIDG) are made in Appendix \ref{APP:Pro_PIIDG}.

\subsection{Separable Function}
Let $\bf{\pi}=\pi(\bf{a})$ be a vector function between two length-$N$ vectors $\bf{\pi}=\{\pi_n\}$ and $\bf{a}=\{a_n\}$. We say that $\pi$ is separable if we can write $\pi_n=\pi_n(a_n), \forall n$. We say that $\pi$ is separable-IID if each $\pi_n$ is IID drawn from an ensemble of random scalar functions. (See \ref{Sec:sep_fun} for more explanations.) 

\subsection{Assumptions}
Throughout this paper, unless stated otherwise, we will assume that (i) the length of a vector is $N$ and (ii) $\bf{\mathcal{X}}$ and  $\bf{x}$ are normalized, i.e., $\frac{1}{N}\mr{E}\{\|\bf{\mathcal{X}}\|^2\}=\frac{1}{N}\mr{E}\{\|{\bf{x}}\|^2\}=1$. Furthermore, we assume that (i) the entries of $\bf{x}$ are IID, (ii) the estimation functions $\gamma_t$ and $\phi_t$ are separable-IID, and (iii) the size of $\bf{V}$ is infinity, i.e., $N\to\infty$. Hence $\bf{\mathcal{X}}$ is PIIDG.

\subsection{Orthogonality under Law of Large Numbers (LLN)}

\begin{definition}[LLN Orthogonal]
 Let $\bf{a}=\{a_n\}$ and $\bf{b}=\{b_n\}$ be two sequences of length $N$, $\mr{E}\{\|\bf{a}\|^2\}\neq 0$,  $\mr{E}\{\|\bf{b}\|^2\}\neq 0$, $ \mr{E}\{a_n b_n\}=0$ and  $ \mr{Var}\{ a_n b_n\}$ finite. For any fixed $\delta>0$ and $\varepsilon>0$, if there is a fixed $N'$ such that
\BE\label{Eqn:chev_IID}
\mr{Pr}\left( \frac{(\bf{a}^{\rm T}\bf{b})^2}{\mr{E}\big\{\|\bf{a}\|^2\big\}\mr{E}\big\{\|\bf{b}\|^2\big\}}<\varepsilon\right) \ge 1-\delta, \;\; \mr{for} \;\; N>N',
\EE
we then say that  $\bf{a}$ and $\bf{b}$  are LLN-orthogonal, denoted as
\BE\label{Eqn:LLN_orth}
 \tfrac{1}{N}\bf{a}^{\rm T}\bf{b} \overset{\rm LLN}{\longrightarrow}0.
\EE
%where LLN indicates the convergence under the law of large numbers. 
\end{definition}

In the sequel, we will use \eqref{Eqn:chev_IID} as a measure for the convergence of $\bf{a}^{\rm T}\bf{b}$. Eq. \eqref{Eqn:LLN_orth}  implies the following, 
\BE\label{Eqn:LLN_ll}
\!\!\mr{E}\{\|\bf{a}\|^2\}\mr{E}\{\|\bf{b}\|^2\} \!\gg\! (\bf{a}^{\rm T}\bf{b})^2 \;\; {\rm in\; probability\; as}\; N\!\to\!\infty.  
\EE
This is useful for our later discussions. Note that  $\mr{E}\{\bf{a}^{\rm T}\bf{b}\}=0 $ alone cannot guarantee \eqref{Eqn:LLN_orth} when $\bf{a}$ and/or $\bf{b}$ are not IID. For example, consider: (i) ${\rm Pr}(a_n = +1) = {\rm Pr}(a_n = -1) = 0.5$ and all $\{a_n\}$ are equal in each trial (i.e., fully correlated), (ii) $\{b_n\}$ follow the same distribution as $\{a_n\}$, and (iii) $\{a_n\}$ and $\{b_n\}$ are independent of each other. Then $\mr{E}\{\bf{a}^{\rm T}\bf{b}\}=0 $  but $\mr{E}\{\|\bf{a}\|^2\|\bf{b}\|^2\}= (\bf{a}^{\rm T}\bf{b})^2=N^2$, so \eqref{Eqn:LLN_ll} does not hold.

\section{ OAMP Principles}\label{Sec:OAMP_Principle}
\LL{In this section, we define OAMP using an orthogonal constraint, based on which we derive the state evolution procedure for the analysis of OAMP.} We will discuss the implementation of the orthogonal constraint in the next section.
  %We mostly focus separable functions in this section.  We will discuss the implementation of the orthogonal constraint in the next section. %The extensions to non-separable functions will be discussed in  Appendix \ref{APP:orth}.

%\subsection{System Model}

AMP algorithms are analysed in a random framework \cite{Bayati2011}: their ``average behaviour'' is found by taking expectations with respect to the Gaussian-distributed matrix $\bf{A}$ in \eqref{Eqn:linear_system}. In this paper, we analyze OAMP in a similar way: the average behaviour of OAMP is studied with respect to the $N \times N$ Haar-distributed matrix $\bf{V}$. Precisely, we study the average behaviour of a sequence of trials, where each trial involves choosing $\bf{x}$ and $\bf{V}$ at random, generating the ``observations'' that determine the constraints $\Gamma$ and $\Phi$, running the OAMP algorithm, and recording the squared-error. Since the analysis studies average behaviour, simulation studies are needed to ascertain whether a specific system exhibits the predicted average behaviour. This is true of all AMP-type algorithms. The primary use of the analysis is for designing AMP-type algorithms: appropriate if not optimal messages to pass are derived under the assumption that the average behaviour holds.

\subsection{Gram-Schmidt (GS) Model}\label{Sec:SGM}
%\JMC{Link to the innovations process, explaining similarities and differences.}
Let $\hat{\bf{x}}$ be an arbitrary observation of $\bf{x}$. We  generate a new vector $\bf{\xi} = \hat{\bf{x}} -\alpha \bf{x}$ using GS orthogonalization \cite{Schmidt1908}, where 
\BE\label{Eqn:GP_modelb}
\alpha= \tfrac{1}{N}\mr{E}\big\{\hat{\bf{x}}^{\rm T}{\bf{x}}\big\}
\EE
 is a scalar. We will call \eqref{Eqn:GP_model} below the GS model of $ \hat{\bf{x}}$ with respect to $\bf{x}$:
\BE\label{Eqn:GP_model}
 \hat{\bf{x}} = \alpha \bf{x} + \bf{\xi}.
\EE 
We treat $\bf{\xi}$ as an error term, referred to as GS error, which is different from the common definition of an error $\hat{\bf{x}}-\bf{x}$. Its average entry-wise power is
\BE\label{Eqn:GS_v}
v=  \tfrac{1}{N}\mr{E}\big\{\|\bf{\xi}\|^2\big\}=\tfrac{1}{N}\mr{E}\big\{\bf{\xi}^{\rm T}\bf{\xi} \big\}.
\EE 
 It can be verified that $\bf{\xi}$ is orthogonal to $\bf{x}$, i.e.
\BE\label{Eqn:GP_modelc}
    \mr{E}\big\{\bf{x}^{\rm T} \bf{\xi}\big\} = 0.
\EE

%In general, \eqref{Eqn:GP_modelc} does not ensure  $\bf{x}^{\rm T}\bf{\xi} \overset{\rm LLN}{\longrightarrow}0$ if $\bf{x}$ and $\hat{\bf{x}}$ are not respectively IID. 

{\color{black}\textit{Normalized Model}: When $\alpha=1$, i.e., $\hat{\bf{x}} = \bf{x} + \bf{\xi}$, we call \eqref{Eqn:GP_model} the \textit{normalized model}. %Typically, the output messages of LE in AMP-type algorithms are normalized.

\textit{MMSE Model}: When $v =\alpha(1-\alpha)$, we call \eqref{Eqn:GP_model} the MMSE model.
In this case, ${\hat {\bm x}}={\rm E}\{\bf{x}|{\hat {\bm x}}\}$ is the MMSE solution. 
%${\rm E}\{{\bm x}|\hat{\bm x}\}=\hat {\bm x}$ the GS model reduces into the MMSE model $\bf{x} = \hat{\bf{x}} + \bf{e}$ with  $\mr{E}\big\{\hat{\bf{x}} ^{\rm T} \bf{e}\big\} = 0$. %Typically, the output messages of NLE in Bayes-optimal AMP-type algorithms are MMSE.

In the existing AMP-type algorithms, the MMSE-NLE is MMSE modeling and their LES are all normalized, which are unified under the GS model. Apart from that, the GS model was used to circumvent the difficulty in the achievable rate analysis and capacity-optimality proof of OAMP \cite{OAMP_ISIT22, OAMP_TCOM}.}

\subsection{Generic Iterative Process}\label{Sec:GIP_ER}  
Return to the system: $\bf{\mathcal{X}} = \bf{V x}$, $\bf{x}\sim\Phi$ and $\bf{\mathcal{X}}\sim\Gamma$.  Our aim is to use the AMP-type iterative approach in Fig.~\ref{Fig:model}(b) to find the \textit{a posteriori} mean of $\bf{x}$,
\BE\label{Eqn:post_mean}
\hat{\bf{x}} \equiv {\mr{E}}\{\bf{x}|\bf{\mathcal{X}}=\bf{V}\bf{x}, {\Gamma}, {\Phi} \},
\EE
which minimises the mean-square error $ \tfrac{1}{N}{\mr{E}}\big\{\|\hat{\bf{x}}-{\bf{x}}\|^2\big\}$. Similarly, we define ${\bf{\mathcal{\hat{X}}}}$. 

We add an iteration index $t$ to the estimates and formally define the iterative process in Fig.~\ref{Fig:model}(b) as follows. Note that, given $\bf{V}$, and $\{\gamma_t\}$ and $\{\phi_t\}$, the messages $\{\bf{\mathcal{\hat{X}}}^{\mr{in}}_t, \hat{\bf{x}}_t^{\mr{in}}, \bf{\mathcal{\hat{X}}}_t^{\mr{out}}, \hat{\bf{x}}_t^{\mr{out}}, \forall t\}$ in \eqref{Eqn:Ite_Proc_para} are completely determined by initial values.  %\footnote{This is a parallel schedule, as opposed to the serial one discussed in \cite{Ma2016}. The parallel schedule is more convenient for analysis while the serial one is more efficient for simulation. The parallel process can be decomposed into two separate serial processes with initialization $\bf{\mathcal{\hat{X}}}^{\mr{in}}_0 \!=\!\bf{0}$ and $\bf{{\hat{x}}}^{\mr{in}}_0\! =\! \bf{0}$, respectively. Define $\bf{\Xi}_t^{\mr{S}}\equiv[\bf{\mathcal{\hat{X}}}_{t}^{\mr{in}}, \bf{\mathcal{\hat{X}}}_{t}^{\mr{out}}, \bf{{\hat{x}}}_{t}^{\mr{in}}, \bf{{\hat{x}}}_{t+1}^{\mr{out}}]$ for the estimations in serial schedule at $t^{\mr{th}}$ iteration in GIP, and $\bf{\Xi}_t^\mr{P}\equiv[\bf{\mathcal{\hat{X}}}_{t}^{\mr{in}},\: \bf{\mathcal{\hat{X}}}_{t+1}^{\mr{out}}, \:\bf{{\hat{x}}}_{t}^{\mr{in}},\bf{{\hat{x}}}_{t+1}^{\mr{out}}]$ the estimations in the parallel schedule. We have $\{\bf{\Xi}_{t}^{\mr{S}}=\bf{\Xi}_{2t}^{\mr{P}}, t=1,2,\cdots\}$. Hence the two schedules are actually equivalent. %See \cite{OMAP_III} for details.}

\textit{Generic Iterative Process (GIP):} \LL{Initializing from arbitrary $\{\bf{\mathcal{\hat{X}}}^{\mr{in}}_0, \bf{{\hat{x}}}^{\mr{in}}_0\}$ and  $t\!=\!1$},%\vspace{-0.1cm}
\BS\label{Eqn:Ite_Proc_para}\begin{alignat}{2}
&\bf{\mathcal{\hat{X}}}_{t}^{\mr{out}}\!=\! \gamma_t\big(\bf{\mathcal{\hat{X}}}^{\mr{in}}_{t-1}\big), &  &  \hat{\bf{x}}^{\mr{out}}_{t} \!=\! \phi_t\big(\hat{\bf{x}}^{\mr{in}}_{t-1}\big),\label{Eqn:Ite_Proc_para1}\\
%& \bf{\Theta}_{t}=\gamma_{\rm SE}(\bf{\Theta}_{t-1}),  & \qquad&  \bf{\theta}_{t}=\phi_{\rm SE}(\bf{\theta}_{t-1}),\label{Eqn:Ite_Proc_para2}\\
& \hat{\bf{x}}_{t}^{\mr{in}}\!=\!\bf{V}^{\rm T} \!\bf{\mathcal{\hat{X}}}^{\mr{out}}_{t},  & \qquad& \bf{\mathcal{\hat{X}}}_{t}^{\mr{in}} \!=\!\bf{V}\hat{\bf{x}}^{\mr{out}}_{t}. \label{Eqn:Ite_Proc_para3}
\end{alignat}\ES  
%\bf{\mathcal{\hat{X}}}_{t}^{\mr{out}}\!=\! \gamma\big(\bf{\mathcal{\hat{X}}}^{\mr{in}}_{t-1},\bf{\Theta}_{t-1}\big), &  &  \hat{\bf{x}}^{\mr{out}}_{t} \!=\! \phi\big(\hat{\bf{x}}^{\mr{in}}_{t-1},\bf{\theta}_{t-1}\big),\label{Eqn:Ite_Proc_para1}\\ 

\LL{The trivial initialization $\bf{\mathcal{\hat{X}}}^{\mr{in}}_0 \!= \bf{{\hat{x}}}^{\mr{in}}_0\! =\! \bf{0}$ are generally adopted. In this case, the iterative process is actually kickstarted using the observations within $\Gamma$ and $\Phi$.}

\LL{In \eqref{Eqn:Ite_Proc_para1}, $\gamma_t\big(\bf{\mathcal{\hat{X}}}^{\mr{in}}_{t-1}\big)$ and $\phi_t\big(\bf{\hat{x}}^{\mr{in}}_{t-1}\big)$
respectively generate refined estimates of $\bf{x}$ and $\bf{\mathcal{X}}$. Proper statistical models of $\bf{\mathcal{\hat{X}}}^{\mr{in}}_{t-1}$ and $\bf{\hat{x}}^{\mr{in}}_{t-1}$ are required for the design of $\gamma_t $ and $\phi_t $, respectively. Tracking such models is in general a prohibitively difficult task due to the correlation problem. We will resolve this difficulty in \ref{Sec:OAMP} by introducing an orthogonal principle, and hence define OAMP.}

We call \eqref{Eqn:Ite_Proc_para} a parallel algorithm. Tracking \eqref{Eqn:Ite_Proc_para}, we obtain two processes as follows:\vspace{-2mm}
\BS\label{Eqn:P2S}\begin{align}
   & \bf{\mathcal{\hat{X}}}^{\mr{in}}_0 \to \bf{\mathcal{\hat{X}}}^{\mr{out}}_1 \to \bf{{\hat{x}}}^{\mr{in}}_1 \to \bf{{\hat{x}}}^{\mr{out}}_2 \to \bf{\mathcal{\hat{X}}}^{\mr{in}}_2 \to\bf{\mathcal{\hat{X}}}^{\mr{out}}_3 \to \cdots, \label{Eqn:P2S1}  \\
   & \bf{{\hat{x}}}^{\mr{in}}_0 \to \bf{{\hat{x}}}^{\mr{out}}_1 \to\bf{\mathcal{\hat{X}}}^{\mr{in}}_1 \to\bf{\mathcal{\hat{X}}}^{\mr{out}}_2 \to \bf{{\hat{x}}}^{\mr{in}}_2 \to\bf{{\hat{x}}}^{\mr{out}}_3\to \cdots.\label{Eqn:P2S2}
\end{align}\ES
There are no common variables (considering both superscripts and subscripts) in these two processes, so they are uncoupled. We call \eqref{Eqn:P2S1}  and \eqref{Eqn:P2S2} two serial algorithms. They are initialized by $\bf{\mathcal{\hat{X}}}^{\mr{in}}_0 \! =\! \bf{0}$ and  $\bf{{\hat{x}}}^{\mr{in}}_0\! =\! \bf{0}$, respectively. It can be verified that the EP/OAMP/VAMP algorithm discussed in \cite{Takeuchi2017} is the one in \eqref{Eqn:P2S1}. Hence, the parallel and serial algorithms are equivalent for the purpose of analysis, and so the results from one can be applied to another. On the other hand, only one of the two serial algorithms is necessary for implementation. Later we will see that \eqref{Eqn:P2S} leads to more concise analysis due to its symmetry.

\subsection{GS Errors in GIP}\label{Sec:GSE}

Note that OAMP defined above is symmetric since $\gamma_t$ and $\phi_t$ in \eqref{Eqn:Ite_Proc_para} are under the same orthogonal constraint. Hence, in Theorem \ref{THE:IIDG} below, we only need to analyze one local estimator and the result is applicable to the other one.   

Let the messages in \eqref{Eqn:Ite_Proc_para} be expressed in their GS models: 
\BS\label{Eqn:errors}\begin{alignat}{2} 
& \bf{\mathcal{\hat{X}}}_{t}^{\mr{out}} = \alpha_{\gamma_t}^{\mr{out}}\bf{\mathcal{X}}+\bf{g}_{t}^{\mr{out}},&&  \hat{\bf{x}}_{t}^{\mr{out}} = \alpha_{\phi_t}^{\mr{out}}\bf{x}+\bf{f}_{t}^{\mr{out}},\label{Eqn:errorsa}\\
&\bf{\mathcal{\hat{X}}}^{\mr{in}}_{t} = \alpha_{\gamma_{t+1}}^{\mr{in}}\bf{\mathcal{X}}+{\bf{g}}^{\mr{in}}_{t}, &\qquad&\hat{\bf{x}}^{\mr{in}}_{t} = \alpha_{\phi_{t+1}}^{\mr{in}}\bf{x}+{\bf{f}}^{\mr{in}}_{t}.\label{Eqn:errorsb}
\end{alignat}\ES 
Let the average powers of the GS errors ${\bf{g}}^{\mr{out}}_{t}$,  ${\bf{g}}^{\mr{in}}_{t}$,  ${\bf{f}}^{\mr{out}}_{t}$ and ${\bf{f}}^{\mr{in}}_{t}$ be $v^{\rm out}_{\gamma_t}, v^{\rm in}_{\gamma_{t+1}}, v^{\rm out}_{\phi_t}$ and $v^{\rm in}_{\phi_{t+1}}$, respectively. {{Since $\bm V$ is orthogonal, combining \eqref{Eqn:Ite_Proc_para3} and \eqref{Eqn:errors}}}, then the following relationships hold:\vspace{-3mm}
\BS\label{Eqn:alpha_SE}\begin{align} 
\alpha_{t}^{\gamma}&=\alpha_{\gamma_t}^{\mr{out}}=\alpha_{\phi_{t+1}}^{\mr{in}},&
  \alpha_{t}^{\phi}&=\alpha_{\phi_t}^{\mr{out}}=\alpha_{\gamma_{t+1}}^{\mr{in}},\label{Eqn:alpha_SEa}\\ 
 v_{t}^{\gamma} &=v^{\rm out}_{\gamma_t}=v^{\rm in}_{\phi_{t+1}}, & v_{t}^{\phi}&=v^{\rm out}_{\phi_t}=v^{\rm in}_{\gamma_{t+1}}.\label{Eqn:alpha_SEb}
\end{align}\ES
We will use \eqref{Eqn:alpha_SE} to simplify the derivation of SE in \ref{Sec:SE}.

% Define 
% \BS\label{Eqn:theta_def}\begin{align} 
% \bf{\Theta}_{t} & \equiv (\alpha_{\gamma_t}, v_{\gamma_t}), & \bf{\theta}_{t}  & \equiv (\alpha_{\phi_t}, v_{\phi_t}),\\
%   \bf{\Theta}^{\rm out}_{t} & \equiv (\alpha_{\gamma_t}^{\mr{out}}, v^{\rm out}_{\gamma_t}), & \bf{\theta}^{\rm out}_{t} & \equiv (\alpha_{\phi_t}^{\mr{out}}, v^{\rm out}_{\phi_t}), \\
%      \bf{\Theta}^{\rm in}_{t} &\equiv (\alpha_{\gamma_{t+1}}^{\mr{in}}, v_{\gamma_{t+1}}^{\mr{in}}), & \bf{\theta}^{\rm in}_{t} &\equiv (\alpha_{\phi_{t+1}}^{\mr{in}}, v^{\rm in}_{\phi_{t+1}}).
% \end{align}\ES
% From \eqref{Eqn:alpha_SE} and \eqref{Eqn:theta_def}, we have
% \BE
%   \bf{\Theta}_{t} = \bf{\Theta}^{\rm out}_{t} = \bf{\theta}^{\rm in}_{t}, \qquad  \bf{\theta}_{t} = \bf{\theta}^{\rm out}_{t} = \bf{\Theta}^{\rm in}_{t}.
% \EE 
% Therefore,  \eqref{Eqn:fun_theta} can be simplified to
% \BS\label{Eqn:fun_theta2}\begin{align}
%  \gamma_t\big(\bf{\mathcal{\hat{X}}}^{\mr{in}}_{t-1}\big) &= \gamma\big(\bf{\mathcal{\hat{X}}}^{\mr{in}}_{t-1},\bf{\theta}_{t-1}\big),  \\ 
%  \phi_t\big(\hat{\bf{x}}^{\mr{in}}_{t-1}\big) &= \phi\big(\hat{\bf{x}}^{\mr{in}}_{t-1},\bf{\Theta}_{t-1}\big). 
% \end{align}\ES
% In Subsection \ref{Sec:SE}, we will discuss the update of $\bf{\theta}_t$ and $\bf{\Theta}_t$ using a state-evolution technique.

\LL{Let $\bf{g}^{\mr{out}}_0=\bf{V}\bf{f}^{\mr{in}}_0$ and $\bf{f}^{\mr{out}}_0 = \bf{V}^{\rm T}\bf{g}^{\mr{in}}_0$.} The following matrices contain the GS errors in \eqref{Eqn:errors} up to iteration $t$:
\BS\label{Eqn:Error_matrix}\begin{alignat}{2}
&\!\!\! \bf{G}^{\mr{in}}_t \equiv \left[\bf{g}_0^{\mr{in}},\dots,\bf{g}_{t-1}^{\mr{in}}\right], &\quad\;\;& \bf{F}^{\mr{in}}_t \equiv \left[\bf{f}_0^{\mr{in}},\dots,\bf{f}_{t-1}^{\mr{in}}\right], \\
& \!\!\!\bf{G}^{\mr{out}}_t \!\equiv\! \left[\bf{g}_0^{\mr{out}}\!,\dots,\bf{g}_{t-1}^{\mr{out}}\right], && \bf{F}^{\mr{out}}_t \!\equiv\! \left[\bf{f}_0^{\mr{out}}\!,\dots,\bf{f}_{t-1}^{\mr{out}}\right].
\end{alignat}
Denote
\BE 
\bf{A}_t \equiv \left[\bf{\mathcal{X}}, \bf{G}^{\mr{in}}_t, \bf{G}^{\mr{out}}_t\right],   \quad \bf{B}_t \equiv \left[ \bf{x}, \bf{F}^{\mr{out}}_t, \bf{F}^{\mr{in}}_t\right].\label{Eqn:Error_matrixc}
\EE\ES
Combining \eqref{Eqn:Ite_Proc_para}, \eqref{Eqn:errors} and \eqref{Eqn:Error_matrix}, we have 
\BE\label{Eqn:G=VF}
  \bf{A}_t =\bf{V} \bf{B}_t.
\EE
We will discuss the behavior of GS errors in \ref{Sec:IID_error} based on the constraint in \eqref{Eqn:G=VF}.  
 
\subsection{Orthogonal AMP (OAMP)}\label{Sec:OAMP}
\begin{definition}\label{Def:OAMP}
 OAMP is a special case of the GIP in \ref{Sec:GIP_ER} when the following orthogonal constraint holds for $N\to \infty$, $t\ge 1$ and $0\le t'<t$,
\BS\label{Eqn:Orthogonality}\begin{alignat}{2}
\tfrac{1}{N}\left(\bf{g}^{\mr{in}}_{t'}\right)^{\rm T}\! \bf{g}^{\mr{out}}_{t} & \overset{\rm LLN}{\longrightarrow} {0},  \qquad &  \tfrac{1}{N}\left(\bf{f}^{\mr{in}}_{t'}\right)^{\rm T}\! \bf{f}^{\mr{out}}_{t} & \overset{\rm LLN}{\longrightarrow} {0}, \;\, \label{Eqn:Orthogonalitya}\\
\tfrac{1}{N}\bf{\mathcal{X}}^{\rm T} \bf{g}^{\mr{out}}_{t}  & \overset{\rm LLN}{\longrightarrow}  {0},   &  \tfrac{1}{N}\bf{x}^{\rm T} \bf{f}^{\mr{out}}_{t}& \overset{\rm LLN}{\longrightarrow}  {0}.\label{Eqn:Orthogonalityb}
\end{alignat}\ES
\end{definition}

In words, in OAMP, the output errors of $\gamma_t$ and $\phi_t$ in \eqref{Eqn:Ite_Proc_para} are LLN-orthogonal to their respective current and previous input errors, as well as to $\bf{\mathcal{X}}$ and $\bf{x}$. The orthogonality in \eqref{Eqn:Orthogonality} is the key to solve the correlation problem, as will be discussed below. 

Here are some intuitions. Return to the GIP in Fig.~\ref{Fig:model}(b) and focus on $\phi_t $. Notice the following:
\begin{itemize}
    \item The error $\bf{f}_{t-1}^{\rm in}$ into $\phi_{t} $  comes from the output error of $\gamma_{t-1} $ in the previous iteration.
    \item The output error $\bf{f}_t^{\rm out}$ of $\phi_{t} $  becomes the error into  $\gamma_{t+1} $ in the next iteration.
\end{itemize} 
The correlation between $\bf{f}_{t-1}^{\rm in}$ and $\bf{f}_t^{\rm out}$ implies error circulation from the output of $\gamma$ back to its input, which may lead to positive feedback and instability.  In general, we should avoid this effect. This gives a heuristic reason for the orthogonality requirement between  $\bf{f}_t^{\rm out}$ and $\bf{f}_{t-1}^{\rm in}$. 

The LLN-orthogonality in \eqref{Eqn:Orthogonalitya} is between $\bf{f}_t^{\rm out}$ and all $\bf{f}_{t'}^{\rm in}$, $t'=1, 2, \dots t-1$, which is referred to as global orthogonality. %If all $\{\bf{f}_{t'}^{\rm in}\}$ are IIDG, the global orthogonality follows from the generalized Stein's Lemma \cite{Stein1981} directly. However, as we will see later, $\{\bf{f}_{t'}^{\rm in}\}$ are not strictly IIDG.  Hence we will examine this issue more carefully.

\LL{Definition \ref{Def:OAMP} is a more general definition of OAMP. Existing AMP-type algorithms are restricted to particular orthogonality realizations, which are sufficient but not necessary conditions of \eqref{Eqn:Orthogonality}.  We will discuss the specific  realizations of \eqref{Eqn:Orthogonality} in \ref{Sec:GS_orth}.} %The specific techniques to realize \eqref{Eqn:Orthogonality} will be discussed in  \ref{Sec:GS_orth}.

\subsection{Average MSE Performance}\label{Sec:aver_MSE}
Return to the original problem of estimating $\bf{x}$ after $t$ iterations in \eqref{Eqn:Ite_Proc_para}. We adopt the following method: 
\begin{enumerate}[(i)]
       \item	We track the GS models in \eqref{Eqn:GP_model} recursively for $t'=1, 2, \dots t$ and obtain (see \eqref{Eqn:errorsa} and \eqref{Eqn:alpha_SEa}):
       \BE\label{Eqn:GS_mod1}
         \hat{\bf{x}}_{t}^{\mr{out}} = \alpha_{t}^\phi \bf{x}+\bf{f}_{t}^{\mr{out}}.
       \EE
    \item We treat $\bf{f}_{t}^{\mr{out}}$ above as a pure error. We use $\omega\hat{\bf{x}}_{t}^{\mr{out}}$ as an estimate of $\bf{x}$, where $\omega$ is a scaling factor. Based on the above discussions, we have
    \BE\label{Eqn:MSE}
        \mr{MSE} \equiv\tfrac{1}{N}\!\mathop\mr{E}\limits_{\bf{r},\bf{x} }\big\{\|\omega \hat{\bf{x}}_{t}^{\mr{out}} - \bf{x}\|^2\big\} = v_{t}^{\phi}/\big((\alpha_{t}^{\phi})^2+v_{t}^{\phi}\big),
    \EE
   where $\omega=\alpha_{t}^{\phi}/\big((\alpha_{t}^{\phi})^2+v_{t}^{\phi}\big)$ minimizes the mean square error (MSE) and $v_{t}^{\phi}$ is defined in \eqref{Eqn:alpha_SEb}. The expectation in \eqref{Eqn:MSE} is over the distribution of the observation $\bf{r}$. The latter is in turn determined by the distributions of $\bf{x}$ and possible distortions in  $\bf{r}$. In particular, if the distortion is an additive noise vector $\bf{\eta}$, then the distribution of $\bf{r}$ is jointly determined by $\bf{x}$ and $\bf{\eta}$. 
   
   %An example of  $\bf{r}$ can be seen in \eqref{Eqn:example_1}. In this case,  $\bf{r}$ is a parameter in $\gamma_t $.  See Section \ref{Sec:alg_opt} for more details. 
\end{enumerate}

The GS model in \eqref{Eqn:GS_mod1} is not an optimal estimation of $\bf{x}$. The GS error $\bf{f}_{t}^{\mr{out}}$  may actually contain useful information about $\bf{x}$, so it is possible to compute a lower MSE as that in \eqref{Eqn:MSE}, but the related cost can be high in iterative processing. The approach based on GS models has a distinguished advantage of low cost. It can be seen from \eqref{Eqn:alpha_SE} that the parameters of the GS models are not affected by transforms involving $\bf{V}$ and $\bf{V}^{\rm T}$. This makes it computationally efficient. Incidentally, the above method may still lead to global optimality in OAMP. See \ref{Sec:Opt_OAMP} for more details.

In general, evaluating \eqref{Eqn:MSE} is still a difficult task.  To overcome this difficulty, we borrow a technique from \cite{Donoho2009}: we redefine the MSE as follows:
\BE\label{Eqn:MSE_ave}
        \mr{MSE} \equiv \tfrac{1}{N}\!\!\mathop\mr{E}\limits_{\bf{r},\bf{x},\bf{V} }\big\{\|\omega \hat{\bf{x}}_{t}^{\mr{out}} - \bf{x}\|^2\big\}.
    \EE
In \eqref{Eqn:MSE_ave}, we compute the GS model of $\hat{\bf{x}}_{t}^{\mr{out}}$ using the joint distribution of $\bf{r}$ and $\bf{V}$. We study the average behavior over a sequence of trials, where each trial involves choosing $\bf{r}$ and $\bf{V}$ at random. (See the second paragraph of this section.)
% running the algorithm, and recording the MSE in \eqref{Eqn:MSE_ave}.  (See Subsection \ref{Sec:IID_error} below.) Clearly, simulation studies are needed to ascertain whether a specific $\bf{V}$ exhibits the predicted average behavior. This approach is employed in almost all AMP-type algorithms. As we will see below, averaging over $\bf{V}$ greatly simplifies the problem. 

\subsection{Error Behavior}\label{Sec:IID_error}
We now consider the details in tracking the GS models in \eqref{Eqn:errors}. As an example, consider the GS model $\hat{\bf{x}}_{t}^{\mr{out}} = \alpha_{t}^\phi\bf{x}+\bf{f}_{t}^{\mr{out}}$. From  \eqref{Eqn:GP_modelb}, \eqref{Eqn:Ite_Proc_para}, \eqref{Eqn:errors} and \eqref{Eqn:alpha_SE}, we generate $\alpha_{t}^\phi$ as follows
\BE\label{Eqn:alpha_gama}
\alpha^\phi_{t} =\tfrac{1}{N}\!\!\mathop\mr{E}\limits_{\bf{r},\bf{x},\bf{V} }\Big\{\phi_{t}(\alpha^\gamma_{t-1} \bf{x} + \bf{f}_{t-1}^{\rm in})^{\rm T} \bf{x} \Big\}.
\EE
where $\alpha^\gamma_{t-1} \bf{x} + \bf{f}_{t-1}^{\rm in}$ is the GS model for $\hat{\bf{x}}_{t-1}^{\mr{in}}$. Assume that $\alpha^\gamma_{t-1} $ is known. In the following, we give the distribution of $\bf{f}_{t-1}^{\rm in}$ in order to evaluate \eqref{Eqn:alpha_gama}. 
  
%Without loss of generality, assume that OAMP start at $t=1$ with initialization (in GS models)  $\bf{\mathcal{\hat{X}}}^{\mr{in}}_{0} =\bf{0}= 0\cdot\bf{\mathcal{X}}+\bf{0}$  and $\bf{ \hat{x}}^{\mr{in}}_{0} =\bf{0}= 0\cdot\bf{x}+\bf{0}$, with GS errors  $\bf{g}^{\mr{in}}_0= \bf{f}^{\mr{in}}_0 = \bf{0}$. Such initialization in general carries no information and is useless. The iterative process actually kick-starts with the observations contained in $\Gamma$ and $\Phi$, which are implicitly used in $\gamma$ and $\phi$.   

Theorem \ref{THE:IIDG} is the main result of this paper. For convenience, we adopt the following notations. \LLC{A matrix is said column-wise PIIDG and row-wise joint-Gaussian (CPIIDG-RJG) if each column is PIIDG and each row is joint Gaussian.} Note two subtle points here: (i) as mentioned for \eqref{Eqn:MSE_ave}, $\bf{V}$ is randomly selected for each trial, and (ii) $\bf{V}$ remains unchanged in all iterations during one iterative process. In other words, $\bf{V}$ is randomly selected in Fig.~\ref{Fig:model}(a) but, once selected, it remains unchanged throughout an iterative process in Fig.~\ref{Fig:model}(b). Point (ii) means that, for $t>0$, $\bf{V}$ is constrained by $\bf{A}_t=\bf{VB}_t$ in \eqref{Eqn:G=VF}. {\color{black} This is referred to as the Bolthausen’s conditioning problem in \cite{Bolthausen2014, Rangan2016, Bayati2011, Takeuchi2019}.} Taking these points into consideration, we prove Theorem \ref{THE:IIDG} in Appendix \ref{Sec:IIDG_proof}. %, which gives the distributions of $\bf{G}_{t}^{\rm in}$ and  $\bf{F}_{t}^{\rm in}$.

\begin{theorem}\label{THE:IIDG}
Assume that $\bf{V}$ is Haar distributed and OAMP is initialized at $t=1$ \LL{with  IIDG $\bf{g}^{\mr{in}}_0$ and $\bf{f}^{\mr{in}}_0 $ independent of $\bf{\mathcal{X}}$ and $\bf{x}$, respectively.} Then, when $t$ is finite, $N\to\infty$ and over the Haar ensemble of $\bf{V}$, the following hold for the errors in \eqref{Eqn:Error_matrix} for OAMP asymptotically:
\begin{enumerate}[(a)]
 \item   $\bf{G}_t^{\rm in}$ is \LLC{CPIIDG-RJG} and independent of  $\bf{\mathcal{X}}$ and any $\bf{z}$ provided that $\bf{z}$ is independent of $\bf{V}$;
 \item   $\bf{F}_t^{\rm in}$ is \LLC{CPIIDG-RJG} and independent of $\bf{x}$ and any $\bf{z}$ provided that $\bf{z}$ is independent of $\bf{V}$.
\end{enumerate} 
\end{theorem}

% \LL{Theorem \ref{THE:IIDG} applies to OAMP  with general initializations $ \bm{\mathcal{\hat{X}}}_{0}^{\mr{in}}$ and $\hat{\bm{x}}_{0}^{\mr{in}}$, which is useful when we have  non-trivial initializations (e.g., independent observations or estimates from another algorithm). The results in \cite{Takeuchi2017, Takeuchi2019} are limited to the trivial initialization $\bm{x}^{\rm in}_0=\bf{0}$, because the initial error $\bm{x}^{\rm in}_0-\bm{x}=-\bm{x}$ is required to satisfy the implicit orthogonalization between $\bm{G}_t^{\rm out}$ and $\bm{x}$. For general $\hat{\bm{x}}_{0}^{\mr{in}}$, the requirement of $\bm{x}^{\rm in}_0-\bm{x} = \alpha \bm{x}$ (where $\alpha\neq 0$) may not hold anymore, which will break the orthogonalization between $\bm{G}_t^{\rm out}$ and $\bm{x}$.}

%   \BE 
%         \tfrac{1}{N}\pi(\bf{N}, \bf{\Xi})^{\rm T} \varpi(\bf{N}, \bf{\Xi}) \overset{\rm P}{\simeq} {\rm E}\big\{\pi_n(\bf{n}_n^{\rm r}, \bf{z}_n^{\rm r}) \varpi_n(\bf{n}_n^{\rm r}, \bf{z}_n^{\rm r})\big\},
%   \EE

%   Let $T$ be a  positive integer, $\{\bf{a}_t,t=1,\dots, T\}$ be independent vectors with $\bf{a}_t$ being PIIDG over $S^N(\rho_t)$. Then for any deterministic separable-IID function $\pi: \mathbb{R}^{N\times T} \to \mathbb{R}^N$, we have
%   \BE\label{Eqn:ALG-SteinLemma}
%         \pi(\bf{a}_1,\dots, \bf{a}_T) \overset{\rm P}{\simeq} {\rm E}\{\pi(\bf{z}_1,\dots, \bf{z}_T)\},
%   \EE
%   where $\{\bf{z}_1,t=1,\dots, T\}$ are independent vectors with $\bf{z}_t\sim \mathcal{N}\big(0, \frac{\rho_t}{N}\bf{I}\big)$. 

Notes: (i) Theorem \ref{THE:IIDG} is defined over a Haar ensemble of $\bf{V}$.  It may not hold for a specific sample of $\bf{V}$. (ii) An example of $\bf{z}$ is additive thermal noise in the system. 

% We now return to GIP in \eqref{Eqn:Ite_Proc_para}. Using 
% Theorem \ref{THE:IIDG}, $\hat{\bf{x}}^{\mr{in}}_{t-1}$ is specified by its GS model $\hat{\bf{x}}^{\mr{in}}_{t-1}=\alpha_{t-1}^\gamma \bf{x} + \bf{f}_{t-1}^{\rm in}$. Hence the $\bf{\theta}_{t-1}^{\rm in}=\bf{\Theta}_{t-1}$ in \eqref{Eqn:phi_theta} consists of
% the GS parameters $\alpha_{t-1}^\gamma$ and $v_{t-1}^\gamma$. The local estimator $\phi$ in \eqref{Eqn:phi_theta} generates a refined estimate of $\bf{x}$ based on $\hat{\bf{x}}^{\mr{in}}_{t-1}$ , $\alpha_{t-1}^\gamma$ and $v_{t-1}^\gamma$. In the next subsection, we will discuss the update of $\bf{\theta}_t$ and $\bf{\Theta}_t$ using a state-evolution technique.

\subsection{State Evolution (SE) for OAMP}\label{Sec:SE} 
% Theorem \ref{THE:IIDG} states that the GS errors $\{\bf{f}_t^{\rm in}\}$ and $\{\bf{g}_t^{\rm in}\}$ are IID for OAMP. We now derive the SE formula for OAMP based on Theorem \ref{THE:IIDG}. Assume that $\phi_{t}$ and $\gamma_{t}$ are both separable-IID. Then using \eqref{Eqn:ALG-SteinLemma}, we can rewrite \eqref{Eqn:alpha_gama} as 
 
% where $z_n^\gamma\sim {\cal{N}}(0, 1)$. Note that, following Theorem \ref{THE:IIDG},  $\sqrt{v_{t-1}^{\gamma}}z_n^\gamma$ in \eqref{Eqn:exp_alp} has the same marginal distribution as $\bf{f}_{t-1}^{\rm in}$ and thus the distribution of $\bf{f}_{t-1}^{\rm in}$ in \eqref{Eqn:alpha_gama} is replaced by the distribution of $\sqrt{v_{t-1}^{\gamma}}z^\gamma_n$ in \eqref{Eqn:exp_alp}. 
 
Return to the error behavior of OAMP for given $\{\bf{f}_t^{\rm in}\}$ and $\{\bf{g}_t^{\rm in}\}$ in the asymptotic case of $N\to \infty$. For this purpose, we track the distributions of the messages in OAMP and evaluate MSEs accordingly, as detailed below. 

Theorem \ref{THE:IIDG} states that any separable-IID function of GS errors $\bf{F}_t^{\rm in}$ or $\bf{G}_t^{\rm in}$ in OAMP are asymptotically \LLC{CPIIDG-RJG}. We still need the GS parameters to determine the GS model of messages.
 
 Recall from \eqref{Eqn:alpha_SE} that the GS parameters are not affected by the transform by an orthogonal $\bf{V}$. This greatly simplifies the problem; we only need to focus on $\phi_t$ and $\gamma_t$. Assume that $\phi_t$ and $\gamma_t$ are both separable-IID. Let $z_n^\gamma\sim {\cal{N}}(0, 1)$. From  Theorem \ref{THE:IIDG}, $\bf{f}_{t-1}^{\rm in}$ is PIIDG. Hence following \LL{Conjecture \ref{Conj:E_Var} (see Appendix \ref{APP:In_Pro})}, we can rewrite \eqref{Eqn:alpha_gama} as 
\begin{align} 
\alpha^\phi_{t} &=  \mathop\mr{E}\limits_{x_n, z_n }\Big\{\phi_{t,n}\big(\alpha^\gamma_{t-1}x_n + \sqrt{v_{t-1}^{\gamma}}z_n^\gamma\big)\cdot  x_n \Big\},\label{Eqn:exp_alp}
\end{align}  
We can apply similar reasoning to generate  $\alpha_{t}^{\!\gamma},v_{t}^{\!\gamma},\alpha_{t}^{\!\phi}$ and  $v_{t}^{\!\phi}$ defined in \eqref{Eqn:errors} and \eqref{Eqn:alpha_SE}. We then obtain the following recursion for the GS parameters. 
\BS\label{Eqn:MSE_phi_IDDG}\begin{align}
 \alpha_{t}^{\!\gamma}  &= \!\!\mathop\mr{E}\limits_{r_n,z_n^{\mathcal{X}},z^\phi_n}\!\Big\{\gamma_{t,n}\big(\alpha_{t-1}^{\!\phi} z_n^{\mathcal{X}} \!+\! \sqrt{\! v^{\phi}_{t-1}}z^\phi_n\big) \cdot z_n^{\mathcal{X}} \Big\},\label{Eqn:MSE_phi_IDDG1}\\ 
 v^{\!\gamma}_{t} & =  \!\!  \mathop\mr{E}\limits_{r_n,z_n^{\mathcal{X}}, z^\phi_n}\!\Big\{\Big( \gamma_{t,n}\big(\alpha_{t-1}^{\!\phi}\! z_n^{\mathcal{X}}\!\!+\!\sqrt{\!v^{\phi}_{t-1}} {z}^\phi_n\big)\!-\!\alpha_{t}^{\!\gamma} z_n^{\mathcal{X}}\Big)^2\Big\},\label{Eqn:MSE_phi_IDDG2} \\
\alpha_{t}^{\!\phi} &= \!\! \mathop\mr{E}\limits_{x_n, z^\gamma_n}\Big\{  \phi_{t,n}\big(\alpha_{t-1}^{\!\gamma} {x}_n \!+\! \sqrt{\! v^{\!\gamma}_{t-1}} {z}^\gamma_n \big)\cdot  {{x}_n}\Big\},\\ 
v^{\phi}_{t}\! &= \!\! \mathop\mr{E}\limits_{x_n, z^\gamma_n}\! \Big\{ \Big( \phi_{t,n}\big(\alpha_{t-1}^{\!\gamma} {x}_n \!+\! \sqrt{\!v^{{\!\gamma}}_{t-1}} {z}^\gamma_n \big)\! -\!\alpha_{t}^{\!\phi} {x}_n \Big)^2 \Big\},
\end{align}\ES   
where $\{z_n^\phi,z_n^\gamma, z_n^{\mathcal{X}}\}$ are IID drawn from ${\cal{N}}(0, 1)$. The expectations are based on the assumption that $\gamma_{t,n}$ contains observation $r_n$ and there is no observation in $\phi_{t,n}$. Similar results can be obtained if $\phi_{t,n}$ also contains observation. The key here is replacing the GS errors using IIDG variables, which makes the problem trackable. %This is allowed for separable functions provided that the entry-wise marginal function remains the same. 
	
We re-write the functions in \eqref{Eqn:MSE_phi_IDDG} into a more concise form in \eqref{Eqn:SE} below, where $\gamma_{\mr{SE}}$ and $\phi_{\mr{SE}}$ may or may not have explicit expressions, but they can always be numerically tabulated.  	 

\textit{State Evolution for OAMP:}  Starting with $t=1$ and $\{\alpha_0^\phi, \alpha_0^\gamma, v_0^\phi, v_0^\gamma\}$,
\BS\label{Eqn:SE} \begin{align}
(\alpha_t^\gamma, v_t^\gamma)&= \gamma_{\mr{SE}}(\alpha_{t-1}^\phi, v_{t-1}^\phi),\\
(\alpha_t^\phi, v_t^\phi) &=\phi_{\mr{SE}}(\alpha_{t-1}^\gamma, v_{t-1}^\gamma). 
% \bf{\Theta}_{t}&= \gamma_{\mr{SE}}(\bf{\theta}_{t-1}),\\
% \bf{\theta}_{t} &=\phi_{\mr{SE}}(\bf{\Theta}_{t-1}). 
\end{align}\ES 
 
Consider the messages in \eqref{Eqn:errors} under their GS models. Assume that the distribution of $\bf{x}$ is given. Lemma \ref{Lem:PIIDG} and Theorem \ref{THE:IIDG} give the distributions of $\bf{\mathcal{X}}$ and the GS errors. Hence we can find the distributions of the messages using the GS parameters generated in \eqref{Eqn:SE}. This answers the question on how to generate the distributions in OAMP.

The performance of an estimation is not determined by a single GS parameter $\alpha$ or $v$. Instead, it is determined by their ratio $\alpha^2/v$, i.e., an effective SNR. As a result, the SE in \eqref{Eqn:SE} consists of dual-input and dual-output (DIDO) functions of GS model parameters. This is slightly different from the conventional single-input and single-output (SISO) SE functions for AMP-type algorithms \cite{Ma2016, Donoho2009}, but the spirits are the same. We can convert \eqref{Eqn:SE} into a SISO recursion via proper scaling so that $\alpha_{t}^{\!\gamma}= \alpha_{t}^{\!\phi}=1$ (scaling does not change the performance of an estimation). However, the singular situations of $\alpha_{t}^{\!\gamma}=0$  or $\alpha_{t}^{\!\phi}=0$ in the \LL{trivial} initialization\footnote{Such initialization ($\alpha=0$) carries no useful information of the true signal. However, once the GIP started, the estimations in general carries useful information of the true signal, i.e., $\alpha\neq 0$. In this case, we can use a normalized GS model (e.g., $\alpha=1$) determined by a single parameter $v$.} may cause problems in normalization.

Using \eqref{Eqn:exp_alp}-\eqref{Eqn:SE}, we can assess the MSE of OAMP via \eqref{Eqn:MSE_ave}. This provides a convenient tool for analysis and optimization. %(See \ref{Sec:alg_opt} for an example of optimization.)

\subsection{Distributions of \texorpdfstring{$\{\phi_n\}$}{TEXT}  and \texorpdfstring{$\{\gamma_n\}$}{TEXT}}\label{Sec:sep_fun}
In \eqref{Eqn:MSE_phi_IDDG} and \eqref{Eqn:SE}, we assume that $\{\phi_n\}$ and $\{\gamma_n\}$ are IID samples from their respective ensembles. We now clarify this assumption using two examples. 

\underline{{\textbf{Example:}}}   Consider $\gamma_n=\gamma, \forall n$. In this special case, the ensemble for $\{\gamma_n\}$ has only one element. 

Next, recall that \eqref{Eqn:MSE_phi_IDDG} and \eqref{Eqn:SE} are derived over the Haar ensemble of $\bf{V}$. Let $\bf{Q}$ be a permutation matrix. Clearly, $\bf{QV}$ is in the Haar ensemble if $\bf{V}$ is. Then we have Lemma \ref{Pro:random_gamma}.

\begin{lemma}\label{Pro:random_gamma}
Let $\bf{Q}$ be a permutation matrix. The SE for GIP in \eqref{Eqn:Ite_Proc_para} remains unchanged if $\bf{V}$ is replaced by $\bf{QV}$, or equivalently, $\gamma(\bf{\mathcal{\hat{X}}})$ is replaced by  $\tilde{\gamma}(\bf{\mathcal{\hat{X}}})=\bf{Q}^{\rm T}\gamma(\bf{Q\mathcal{\hat{X}}})$. 
\end{lemma}

\underline{{\textbf{Example:}}} Consider  ${\bf{r}}=\bf{D}\bf{\mathcal{X}} + \bf{\eta}$, where $\bf{D}$ is diagonal. We construct a separable $\gamma$ using the so-called LMMSE estimator (see (12) in \cite{Ma2016}):
\BE\label{Eqn:LMMSE_SVD}
\gamma_n(\mathcal{\hat{X}}_n) \!=\! \tfrac{N}{M}D_{nn}^2r_n\!+\!\left(1\!+\! \tfrac{N}{M}\tfrac{D_{nn}^2}{vD_{nn}^2+\sigma^2} \right)\!\mathcal{\hat{X}}_n, \; \forall n.
\EE
It can be verified that such $\gamma$ is orthogonal. We further consider a special case of $\bf{D}$ with its diagonal entries given by
\BE\label{Eqn:D}
\{D_{nn}\}=\{\underbrace{1,1,\dots, 1}_{M\; {\rm ones}}, \underbrace{0,0, \dots, 0}_{N-M\; {\rm zeros}}\}.
\EE
We cannot regard  $\{D_{nn}\}$ as a random sequence. However, from Lemma \ref{Pro:random_gamma}, we can randomly permute $\{D_{nn}\}$ without affecting the related SE. In this sense, when $M\to\infty, N\to\infty$ and ratio $M/N$ fixed, the entries in a randomly permuted $\{D_{nn}\}$ asymptotically follow a binary distribution $D_{nn}=1$ of probabilities $M/N$ and $D_{nn}=0$ of probabilities $(N-M)/N$.

In conclusion, the separable-IID assumption is applicable to $\gamma $ in \eqref{Eqn:LMMSE_SVD}. Incidentally, \eqref{Eqn:D} corresponds to $\bf{A}$ formed by the first $M$ rows in an $N\times N$ Haar distributed matrix. 

\subsection{A Brief Summary} 
The following is a summary on the discussions so far. 
\begin{itemize}
    \item We presented a message passing process named GIP (see Fig.~\ref{Fig:model}(b) and \eqref{Eqn:Ite_Proc_para}) to solve the problem in Fig.~\ref{Fig:model}(a). 
    \item We introduced the GS model to characterize the messages in GIP. The related GS parameters are not affected by transform by an orthogonal matrix $\bf{V}$, which facilitates a SE technique to track GS errors.
    \item We defined OAMP as an orthogonalized GIP.  
    \item We showed that the separable-IID function of GS errors in OAMP is asymptotically the same as that of PIIDG random vectors when $N \to \infty$. We developed a SE technique to track GS parameters in OAMP. 
\end{itemize} 
Hence, using SE, we can approximately determine the MSE  of OAMP for the system in Fig.~\ref{Fig:model}(a) when $N$ is large.

\section{Gram-Schmidt Orthogonalization}\label{Sec:GS_orth}
In this section, we discuss techniques to realize the orthogonality required in \eqref{Eqn:Orthogonality}. For simplicity, we may sometimes omit the subscript $t$.

\subsection{Gram-Schmidt Orthogonalization}\label{Sec:Bussgang_Orho}
\begin{definition}
 Consider the GS models:  $\bf{\pi}=\alpha^{\rm out}\bf{x}+\bf{\xi}^{\rm out}$ and $\hat{\bf{x}}=\alpha^{\rm in}\bf{x}+\bf{\xi}^{\rm in}$. We say that $\bf{\pi}=\pi(\hat{\bf{x}})$ is an orthogonal estimator if ${\rm E}\big\{\big(\bf{\xi}^{\rm in}\big)^{\rm T} \bf{\xi}^{\rm out}\big\}=0$ \cite{Schmidt1908}. 
\end{definition}

Let $\hat{\bf{\pi}}=\hat{\pi}(\hat{\bf{x}})$ be an arbitrary prototype. We construct an orthogonal $\pi(\hat{\bf{x}})$ as follows
\BE\label{Eqn:pi_orth}
\bf{\pi}=\pi(\hat{\bf{x}})=\hat{\pi}(\hat{\bf{x}})- B \hat{\bf{x}}.
\EE
Then we can rewrite the orthogonal requirement as
\BS\label{Eqn:GSO_ort}\begin{align}
{\rm E}\big\{(\bf{\xi}^{\mr{in}})^{\rm T}\bf{\xi}^{\rm out}\big\} & \mathop  = \limits^{({\rm{a}})}{\rm E}\big\{(\bf{\xi}^{\mr{in}})^{\rm T}(\bf{\pi}\!-\!\alpha^{\rm out}\bf{x})\big\}  \\ & \!\mathop  = \limits^{({\rm{b}})} \! {\rm E}\big\{(\bf{\xi}^{\mr{in}})^{\rm T}{\bf{\pi}}\big\}  \\ 
&\!\mathop  = \limits^{({\rm{c}})}   {\rm E}\big\{(\bf{\xi}^{\mr{in}})^{\rm T}\big(\hat{\pi}(\hat{\bf{x}})- B \hat{\bf{x}}\big) \big\}\! \!=\!0,
\end{align}\ES 
where  (a) follows the definition of $\bf{\xi}^{\rm out}$, (b) follows \eqref{Eqn:GP_modelc} for the GS model of $\hat{\bf{x}}$, and (c) due to \eqref{Eqn:pi_orth}. Noting that $  {\rm E}\big\{(\bf{\xi}^{\mr{in}})^{\rm T}\hat{\bf{x}}\big\} \! = \! {\rm E}\big\{(\bf{\xi}^{\mr{in}})^{\rm T}{(\alpha^{\mr{in}}\bf{x}+\bf{\xi}^{\mr{in}})} \big\} \!= \! {\rm E}\big\{\|{\bf{\xi}}^{\mr{in}}\|^2\big\} $ (recalling from \eqref{Eqn:GP_modelc} that ${\rm E}\big\{\bf{x}^{\rm T} \bf{\xi}^{\rm in}\big\} \!=\!0$ for a GS model), from \eqref{Eqn:GSO_ort},
\BE\label{Eqn:GSO_B}
B= \mr{E}\big\{ (\bf{\xi}^{\mr{in}})^{\rm T}\hat{\pi}(\hat{\bf{x}})\big\}/\mr{E}\big\{\|{\bf{\xi}}^{\mr{in}}\|^2\big\}.
\EE

 It's interesting to note that GSO is a fundamental component of the conjugate gradient approach, which is used to solve \LL{the high-complexity challenges} of the LMMSE estimator in OAMP \cite{TakeuchiCG2017}. We think there are some similarities between the GSO for OAMP in this paper and the GSO for the conjugate gradient approach.   

The orthogonality discussed above is looser than LLN-orthogonality required in \eqref{Eqn:LLN_orth}. We will consider the latter next.

\subsection{LLN-Orthogonality}
The GSO in \ref{Sec:Bussgang_Orho} establishes the orthogonality between the current input and output errors. The following theorem gives a sufficient condition for the LLN-orthogonality in \eqref{Eqn:Orthogonality}, where the current output errors are orthogonal to the current and previous input errors.

\begin{theorem}\label{THE:orth}
The orthogonality in \eqref{Eqn:Orthogonality} is satisfied if $\gamma_t$ and $\phi_t$, $\forall t$, are orthogonal and separable-IID.
\end{theorem}

\begin{IEEEproof}
See Appendix \ref{APP:orth}.
\end{IEEEproof}

GSO and Theorem \ref{THE:orth} together show a way to construct an OAMP algorithm based on two arbitrary separable $\gamma_t$ and $\phi_t$. % In Appendix \ref{APP:orth}, we will show that Theorem \ref{THE:orth} can be extended to the so-called group-separable functions. The latter covers a wide class of detectors including Wiener filtering, Viterbi decoding and BCJR algorithm. Hence OAMP can be constructed using a wider range of local estimators.  

\subsection{Computational Aspects for GSO}\label{Sec:B_methods}
The key to GSO is to find $B$ in \eqref{Eqn:GSO_B}. We first consider a special case when $\hat{\phi}$ is separable and integrable, and $\bf{x}$ IID. From Theorem \ref{THE:IIDG} and \LL{Conjecture \ref{Conj:E_Var} (see Appendix \ref{APP:In_Pro})}, \eqref{Eqn:GSO_B} reduces to
\BE\label{Eqn:inte_B}
\!\!\!B\!=\!\frac{\int_{-\infty}^{+\infty}\int_{-\infty}^{+\infty} z_f\,\hat{\phi}_t(\alpha^{\mr{in}}x\!+\!z_f)\,p_{z_f}(z_f)\,p_x(x) \,d z_f\, dx}{v_{f^{\mr{in}}}},
\EE
where $z_f\sim \cal{N}(0, v_{f^{\mr{in}}})$ and $v_{f^{\mr{in}}}=\frac{1}{N}\|\bf{f}^{\rm in}\|^2$. Assume that $p_x(x)$ is known. Then \eqref{Eqn:inte_B} can be evaluated numerically. We may pre-calculate a table for $B$ as a function of $v_{f^{\mr{in}}}$ off-line. Then the cost is low for on-line processing.  

%In the general case when $\hat{\phi}$ is not separable, \eqref{Eqn:GSO_B} involves multi-dimensional integrals but the principle is similar to that for \eqref{Eqn:inte_B}. 

In some cases, $\hat{\phi}$ (or $\hat{\gamma}$ or both) may not be explicitly given. For example, $\hat{\phi}$ can be a black-box type estimator in a software package. In this case, we can still generate $B$, $v^{\gamma}_t$ and $v^{\phi}_t$ numerically by the Monte-Carlo method. %See \ref{Sec:simulation} for an example. Incidentally, such a black-box approach is potentially useful to the deep learning applications involving OAMP where orthogonality can be used to remove correlation of errors in consecutive neuron layers  \cite{He2018AI, Zhang2019AI,  Takabe2019}.

\LL{We call \eqref{Eqn:inte_B} the integral approach to $B$. The advantage of this approach over the derivative one can be found in \cite{Yiyao_integral} for an application in image processing. 
 The integral approach requires $p_X(x)$. This condition is usually met in communication applications, but not in some signal-processing applications. The derivative approach discussed next provides an alternative approach.}

\subsection{Derivative Approach to \texorpdfstring{$B$}{B}}\label{Sec:B_methods2}
We next consider a special case when $\hat{\phi}$ is separable and derivable, and $\bf{x}$ IID. The following is an alternative derivative approach to OAMP, first introduced in \cite{Ma2016} and  inspired by \cite{Donoho2009}. %We start with a definition of the derivative of a complex function.
 
%  \begin{definition}[Complex Derivative]
%  For a complex number $z=x+\mr{i}y$, the complex derivative is defined as 
%  \BE
%  {\partial}/{\partial z} ={1}/{2}\left( {\partial}/{\partial x} - \mr{i} \cdot {\partial}/{\partial y}\right).
%  \EE
%  For a complex function $f$: $\bb{R}\to \bb{R}$, we write $(\partial/\partial z)( \mr{Re}[f]+\mr{i}\cdot\mr{Im}[f])$ as $\partial f/\partial z$.
%  \end{definition}
 
From Theorem \ref{THE:IIDG}, \LL{Conjecture \ref{Conj:E_Var} (see Appendix \ref{APP:In_Pro})} and Stein’s Lemma \cite{Stein1972},  
\BE \label{Eqn:stein_lemma}
 \tfrac{1}{N}\mr{E}\{ (\bf{f}^{\mr{in}})^{\rm T}\hat{\phi}(\hat{\bf{x}}^{\mr{in}})\} =  v_{f^{\mr{in}}}   \mr{E}\{\partial \hat{\phi}({\hat{x}^{\mr{in}}})/\partial {\hat{x}^{\mr{in}}}\},
\EE
where $v_{f^{\mr{in}}}\equiv\tfrac{1}{N} \|{\bf{f}}^{\mr{in}}\|^2$. Hence, \eqref{Eqn:GSO_B} can be rewritten as
\BE\label{Eqn:B_derivationa}
B=\mr{E}\{{\partial\hat{\phi}({\hat{x}^{\mr{in}})}}/{\partial{\hat{x}^{\mr{in}}}}\}\equiv  \mr{E}\{\hat{ {\phi}}'\}.
\EE 
For an orthogonal $\phi$, substituting \eqref{Eqn:B_derivationa} into $\phi(\hat{\bf{x}}^{\mr{in}})= \hat{\phi}(\hat{\bf{x}}^{\mr{in}})\\-B \hat{\bf{x}}^{\mr{in}}$, the derivative of $\phi$ \LL{is equal to} zero. On the other hand, if the derivative of $\phi$ is zero, we have $B=0$ from \eqref{Eqn:B_derivationa} and also $\mr{E}\{(\bf{f}^{\mr{in}})^{\rm T}\phi(\hat{\bf{x}}^{\mr{in}})\}$ from \eqref{Eqn:GSO_B}, i.e., $\phi$ is orthogonal. Thus, we have a necessary and sufficient condition below for the orthogonal requirement in OAMP. 

\begin{proposition}\label{Pro:GSO_derive_zero}
Under the IID $\bf{x}$  and the GS error $\bf{f}^{\mr{in}}$ in OAMP, an estimator $\phi$ is orthogonal if and only if 
\BS\BE\label{Eqn:deriv_0}
\mr{E}\{\partial {\phi}({\hat{x}^{\mr{in}}})/\partial {\hat{x}^{\mr{in}}}\}=0.
\EE
In particular, if $\phi$ can be expressed as  $\phi=\bf{\Lambda}\bf{{\hat{x}}}^{\mr{in}}$, where $\bf{\Lambda}$ is a diagonal matrix, then \eqref{Eqn:deriv_0} is equivalent to
 \BE 
\mr{tr}\{\bf{\Lambda}\}=0. 
\EE\ES
\end{proposition}

Finally, when $\phi$ is separable, \eqref{Eqn:B_derivationa} can be rewritten to
\BE\label{Eqn:B_derivation}
B \equiv  \mr{E}\{\hat{ {\phi}}'\} \approx \tfrac{1}{N}{\textstyle\sum}_{i=1}^{N} {\partial [\hat{\phi}(\hat{\bf{x}}^{\mr{in}})]_i} /{\partial [\hat{\bf{x}}^{\mr{in}}]_i }.
\EE
Eqn. \eqref{Eqn:B_derivation} does not require the distribution
of $\bf{x}$, which makes it attractive in many signal processing problems. This advantage was first pointed out in  \cite{Donoho2009}.

\subsection{Alternative Form of OAMP}
The original OAMP algorithm was derived in \cite{Ma2016} for solving \eqref{Eqn:linear_system} via the recursion of a linear estimator (LE) and a non-linear estimator (NLE): 
\BS\label{Eqn:OAMP_exm}\begin{alignat}{2}
& \mr{LE:} & \quad& \bf{{r}}_t={\gamma}_t(\bf{{s}}_t)=\bf{{s}}_t+ \tfrac{N}{\mr{tr}\{\bf{{W}}_t\bf{A}\}}{\bf{W}}_t(\bf{y}-\bf{A}\bf{{s}}_t), \label{Eqn:OAMP_exma}\\
& \mr{NLE:} & & \bf{{s}}_{t+1}= \phi_t(\bf{{r}}_t)= C_t\big(\hat{\phi}_t(\bf{{r}}_t)-  \mr{E}\{\hat{ {\phi}}'_t\}\cdot \bf{{r}}_t\big), \label{Eqn:OAMP_exmb}
\end{alignat}\ES	
where  $\bf{{W}}_{\!t}$ is an arbitrary prototype matrix and $C_t$ a proper scalar. Eqn. \eqref{Eqn:OAMP_exm} does not involve SVD and so is usually convenient for implementation. It is equivalent to the OAMP in \ref{Sec:OAMP}, which can be seen from  the discussions below. 

From \eqref{Eqn:linear_system},  $\bf{A}=\bf{U}^{\rm T}\bf{DV}$ and $\bf{r}=\bf{U}\bf{y}$. Denote  $\hat{\bf{x}}_{t}^{\mr{in}}=\bf{q}_t$,  $\hat{\bf{x}}_{t+1}^{\mr{out}}=\bf{s}_{t+1}$,  $\hat{\bf{\mathcal{X}}}_{t}^{\mr{in}}=\bf{V}\bf{s}_t$,  $\hat{\bf{\mathcal{X}}}_{t}^{\mr{out}}=\bf{V}\bf{q}_t$,  $\bf{\Delta}_t=\bf{VW}_t\bf{U}^{\rm T}$ and $\lambda_t=\frac{N}{\mr{tr}\{\bf{\Delta}_t\bf{D}\}}$. We can rewrite \eqref{Eqn:OAMP_exm} as 
\BS\label{Eqn:OAMP_exm2}\begin{align}
\mr{LE:}  \; \; & \bf{\mathcal{\hat{X}}}_{t}^{\mr{in}}=\bf{V}\hat{\bf{x}}^{\mr{out}}_t,\\
& \bf{\mathcal{\hat{X}}}_{t}^{\mr{out}}\!=\!\gamma_t( \bf{\mathcal{\hat{X}}}_{t}^{\mr{in}})\!=\!\lambda_t\bf{\Delta}_t\bf{r} \!+\! \left(\bf{I} \!-\!\lambda_t\bf{\Delta}_t\bf{D} \right) \bf{\mathcal{\hat{X}}}_{t}^{\mr{in}},\\  
&\hat{\bf{x}}^{\mr{in}}_t=\bf{V}^{\rm T}\bf{\mathcal{\hat{X}}}_{t}^{\mr{out}},\\
\mr{NLE:} \;\;& \hat{\bf{x}}^{\mr{out}}_{t+1}= \phi_t(\hat{\bf{x}}^{\mr{in}}_t)= C_t\big(\hat{\phi}_t(\hat{\bf{x}}^{\mr{in}}_t)- \mr{E}\{\hat{\phi}'_t\}\!\cdot\! \hat{\bf{x}}^{\mr{in}}_t\big).
\end{align}\ES	
It can be verified that $\gamma_t$ and $\phi_t$ in \eqref{Eqn:OAMP_exm2} are both orthogonal, $\alpha_t^\gamma=1, \forall t$ and $\bf{g}_t^{\rm out}= \bf{\mathcal{\hat{X}}}_{t}^{\mr{out}}-\bf{x}$. Hence  \eqref{Eqn:OAMP_exm2} is consistent with OAMP in \ref{Sec:OAMP}. 

Eqn. \eqref{Eqn:OAMP_exm} and  \eqref{Eqn:OAMP_exm2} give two equivalent forms of OAMP. Eqn. \eqref{Eqn:OAMP_exm} does not involve SVD so it is convenient for implementation.  Eqn. \eqref{Eqn:OAMP_exm2} is convenient for performance analysis, as seen \LL{in Section \ref{Sec:OAMP_Principle}}. Its implementation involves SVD, which is very costly when the size of $\bf{A}$ is large. 

\subsection{Optimality of OAMP}\label{Sec:Opt_OAMP}
The MMSE optimality of OAMP for the system in \eqref{Eqn:linear_system} is analyzed in \cite{Ma2016} based on the assumption that $\hat{\phi}_t$ and $\hat{\gamma}_t$ are respectively linear and nonlinear, and they are separable, Lipschitz continuous and locally optimal. The SE fixed point of   OAMP satisfies the following replica equation \cite{Replica_Guo2005}:
\BE \label{Eqn:Replica}
  {v_{\infty}^{-1}} =  {\sigma^{-2}} \cdot R_{\bf{A}^{\rm T}\bf{A}} \left( -  {\sigma^{-2}} \cdot \mathrm{mmse}(x|x+\sqrt{v_{\infty}}\cdot z) \right),
\EE
where $\sigma^2$ is the noise variance in \eqref{Eqn:linear_system} and $z \sim \cal{N}(0,1)$. $R_{\bf{A}^{\rm T}\bf{A}}$ denotes the $R$-transform w.r.t. the eigenvalue distribution of $\bf{A}^{\rm T}\bf{A}$ \cite{Tulino2004}. Following \cite{Replica_Guo2005}, OAMP achieves MMSE when \eqref{Eqn:Replica} has only one solution.

In a similar way, the replica optimality of VAMP for the system \eqref{Eqn:linear_system} is shown in \cite{Rangan2016}. The result is extended to a system with both $\hat{\phi}_t$ and $\hat{\gamma}_t$ being nonlinear, separable and locally MMSE-optimal \cite{Fletcher2016,Cakmak2018}. This conclusion also applies to OAMP under the unified orthogonal framework discussed in \ref{Sec:comp_other_alg}. These results are summarized in Proposition \ref{OptOAMP}.

\begin{proposition} \label{OptOAMP}
For the system in Fig.~\ref{Fig:model}(a), OAMP converges to the MMSE provided that (i) both $\hat{\phi}_t$ and $\hat{\gamma}_t$ are separable and locally MMSE-optimal, and (ii) the fixed-point equation \eqref{Eqn:Replica} for $\phi_t$ and $\gamma_t$ has only one solution. 
\end{proposition}

%For non-separable or sub-optimal local processors, the optimality of AMP type algorithm including OAMP is still an open problem \cite{Fletcher2018,Hannak2018}.   

\section{Comparison with Related Algorithms}\label{Sec:comp_algs}
In this section, we compare OAMP-GSO with other related algorithms in \cite{Cakmak2018, Minka2001, Donoho2009, Rangan2016, Ma2016}. We will show that orthogonality is a common feature underpinning these algorithms. This observation provides useful insights into the turbo/EP/AMP family of iterative signal processing techniques. 

%In this section, we compare OAMP-GSO with other EP and AMP type of algorithms, including EP \cite{Cakmak2018, Minka2001}, AMP \cite{Donoho2009}, VAMP \cite{Rangan2016} as well as the original OAMP \cite{Ma2016}. In \cite{Meng2018SPL, Meng2015}, these algorithms were unified under an EP under the assumption that local processors are optimal. Below we extend this result to arbitrary local processors, optimal or sub-optimal. We will show that orthogonality is a common feature in these algorithms. This observation provides useful insights into all iterative signal estimation. We will also show that OAMP is different from EP in its original form. 

\subsection{Connection to EP}\label{Sec:EP}
When $\hat{\phi} $ achieves MMSE, $B$ in \eqref{Eqn:GSO_B} can be calculated using Proposition \ref{Pro:EP}, which is proved in \cite[Appendix B]{Yiyao_mmv}.

\begin{proposition}\label{Pro:EP}
Assume that $\hat{\bf{x}}^{\mr{in}}=\bf{x}+{\bf{f}}^{\mr{in}}$ (GS model) and $\hat{\phi} $ achieves local MMSE. Then 
\BE\label{Eqn:MMSE_df}
B  = v_{\hat{\phi}}/v_{\hat{x}^{\mr{in}}},
\EE
where $v_{\hat{\phi}}\!\equiv \!\tfrac{1}{N}\mr{E}\{\| \hat{\phi}(\hat{\bf{x}}^{\mr{in}}) \!-\! \bf{x}\|^2\}$ and $v_{\hat{x}^{\mr{in}}}\!\equiv\!\tfrac{1}{N} \mr{E}\{\|{\bf{f}}^{\mr{in}}\|^2\}$.
\end{proposition}

Recall the EP updating rule \cite{Cakmak2018, Minka2001}:
\BE\label{Eqn:MMSE_EP}
	\hat{\bf{x}}^{\mr{out}} ={ {\phi}}(\hat{\bf{x}}^{\mr{in}}) =\frac{1}{1-v_{\hat{\phi}}/v_{\hat{x}^{\mr{in}}}} \left( \hat{ {\phi}}(\hat{\bf{x}}^{\mr{in}}) -\frac{v_{\hat{\phi}}}{v_{\hat{x}^{\mr{in}}}}\hat{\bf{x}}^{\mr{in}}\right).
\EE
Clearly,  \eqref{Eqn:MMSE_EP} is equivalent to \eqref{Eqn:pi_orth} with $B$ given in \eqref{Eqn:MMSE_df}. Therefore EP and OAMP are equivalent when local estimators achieve MMSE. Otherwise, they do not. For example, $[\hat{ {\phi}}(\hat{\bf{x}}^{\mr{in}})]_i = (\hat{x}_i^{\mr{in}})^2$, where $\hat{\bf{x}}^{\mr{in}}={\bf{x}}+\bf{n}$ and  $\bf{x}$ is zero mean and  independent of the Gaussian noise $\bf{n}$. In this case, following  \eqref{Eqn:GSO_B} or \eqref{Eqn:B_derivationa}, we have $B=0$, i.e., \eqref{Eqn:MMSE_df} does not hold. Therefore, OAMP provides a new treatment for sub-optimal local estimators.
 
After the introduction of OAMP in \cite{Ma2016}, a derivative form of EP is discussed in  \cite{Fletcher2016}. These two algorithms are equivalent. 
 
\subsection{\LL{Connection to Conventional Turbo or Belief Propagation}}\label{Sec:BP}
\LL{The celebrated turbo principle \cite{TurboCode, Wang1999, Tuchler2002, Douillard1995}, which has been regarded as the de facto solution to coded linear systems \cite{Loeliger2007,  Yuan2014, LiuLei2019TSP}, is based on the concept of belief propagation (BP), where extrinsic messages are used to avoid the correlation problem of an iterative process. Intuitively speaking, the conventional turbo requires the input and output errors to be independent, while OAMP requires the input and output errors to be orthogonal. Therefore, turbo is a special case of OAMP because independence is a subset of orthogonality. OAMP may potentially outperform turbo for non-Gaussian constraints. For more details, refer to \cite[Section III-D]{MaLiu2018}.}

\LL{Since EP is also a special case of OAMP (see Subsection \ref{Sec:EP}), the results in this paper are  applicable to the combined BP-EP message-passing algorithms in \cite{Sun2015}.}

\subsection{Connection to AMP}\label{Sec:relat2AMP}
Early discussions on AMP did not involve orthogonality, but new insight can be gained from the recent results in \cite{Takeuchi2019}. In this part, we outline the connection of OAMP and AMP in \cite{Takeuchi2019},  explicitly emphasizing the orthogonality in AMP. 

AMP can be rewritten as the following iteration \cite{Donoho2009} for solving \eqref{Eqn:linear_system}. Starting with $t=1$, $ \bf{s}_1 =\bf{s}_0^{\mr{Onsager}} = \bf{0}$,
\BS\label{Eqn:AMP}\begin{alignat}{2} 
&\mathrm{LE:}\quad\; & \hat{\bf{x}}^{\mr{in}}_{t}&= \bf{s}_{t} + \bf{A}^{\rm T}(\bf{y}\!-\!\bf{A}\bf{s}_{t}) +   {\bf{s}}_{t-1}^{\mr{Onsager}},\label{Eqn:LD}\\
&\mathrm{NLE:} & \bf{s}_{t+1} &= \hat{\phi}_t(\hat{\bf{x}}^{\mr{in}}_t),\label{Eqn:NLD}
\end{alignat}
where $ {\bf{s}}_{t-1}^{\mr{Onsager}}$ is an ``Onsager term"  \cite{Donoho2009} defined as
\begin{align}
{\bf{s}}_{t-1}^{\mr{Onsager}} &= B_{t-1}  \cdot(\hat{\bf{x}}^{\mr{in}}_{t-1}-\bf{s}_{t-1}),\label{Eqn:Onsager_AMP}\\
 B_{t-1}&=\mr{E}\left\{{\partial [\hat{\phi}_{t-1}(\hat{\bf{x}}_{t-1}^{\mr{in}})]_i} /{\partial [\hat{\bf{x}}_{t-1}^{\mr{in}}]_i } \right\} \label{Eqn:Onsager_AMPa}\\
    &\approx \tfrac{1}{N} {\textstyle\sum}_{i=1}^N {\partial [\hat{\phi}_{t-1}(\hat{\bf{x}}_{t-1}^{\mr{in}})]_i} /{\partial [\hat{\bf{x}}_{t-1}^{\mr{in}}]_i }.  \label{Eqn:Onsager_AMPb}
\end{align}\ES
%The subscript ``t" in $\hat{\phi}_t$ indicates the involved distribution $\mathfrak{V}^{\mr{in}}_{\bf{\hat{x}}_{t}}$. Furthermore, $B_{t-1}$ in \eqref{Eqn:Onsager_AMPa} is a function of distribution $\mathfrak{V}^{\mr{in}}_{\bf{\hat{x}}_{t}}$, and \eqref{Eqn:Onsager_AMPb} is just an approximation (by law of large numbers). See \ref{Sec:B_methods2} and \ref{Sec:B_methods} for more details. 

Let $\bf{A} = \bf{U}^{\rm T} \bf{D V}$ and $\hat{\bf{y}}\equiv\bf{U}\bf{y}$, we rewrite \eqref{Eqn:AMP} as
\BS\label{Eqn:AMP_V}\begin{alignat}{2}
&{\bf{\mathcal{{S}}}}_{t} = \bf{V}\bf{s}_{t}, &\quad\;\;\;&\bf{\mathcal{\hat{X}}}^{\mr{out}}_{t}\!=\! \hat{\gamma}_t({\bf{\mathcal{{S}}}}_{1},\cdots,{\bf{\mathcal{{S}}}}_{t}),\label{Eqn:AMP_Va}\\
&  \hat{\bf{x}}^{\mr{in}}_t = \bf{V}^{\rm T}\bf{\mathcal{\hat{X}}}^{\mr{out}}_t, && \bf{s}_{t+1} = \hat{\phi}_t(\hat{\bf{x}}^{\mr{in}}_t),\label{Eqn:AMP_Vb}
\end{alignat}
where 
\begin{align}
\hat{\gamma}_t({\bf{\mathcal{{S}}}}_{1},\cdots,{\bf{\mathcal{{S}}}}_{t})&\equiv {\bf{\mathcal{{S}}}}_{t} + \bf{D}^{\rm T}(\hat{\bf{y}}\!-\!\bf{D}{\bf{\mathcal{{S}}}}_{t}) + {\bf{\mathcal{{S}}}}_{t-1}^{\mr{Onsager}}, \label{Eqn:gamma_hat_AMP}\\
{\bf{\mathcal{S}}}_{t-1}^{\mr{Onsager}}&\equiv\bf{V}  {\bf{s}}_{t-1}^{\mr{Onsager}}.\label{Eqn:gamma_hat_AMPb}
\end{align}\ES
% \begin{framed}
% Here, we explain why $\hat{\gamma}_t$ is a function of $\{{\bf{\mathcal{{S}}}}_{1},\cdots,{\bf{\mathcal{{S}}}}_{t}\}$. Based on $\bf{V}\hat{\bf{x}}^{\mr{in}}_{t-1} =\bf{\mathcal{\hat{X}}}^{\mr{out}}_{t-1}=\hat{\gamma}_{t-1}$ (see \eqref{Eqn:AMP_Va} and \eqref{Eqn:AMP_Vb}) and \eqref{Eqn:Onsager_AMP}, we rewrite  \eqref{Eqn:gamma_hat_AMPb} to 
% \BE\label{Eqn:Onsagert_1}
% {\bf{\mathcal{S}}}_{t-1}^{\mr{Onsager}} =B_{t-1}( \hat{\gamma}_{t-1}  - {\bf{\mathcal{{S}}}}_{t-1}).
% \EE
% Substituting \eqref{Eqn:Onsagert_1} into \eqref{Eqn:gamma_hat_AMP}, we have
% \BE\label{Eqn:Onsagert_recursion}
% \hat{\gamma}_t  \!= \! \bf{\Lambda}{\bf{\mathcal{{S}}}}_{t} \!+ \!B_{t-1} ( \hat{\gamma}_{t\!-\!1}  \!- \!{\bf{\mathcal{{S}}}}_{t\!-\!1}) \!+ \bf{b},
% \EE
% where $ \bf{\Lambda}=\bf{I}\!-\!\bf{D}^{\rm T}\bf{D}$ and $\bf{b}=\bf{D}^{\rm T}\hat{\bf{y}}$. Since $\hat{\gamma}_0= {\bf{\mathcal{{S}}}}_0 =\bf{0}$, we have
% \BE\label{Eqn:gamma1}
% \hat{\gamma}_1  \!= \! \bf{\Lambda}{\bf{\mathcal{{S}}}}_1 + \bf{b}.
% \EE
% Then, using \eqref{Eqn:Onsagert_recursion} and \eqref{Eqn:gamma1}, we can prove by induction that $\hat{\gamma}_t$ is a function of $\{{\bf{\mathcal{{S}}}}_{1},\cdots,{\bf{\mathcal{{S}}}}_{t}\}$ .  
% \end{framed}

We now define an intermediate variable as follows \cite{Takeuchi2019}.
\BE\label{Eqn:intermediate}
\hat{\bf{x}}^{\mr{out}}_{t+1} \equiv  \phi_t (\bf{\hat{x}}^{\mr{in}}_{t})=   A_t \Big(\hat{\phi}_t(\hat{\bf{x}}^{\mr{in}}_{t}) -B_t \hat{\bf{x}}^{\mr{in}}_{t} \Big),
\EE
where $A_t=1/(1-B_t)$ and $B_t$ is defined in \eqref{Eqn:Onsager_AMPb}. Define $\bf{\mathcal{\hat{X}}}^{\mr{in}}_{t}\equiv\bf{V}\hat{\bf{x}}^{\mr{out}}_t$. Furthermore, we can express $\bf{\mathcal{\hat{X}}}_{t}^{\mr{out}}$ by a function of $\gamma_t(\bf{\mathcal{\hat{X}}}^{\mr{in}}_{1}\!, \cdots\!,\bf{\mathcal{\hat{X}}}^{\mr{in}}_{t})$. Then, we rewrite \eqref{Eqn:AMP} (i.e. AMP) into a GIP form with memory (GIP-M) as follows.

\textit{GIP-M:} Starting with $t=1$ and $ \bf{{\hat{x}}}^{\mr{out}}_1 = \bf{0}$,%\vspace{-0.1cm}
\BS\label{Eqn:Ite_Proc_AMP}\begin{alignat}{2}
&\!\bf{\mathcal{\hat{X}}}_{t}^{\mr{in}} \!=\!\bf{V}\hat{\bf{x}}^{\mr{out}}_{t}\!\!,  & \qquad& \bf{\mathcal{\hat{X}}}_{t}^{\mr{out}}\!\!=\! \gamma_t\big(\bf{\mathcal{\hat{X}}}^{\mr{in}}_{1}, \cdots,\bf{\mathcal{\hat{X}}}^{\mr{in}}_{t}\big),  \\
&\hat{\bf{x}}_{t}^{\mr{in}}\!=\!\bf{V}^{\rm T} \!\bf{\mathcal{\hat{X}}}^{\mr{out}}_{\!t}\!\!, & \;\;\;&  \hat{\bf{x}}^{\mr{out}}_{t+1} \!\!=\! \phi_t\big(\hat{\bf{x}}^{\mr{in}}_{t}\big).
\end{alignat}\ES 

The above can be seen as an extension of GIP in \eqref{Eqn:Ite_Proc_para} by allowing memories in $\gamma_t$ and $\phi_t$. Corresponding to Theorems \ref{THE:IIDG} and \ref{THE:orth}, we have the following properties for GIP-M. 

\begin{proposition}\label{Pro:AMP_orth}
The followings hold for the equivalent form of AMP in \eqref{Eqn:Ite_Proc_AMP} if $\bf{A}$ is IIDG and $\phi_t$ is separable-IID.
\begin{itemize}
    \item[(i)] The error orthogonality in \eqref{Eqn:Orthogonality} holds.
    \item[(ii)] All the conclusions of Theorem \ref{THE:IIDG}  apply.  
\end{itemize} 
\end{proposition}
 
Proposition \ref{Pro:AMP_orth} can be verified as follows. It is shown in \cite{Takeuchi2019} that $\mr{E}\{\partial \big[\gamma_t( \bf{\mathcal{\hat{X}}}^{\mr{in}}_{t})\big]_i/\partial [\bf{\mathcal{\hat{X}}}^{\mr{in}}_{t}]_i \} =0$ (see (4) and Theorem  2 in \cite{Takeuchi2019}). Furthermore, $\mr{E}\{\partial \big[\phi_t( \bf{\hat{x}}^{\mr{in}}_{t})\big]_i/\partial [\bf{\hat{x}}^{\mr{in}}_{t}]_i \} =0$ can be derived strictly from \eqref{Eqn:Onsager_AMPb} and \eqref{Eqn:intermediate}. Then (i) follows from Proposition \ref{Pro:GSO_derive_zero}. The proof of Theorem \ref{THE:IIDG}  requires the orthogonality in \eqref{Eqn:Orthogonality} on the input and output errors of $\gamma$ and $\phi$ only, which does not involve their internal structures with or without memory. Hence (ii) holds.  

The above shows some interesting connections between AMP and OAMP as follows.
\begin{itemize}
    \item There are different ways to establish orthogonality such as using GSO or Onsager term. 
    \item Orthogonality provides useful insights into the mechanism of the AMP-family of algorithms, based on which many findings can be derived more concisely. This can be seen by comparing the proof of SE for AMP in \cite{Bayati2011} and that of OAMP in this paper. (Note: Proposition \ref{Pro:AMP_orth} together with Theorems \ref{THE:IIDG} and \ref{THE:orth} in this paper form an alternative proof for the SE for AMP.) 
\end{itemize} 

Despite that they are both underpinned by orthogonality, AMP is not equivalent to OAMP without memory. \LL{In particular}, Proposition \ref{Pro:AMP_orth} holds for AMP only when the sensing matrix $\bf{A}$ in \eqref{Eqn:linear_system} has IIDG entries \cite{Donoho2009, Bayati2011}. Otherwise, the performance of AMP may degrade noticeably. OAMP is not subjected to these restrictions. %This issue has been recently addressed in \cite{Takeuchi2020}.
 
\subsection{Connection to OAMP in the Differentiation Form} \label{Sec:comp_OAMP_DF}
OAMP in the differentiation form (OAMP-DF) is introduced in \cite{Ma2016}. Its equivalence to OAMP-GSO is clear from the discussions in \ref{Sec:comp_other_alg}. OAMP-DF can be implemented using \eqref{Eqn:OAMP_exm} without knowing the distribution of $\bf{x}$. This has an advantage in many signal processing problems where the distributions of $\bf{x}$ are not known a prior. (In communication problems, the distributions of $\bf{x}$ are usually determined by modulation methods and can be assumed to be known.)

\subsection{Connection to VAMP}\label{Sec:comp_other_alg}
The derivative version of OAMP was originally derived in \cite{Ma2016} in the form of \eqref{Eqn:OAMP_exm} and \eqref{Eqn:OAMP_exm2}. Its equivalence\footnote{VAMP and OAMP are equivalent when OAMP uses the variance updating formula given in (23) in \cite{Ma2016}. It is mentioned in \cite{Rangan2016} that VAMP and OAMP are different when OAMP is based on an approximate updating technique.} to VAMP can be seen by  comparing \eqref{Eqn:OAMP_exm2} with Algorithm 2 in \cite{Rangan2016}. The latter is a special case of \eqref{Eqn:OAMP_exm} using $C_t=(1-\mr{E}\{\hat{\phi}'_t\})^{-1}$ and an MMSE form of \eqref{Eqn:OAMP_exma} with  ${\bf{W}}_t=v_{s_t}\bf{A}^{\rm T}(v_{s_t}\bf{A}\bf{A}^{\rm T} \!+\! \sigma^2\bf{I})^{-1}$, where $v_{s_t}=\frac{1}{N}{\rm E}\{\|\bf{s}_t-\bf{x}\|^2\}$. 

The SE for OAMP was conjectured based on the Haar distribution of $\bf{V}$ in \cite{Ma2016}. (See Definition 1 and Proposition 1 in \cite{Ma2016}.) The key contribution in \cite{Rangan2016} is an elegant proof of SE for VAMP based on the same assumption on $\bf{V}$.  The proof of SE for OAMP in this paper is inspired by the work in \cite{Rangan2016}. There is a difference though. The proof in this paper relies solely on the orthogonality of local estimators. It is easier to extend the proof in this paper to a broader class of applications, which will be discussed elsewhere \cite{Lei_TSP_2_2019}.

 \subsection{Summary}
 We summarize the characteristics of the existing EP, turbo, AMP, OAMP and VAMP below.  
\begin{itemize}
    \item EP and OAMP are equivalent when local estimators are optimal. They are different otherwise.
    \item  \LL{Turbo is a special case of OAMP since orthogonality includes independence}. 
    \item  VAMP is equivalent to the derivative form of OAMP originally proposed in \cite{Ma2016}.
    \item AMP is not equivalent to the standard form of OAMP without memory. However, AMP falls in the general class of OAMP with memory. 
\end{itemize}

Finally, the behavior of all these algorithms can be analyzed using the orthogonal principle. In general, the performance of an algorithm deteriorates if such orthogonality no longer holds. This is the case, e.g., for AMP when the sensing matrix is not IID, or for EP when the  estimators are not locally optimal.

{\color{black} \section{Optimization via GS Model and GSO}\label{Sec:Optimization_OAMP}

\subsection{Motivation}

This section provides an example to demonstrate an advantage of the GS model and GSO. It results from the freedom in tuning $\gamma$ and $\phi$ in \eqref{Eqn:Ite_Proc_para} under the orthogonal principle in \eqref{Eqn:Orthogonality}. Specifically, we consider the standard linear model in \eqref{Eqn:linear_system}. Assume that $\bf{A}$ in \eqref{Eqn:linear_system} is not IIDG, such as in the applications in \cite{He2018AI, Zhang2019AI, Takabe2019, OAMP_ISIT22, OAMP_TCOM, Yiyao_integral, Lei_TSP_2_2019, KhaniANSD2020, IC-SRC2021, Yiyao_ofdm, HWang2019, XZhou2022, ZhangOAMP2017, SLiu2022, LiOTFS2022,Yiyao_mmv, Fletcher2016}. Then \eqref{Eqn:linear_system} can be solved by EP, OAMP and VAMP, but not AMP. The standard EP/OAMP/VAMP solution is given in \eqref{Eqn:OAMP_exm}. Below, we  modify \eqref{Eqn:OAMP_exma} for performance optimization facilitated by the GS model and GSO. This is useful when sub-optimal $\hat{\phi}$ or $\hat{\gamma}$ or both are used due to complexity concerns \cite{Donoho2009}. Significant performance improvement over standard EP/OAMP/VAMP is demonstrated.}

% In this section, we consider the linear system as an example:
% \BE\label{Eqn:linear_opt}
% \bf{y}=\bf{A}\bf{x} + \bf{n}.
% \EE	
%  The standard OAMP for \eqref{Eqn:linear_opt} is provided in \eqref{Eqn:OAMP_exm} with $C_t=1/\big(1-\mr{E}\{\hat{ {\phi}}'_t\}\big)$ (commonly used in EP, OAMP and VAMP). %The MSE performance of the standard LMMSE-OAMP is shown in Fig. \ref{Fig:Sim_PM}.  %which applies to various systems, including compressed sensing \cite{Donoho2009, Ma2016, Rangan2016}, grant-free MTC \cite{Senel2018,SLiu2022}, massive random access \cite{ChenAMP2019}, MIMO \cite{KhaniANSD2020, MaLiu2018}, coding \cite{LiangAMP2020,IC-SRC2021}, OFDM \cite{Yiyao_ofdm, HWang2019, XZhou2022, ZhangOAMP2017},  massive connectivity \cite{SunAMP2019,ZhuAMP2021}, RIS-MIMO \cite{Ruan2022}, and OTFS modulation \cite{LiOTFS2022}.

%In the previous sections, we demonstrated that the MSE of OAMP can be characterized by SE when $\gamma$ and $\phi$ are orthogonal. However, SE is not a guarantee of good performance. When $\hat{\phi}$ and $\hat{\gamma}$ are locally optimal, the standard OAMP in \eqref{Eqn:OAMP_exm} is potentially optimal for the separable case \cite{Ma2016, Takeuchi2017, Rangan2016}. When they are not (e.g. the thresholding estimator in \eqref{Eqn:Thresh}), we may have extra room for optimization by tuning $\gamma$ and $\phi$ using the GS model. }

{\color{black}\subsection{Tuning OAMP via GS Model and GSO} 
Replace \eqref{Eqn:OAMP_exma} by the following linear operation:
\BE\label{Eqn:OAMP_LE_new}
  \bf{{r}}_t={\gamma}_t(\bf{{s}}_t) = C_t\big[\hat{\gamma}_t( \xi\bf{{s}}_t) -  B_t \bf{{s}}_t\big], 
\EE
where $ \hat{\gamma}_t(\bf{{s}}_t)=\bf{W}_{\!t}\bf{y} + (\bf{I} - \bf{W}_{\!t}\bf{A})\bf{s}_t$, the GSO coefficient $B_t$ is defined in \eqref{Eqn:pi_orth} and is given by $B_t=\tfrac{1}{N}{\mr{Tr}\{\xi (\bf{I} - \bf{{W}}_t\bf{A})\}}$ (following \eqref{Eqn:GSO_B}), $C_t$ is a normalization coefficient given by $ C_t= \tfrac{N}{\mr{Tr}\{\bf{{W}}_t\bf{A}\}}$, and $\xi$ is a variable for optimization. Then, \eqref{Eqn:OAMP_LE_new} can be rewritten as
\BE\label{Eqn:OAMP_LE_opt}
\bf{{r}}_t={\gamma}_t\big(\bf{{s}}_t)=\xi\bf{{s}}_t+ \tfrac{N}{\mr{tr}\{\bf{{W}}_t\bf{A}\}}{\bf{W}}_t(\bf{y}-\xi\bf{A}\bf{{s}}_t\big).
\EE
 We can see that \eqref{Eqn:OAMP_exma} is a special case of \eqref{Eqn:OAMP_LE_opt} when $\xi=1$. More generally, the MSE of $\bf{{r}}_t$ (i.e., $v_{r_t}$) is a concave function of $\xi$. Hence, the optimal $\xi$ that minimizes $v_{r_t}$ can be obtained by $d\,v_{r_t}(\xi)/d\,\xi=0$. 

Assume that ${\bm s}_t$ is characterized by the GS model: ${\bm s}_t={\alpha_t}{\bm x}+{\bm f}_t$ with $v_{f_t}$ the variance of ${\bm f}_t$. The optimal $\xi$ is
\BS\BE
%\xi = 1/(v_{s_t}+1),
\xi= \alpha_t/(\alpha_t^2+v_{f_t}),
\EE
and the corresponding MSE is given by
\BE\label{Eqn:var_opt}
v_{r_t} = \tfrac{1}{N}\mr{Tr}\{\bf{B}_t\bf{B}_t^{\mr{T}}\}\tfrac{v_{f_t}}{\alpha_t^2+v_{f_t}}  +\tfrac{1}{N}\mr{Tr}\{\tilde{\bf{W}}_t\tilde{\bf{W}}_t^{\mr{T}}\}\sigma^2,
\EE\ES
where $\bf{B}_t=\bf{I}-\tilde{\bf{W}}_t\bf{A}$ and $\tilde{\bf{W}}_t=\tfrac{N}{\mr{Tr}\{\bf{{W}}_t\bf{A}\}}\bf{W}_{\!t}$. Some interesting observations are as follows.
\begin{itemize}
\item[(i)]  The $v_{r_t} $ in \eqref{Eqn:var_opt} is not larger than the MSE of \eqref{Eqn:OAMP_exma} (see (32a) in \cite{Ma2016}) in the standard EP/OAMP/VAMP. 

\item[(ii)]  The complexity of the optimized OAMP in \eqref{Eqn:OAMP_LE_new} is the same as the standard EP/OAMP/VAMP in \eqref{Eqn:OAMP_exma}.

\item[(iii)] For MMSE $\hat{\phi}$, we can show that \eqref{Eqn:OAMP_exma} equivalents to \eqref{Eqn:OAMP_LE_new}. We omit the proof as it is quite straightforward.  %The key point is that for MMSE $\hat{\phi}_t$, $\bf{s}_t$ can be modeled as $\bf{s}_t=\alpha\bf{x}+\bf{\xi}$ with ${v_{{s}_t}}=\alpha(1-\alpha)$.

%\item[(iii)] It is interesting to review the GS model. We applied this model to $\{\bf{\hat{x}}_t^{\mr{in}}\!\!, \bf{\hat{x}}_{t}^{\mr{out}}\!\!,  \bf{\mathcal{\hat{X}}}_t^{\mr{in}}\!\!, \bf{\mathcal{\hat{X}}}_{t}^{\mr{out}}\}$ that are the inputs and outputs of $\gamma_t$ and $\phi_t$. This is necessary since the GS model is preserved over a unitary transform. For example, for $\bf{\mathcal{\hat{X}}}_t^{\mr{in}}=\bf{V}\bf{\hat{x}}_t^{\mr{out}}$, $\bf{\mathcal{\hat{X}}}_t^{\mr{in}}$ and $\bf{\hat{x}}_t^{\mr{out}}$ have the same GS model, which makes it easier to track error behavior during iterative processing. On the contrary, in deriving \eqref{Eqn:OAMP_LE_new}, $\bf{r}_t$ is not expressed in its GS model. It can be verified that using the GS model of $\bf{r}_t$ in \eqref{Eqn:OAMP_LE_new} may lead to poor performance.  As a rule, we use the GS model only when necessary (e.g. for the iterative inputs and outputs. Otherwise, we should use more accurate models whenever possible. 
\end{itemize}

Since MMSE $\hat{\phi}$ may be hard to obtain or the computational complexity is practically prohibitive, sub-optimal solutions are commonly employed.  The optimized LE in \eqref{Eqn:OAMP_LE_new} for OAMP is useful in this case.

\subsection{Numerical Example} \label{Sec:simulation} 
Let us consider a special case of \eqref{Eqn:linear_system} for the following compressed sensing problem \cite{Donoho2006}. Let the SVD of $\bf{A}$  be $\bf{A} = \bf{U}^{\mr{H}}\bf{D V}$. The eigenvalues $\{d_i\}$, i.e. diagonal elements} \LL{of $\bf{D}$, are generated as: $d_i/d_{i+1}=\kappa^{1/M}$ for $i = 1,\ldots, M-1$ and $\sum_{i=1}^Md_i^2=N$ \cite{Vila2015}. Here, $\kappa$ is the condition number of $\bf{A}$. The entries of $\bf{x}$ are IID and follow the Bernoulli-Gaussian (i.e., sparse) distribution, i.e. $\forall i$,
\BE
{x_i}\sim\left\{ \begin{array}{l}
0, \qquad\qquad\;\;\; \mathrm{probability} = 1-\lambda,\\
\mathcal{CN}(0,{\lambda ^{ - 1}}),\quad \mathrm{probability} = \lambda.
\end{array} \right.
\EE
The variance of $x_i$ is normalized to 1. The applications of this example can be seen \cite{Ahn2019}.}
\begin{figure}[t]
  \centering
  \includegraphics[width=5.5cm]{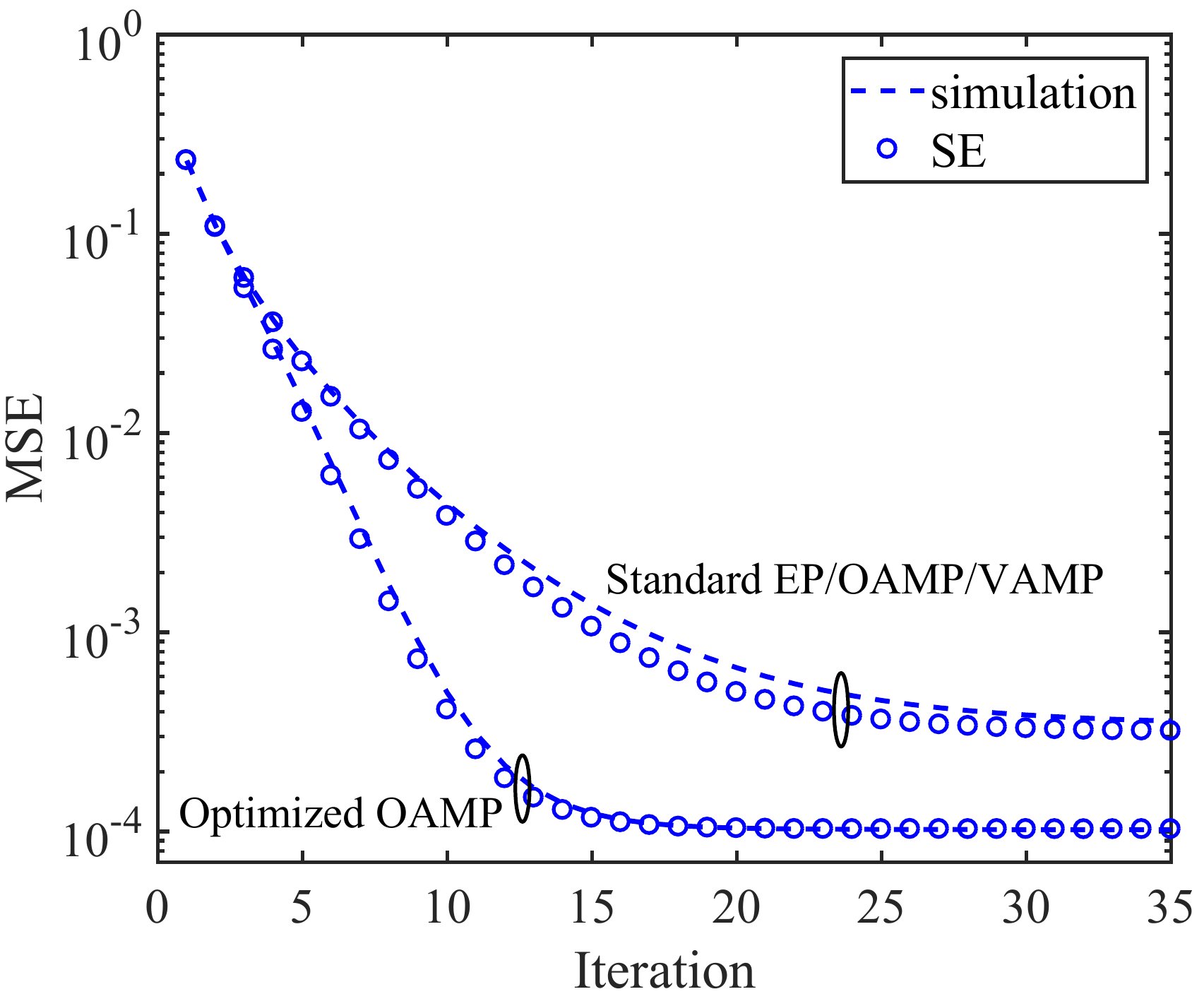}\\
  \caption{\color{black}MSE comparison between the standard LMMSE-EP/OAMP/VAMP and the optimized LMMSE-OAMP, where $N\!=\!1500$, $M/N\!=\!0.65$, $\kappa\!=\!10$, $\lambda\!=\!0.25$, $\vartheta_t=v_{r_t}$ and $\mr{SNR}=\sigma_n^{-2}=45$~dB.}\label{Fig:Sim_PM}
\end{figure}

\LL{Consider employing the prototypes in \eqref{Eqn:OAMP_exm2} for the above problem. To reduce complexity, assume that $\hat{\phi}_t$ is constructed using a low-cost thresholding estimator \cite{Donoho1995}:
\BE\label{Eqn:Thresh}
\hat{\phi}_t(r_t)= \max\big(\|r_t\|-\vartheta_t, 0\big)\cdot \mr{sign}(r_t),
\EE
where $\vartheta_t$ is a threshold.}

\LL{Assuming that ${\phi}$ is obtained by plugging  \eqref{Eqn:Thresh} into \eqref{Eqn:OAMP_exm2} with $C_t=1/\big(1-\mr{E}\{\hat{ {\phi}}'_t\}\big)$ (commonly used in EP, OAMP and VAMP), we consider two LMMSE-LEs \eqref{Eqn:OAMP_exma} and \eqref{Eqn:OAMP_LE_new} (with ${\bf{W}}_t=v_{s_t}\bf{A}^{\rm T}(v_{s_t}\bf{A}\bf{A}^{\rm T} \!+\! \sigma^2\bf{I})^{-1}$)  to get $\gamma_t$ for the standard EP/OAMP/VAMP and optimized OAMP, respectively. (Note that $v_{s_t}=N^{-1}\|{\bm s}_t-{\bm x}\|^2$ is used in the standard EP/OAMP/VAMP, while in the optimized OAMP, $v_{s_t}$ is set as $N^{-1}\|\xi{\bm s}_t-{\bm x}\|^2$.) Fig. \ref{Fig:Sim_PM} shows that the optimized LE via the GS model and GSO can bring significant performance improvement over the standard EP/OAMP/VAMP in this example. Additionally, there is a good agreement between simulation results for standard/optimized OAMPs and their SEs.} %{\color{red}Besides, the sub-optimal LE based on the GS model of $\bf{r}$ has a very poor performance and the iteration fails.} 

\section{Conclusion}
\LL{This paper studied the impact of orthogonality in OAMP and other AMP-type algorithms. Specifically, orthogonality suppresses the correlated error component in the iterative process and hence ensures the correctness of SE. This provides useful insights into the mechanism of OAMP. We also developed a GSO procedure to establish orthogonality. GSO offers a simple and versatile realization technique for OAMP. Various turbo/EP/AMP-type algorithms can be transformed into equivalent forms with orthogonal local estimators. Hence they can be unified under a common orthogonal framework.} 
	
%\LL{The findings in this paper can be extended to a wider range of applications. Some of such applications have been  discussed in \cite{OAMP_ISIT22, OAMP_TCOM, Yiyao_integral, Lei_TSP_2_2019, KhaniANSD2020, IC-SRC2021, Yiyao_ofdm, HWang2019, XZhou2022, ZhangOAMP2017, SLiu2022, LiOTFS2022,Yiyao_mmv, Fletcher2016}. OAMP has also been applied to deep learning problems \cite{He2018AI, Zhang2019AI,  Takabe2019}. GSO can be naturally incorporated into the training process in such problems, which is interesting for future work.}  

A householder dice method has recently been proposed for effectively simulating the dynamics on Haar random matrices \cite{Lu2021}. It will be fascinating to see how the householder dice technique may be used to demonstrate the SE of OAMP. 
 
\appendices  
\section{Properties/Conjectures of IID/IIDG and PIID/PIIDG Variables}\label{APP:Pro_PIIDG}

%This appendix contains useful properties on IID/IIDG and PIID/PIIDG variables. We are not able to prove those related to PIID/PIIDG variables yet, so we only prove those related to IID/IIDG variables. We list the former as conjectures.

\subsection{LLN Orthogonality} 
Let $\bf{a}=\{a_n\}$ and $\bf{b}=\{b_n\}$ be two sequences of length $N$.  \LLC{Assume that both $\bf{a}$ and $\bf{b}$ are IID.} Then $\{a_n b_n\}$ are also IID and so we have 
\BS\label{Eqn:IID_ab}\begin{align}
    \mr{E}\{\|\bf{a}\|^2\} &= N\,\mr{E}\big\{a_n^2\big\},\\
     \mr{E}\{\|\bf{b}\|^2\} &= N\,\mr{E}\big\{b_n^2\big\},\\
     \mr{Var}\big(\tfrac{1}{N}\bf{a}^{\rm T}\bf{b}\big)&= \mr{Var}\{ a_n b_n\}/N.
\end{align}\ES
The following is related to the convergence of $\bf{a}^{\rm T}\bf{b}$ when $N\to\infty$, which follows the law of large numbers \cite{Dekking2005}.  

\begin{lemma}\label{Pro:Chev_IID}
Let $\mr{E}\{\|\bf{a}\|^2\}\neq 0$,  $\mr{E}\{\|\bf{b}\|^2\}\neq 0$ and \LLC{both $\bf{a}$ and $\bf{b}$ be IID} with $ \mr{E}\{a_n b_n\}=0$ and  $ \mr{Var}\{ a_n b_n\}$ finite. Then  $\tfrac{1}{N}\bf{a}^{\rm T}\bf{b} \overset{\rm LLN}{\longrightarrow}0$.
\end{lemma}

\begin{IEEEproof}
From Chebyshev's inequality \cite{Dekking2005}, we have
\BE\label{Eqn:Cheb_ine}
\mr{Pr}\Big( (\tfrac{1}{N}\bf{a}^{\rm T}\bf{b})^2 < \mr{Var} \big\{\tfrac{1}{N}\bf{a}^{\rm T}\bf{b}\big\}/\delta \Big) \ge 1-\delta.
\EE
Using \eqref{Eqn:IID_ab}, we rewrite \eqref{Eqn:Cheb_ine} as
\BE\label{Eqn:P8}
\mr{Pr}\!\left(\! \frac{(\bf{a}^{\rm T}\bf{b})^2}{\mr{E}\big\{\|\bf{a}\|^2\big\}\mr{E}\big\{\|\bf{b}\|^2\big\}} \!<\! \frac{\mr{Var}\{ a_n b_n\}}{N\mr{E}\big\{a_n^2\big\}\!\cdot\!\mr{E}\big\{b_n^2\big\}\!\cdot\!\delta}\!\right)\! \ge \!1\!-\!\delta.
\EE
Then \eqref{Eqn:chev_IID} holds provided that
\BE
N\ge N' \equiv  \frac{\mr{Var}\{ a_n b_n\}}{ \mr{E}\big\{a_n^2\big\}\!\cdot\!\mr{E}\big\{b_n^2\big\}\!\cdot\!\delta\!\cdot\! \varepsilon}.
\EE
This completes the proof.
\end{IEEEproof}

The following lemma follows Lemma \ref{Pro:Chev_IID}.
 
\begin{lemma}[LLN Orthogonality of Separable-IID Function]\label{Pro:LLN_F_IID}
Assume that $\bf{\pi}=\{\pi_n\}$ is separable-IID, and  $\bf{a}=\{a_n\}$ IID, both of length $N\to \infty$. Then (i) $\{a_n \pi_n(a_n)\}$ are IID and (ii) $\tfrac{1}{N}\bf{a}^{\rm T}\pi(\bf{a})\overset{\rm LLN}{\longrightarrow} 0$ if ${\rm E}\{a_n \pi(a_n)\}=0$. 
\end{lemma} 

%Similar to Lemma \ref{Pro:Chev_IID} and Lemma \ref{Pro:LLN_F_IID} for IID sets, we have the following properties for PIID sets.

Similar to Lemma \ref{Pro:Chev_IID}, we can verify Property \ref{Pro:PIID_LLN} for PIID  variables using the concept of group IID variables. The details are omitted due to space limitations.  %See Appendix \ref{APP:orth}. 

\begin{property}\label{Pro:PIID_LLN}
 \LLC{Let both $\bf{a}$ and $\bf{b}$ be PIID} with $ \mr{E}\{a_n b_n\}=0$ and  $ \mr{Var}\{ a_n b_n\}$ finite. Then, $\tfrac{1}{N}\bf{a}^{\mr{T}}\bf{b} \overset{\rm LLN}{\longrightarrow}0$. 
\end{property}

Property \ref{Pro:LLN_F_PIID} follows Property \ref{Pro:PIID_LLN}. 

\begin{property}\label{Pro:LLN_F_PIID}
Assume that $\bf{\pi}=\{\pi_n\}$ is separable-IID, and  $\bf{a}=\{a_n\}$ PIID, both of length $N\to \infty$. Then (i) $a_n^{\rm T}\pi_n(a_n)$ are PIID and (ii) $\tfrac{1}{N}\bf{a}^{\rm T}\pi(\bf{a})\overset{\rm LLN}{\rightarrow} 0$ if ${\rm E}\{a_n^{\rm T}\pi(a_n)\}=0$. 
\end{property}

\subsection{Inner Product of Separable-IID Functions}\label{APP:In_Pro}  
 
\begin{property}\label{Pro:Exp}
  Let $\pi_1$ and $\pi_2$ be separable-IID,  \LLC{$\bf{\xi}=\{\xi_{1n}\}$ and $\bf{\xi}=\{\xi_{2n}\}$ be IID  variables}. It is easy to verify the following. 
\BE\label{Eqn:ALG-SteinLemma}
\tfrac{1}{N}\!\! \mathop\mr{E}\limits_{\bf{r},\bf{\xi}_1,\bf{\xi}_2}\!\!\big\{  \pi_1(\bf{\xi}_1)^{\rm T} \pi_2(\bf{\xi}_2)\big\} \!= \!\!\! \mathop\mr{E}\limits_{r_n,\xi_{1n},\xi_{2n}}\!\!\!\big\{ \pi_{1n}(\xi_{1n}) \pi_{2n}(\xi_{2n}) \big\},
\EE
where $\bf{r}$ denotes the potential IID variables except for $\bf{\xi}_1$ and $\bf{\xi}_2$ in function $\pi_1 $ or $\pi_2 $. 
\end{property}
  
  The conjecture below extends Property \ref{Pro:Exp} to the PIIDG case. 

\begin{conjecture}\label{Conj:E_Var}
Let $\pi_1$ and $\pi_2$ be separable-IID,  \LLC{$\bf{\xi}=\{\xi_{1n}\}$ and $\bf{\xi}=\{\xi_{2n}\}$ be PIID  variables}. Then, 
\BE 
\tfrac{1}{N}\!\! \mathop\mr{E}\limits_{\bf{r},\bf{\xi}_1,\bf{\xi}_2}\!\!\big\{  \pi_1(\bf{\xi}_1)^{\rm T} \pi_2(\bf{\xi}_2)\big\} \!= \!\!\! \mathop\mr{E}\limits_{r_n,\xi_{1n},\xi_{2n}}\!\!\!\big\{ \pi_{1n}(\xi_{1n}) \pi_{2n}(\xi_{2n}) \big\},
\EE
where $\bf{r}$ denotes the potential random variables except for $\bf{\xi}_1$ and $\bf{\xi}_2$ in function $\pi_1 $ or $\pi_2 $. 
\end{conjecture}

From Conjecture \ref{Conj:E_Var}, in evaluating $\tfrac{1}{N}\!\! \mathop\mr{E}\limits_{\bf{r},\bf{x},\bf{\xi}}\!\!\big\{  \pi(\alpha \bf{x} +  \bf{\xi})^{\rm T} {\bf{x}}\big\}$ and $\tfrac{1}{N}\!\! \mathop\mr{E}\limits_{\bf{r},\bf{x},\bf{\xi}}\!\!\big\{  \pi(\alpha \bf{x} +  \bf{\xi})^{\rm T} {\bf{\xi}}\big\}$, we can replace the PIIDG $\bf{\xi}$ by an IIDG vector $\sqrt{\! {\rm Var} (\xi_n) }\,\bf{z}$, where $\bf{z}\sim \mathcal{N}(\bf{0},\bf{I})$. This is useful in deriving the state evolution method for OAMP.

\subsection{Incomplete Orthogonal Transform}\label{APP:orth_trans} 

\begin{property}\label{Lem:VT_Gau}
  Let $\bf{a}=\bf{Ub}$ where (i) $\bf{U}$ is orthogonal, (ii) $\bf{b}$ IIDG, and (iii) $\bf{U}$ and $\bf{b}$ are mutually independent. Then (i) $\bf{a}$ is IIDG, and (ii) $\bf{a}$ and $\bf{U}$ are mutually independent.
\end{property}

\begin{property}\label{Lem:part_U_IIDG}
    Let $\bf{U}_1$ be an $N\times(N-m)$ block in $\bf{U}\in\bf{\cal{U}}^N$, $\bf{b}_1\in\bb{R}^{N-m}$ be IIDG, and $\bf{U}$ and $\bf{b}_1$ are mutually independent. Then the following claims hold when $m/N\to0$: (i) $\bf{U}_1\bf{b}_1\to$IIDG; (ii) ${\rm E}\big\{\|\bf{U}_1\bf{b}_1\|^2\big\} \to {\rm E}\big\{\|\bf{b}_1\|^2\big\}$; and (iii) $\bf{U}_1\bf{b}_1$ is asymptotically independent of $\bf{U}_1$. 
\end{property}

\begin{IEEEproof}
Consider partitions $\bf{U}=[\bf{U}_1,\bf{U}_2]$ and $\bf{b}=\left[\begin{array}{cc}
     \bf{b}_1  \\
     \bf{b}_2 
\end{array}\right]$ such that $\bf{Ub}=\bf{U}_1\bf{b}_1+\bf{U}_2\bf{b}_2$, where $\bf{U}_1$ has $N-m$ columns and $\bf{b}$ is IIDG. Then
\BS\label{Eqn:b1_b2}\begin{align}
    {\rm E}\big\{\|\bf{U}_1\bf{b}_1\|^2\big\}&=\tfrac{N-m}{N}\cdot {\rm E}\big\{\|\bf{b}\|^2\big\},\\
    {\rm E}\big\{\|\bf{U}_2\bf{b}_2\|^2\big\}&=\tfrac{m}{N}\cdot {\rm E}\big\{\|\bf{b}\|^2\big\}.
\end{align}\ES
When $m\ll N$, \eqref{Eqn:b1_b2} leads to
\BS\label{Eqn:b1_2_b}\begin{align}
    \frac{{\rm E}\big\{\|\bf{U}_2\bf{b}_2\|^2\big\}}{{\rm E}\big\{\|\bf{U}_1\bf{b}_1\|^2\big\}}&= \frac{m}{N-m}\to 0,\\
    \bf{U}_1\bf{b}_1&\overset{\rm d}{\to}\bf{U}\bf{b},\\
     {\rm E}\big\{\|\bf{U}_1\bf{b}_1\|^2\big\}&\to  {\rm E}\big\{\|\bf{b}\|^2\big\}\to {\rm E}\big\{\|\bf{b}_1\|^2\big\}.
\end{align}\ES
When $\bf{b}$ is IIDG, both $\bf{Ub}$ and $\bf{U}_1\bf{b}_1$ are  IIDG and independent of $\bf{U}$ and $\bf{U}_1$. Therefore, we obtain Property \ref{Lem:part_U_IIDG}.
\end{IEEEproof} 

Intuitively, Property \ref{Lem:part_U_IIDG} hinges on the factors that $\|\bf{b}_2\|$ and $\|\bf{U}_2\bf{b}_2\|$ are statistically negligible compared with $\|\bf{b}_1\|$ and $\|\bf{U}_1\bf{b}_1\|$  when $m\ll N$. 

The conjectures below extend the above two properties to the PIIDG case. 

\begin{conjecture}\label{Conj:VT_Gau}
  Let $\bf{a}=\bf{Ub}$ where $\bf{U}$ is orthogonal, $\bf{b}$ PIIDG, and $\bf{U}$ and $\bf{b}$ are mutually independent. Then $\bf{a}$ is PIIDG, and $\bf{a}$ and $\bf{U}$ are mutually independent.
\end{conjecture}

\begin{conjecture}\label{Conj:part_U_IIDG}
    Let $\bf{U}_1$ be an $N\times(N-m)$ block in $\bf{U}\in\bf{\cal{U}}^N$, $\bf{b}_1\in\bb{R}^{N-m}$ be PIIDG, and $\bf{U}$ and $\bf{b}_1$ are mutually independent. Then the following claims hold when $m/N\to0$: (i) $\bf{U}_1\bf{b}_1\to$PIIDG; (ii) ${\rm E}\big\{\|\bf{U}_1\bf{b}_1\|^2\big\} \to {\rm E}\big\{\|\bf{b}_1\|^2\big\}$; and (iii) $\bf{U}_1\bf{b}_1$ is asymptotically independent of $\bf{U}_1$. 
\end{conjecture}

\subsection{Extended Stein’s Lemma} 

The following is a variation of the generalized Stein’s Lemma \cite{Stein1981}. It will be useful in the proof of Theorem \ref{THE:orth}. We say that $\bf{a}$ and $\bf{u}$ are entry-wise jointly Gaussian if every $\{u_i, a_i\}$ pair are jointly Gaussian.

\begin{lemma}[Extended Stein's Lemma]\label{Pro:ESteinLemma}
    Let $\bf{\pi}=\pi(\bf{a})$ where $\bf{a}$ is IIDG and $\pi(\bf{a})$ separable-IID. \LLC{Let $\bf{u}$ be any IIDG vector that is entry-wise jointly Gaussian with $\bf{a}$.} Let ${\rm E}\{\bf{a}\}= {\rm E}\{\bf{u}\}=\bf{0}$ and ${\rm E}\{\bf{a}^{\rm T}\bf{\pi}\}=0$. Then (i) $\tfrac{1}{N}\bf{a}^{\rm T}\bf{\pi}\overset{\rm LLN}{\longrightarrow} 0$; and (ii)  $\tfrac{1}{N}\bf{u}^{\rm T}\bf{\pi}\overset{\rm LLN}{\longrightarrow} 0$.
\end{lemma}

\begin{IEEEproof}
Since $\bf{a}$ is IIDG and $\pi(\bf{a})$ is separable-IID, $\bf{\pi}$ is IID (but not necessarily Gaussian). Then claim (i) follows Lemma \ref{Pro:Chev_IID} directly. For claim (ii), denote 
\BE\label{Eqn:u_GS}
\bf{\xi} =\bf{u}-\alpha \bf{a},
\EE
where $\alpha = {\mr E}\{\bf{u}^{\rm T}\bf{a}\}/{\mr E}\{\bf{a}^{\rm T}\bf{a}\}$ and so ${\mr E}\{\bf{\xi}^{\rm T}\bf{a}\}=0$. (This can be compared to \eqref{Eqn:GP_modelb} for the GS model \eqref{Eqn:GP_model}.) \LLC{Then $\bf{\xi}$ and $\bf{a}$ are uncorrelated and so mutually independent due to their Gaussianity. (When $\bf{u}$  and $\bf{a}$ are both IIDG, so is $\bf{\xi}$.)} Then $\bf{\xi}$  is independent of $\bf{\pi}$ due to the Markov chain $\bf{\xi}\to\bf{a}\to\bf{\pi}$. Therefore, ${\mr E}\{\bf{\xi}^{\rm T}\bf{\pi}\}=0$. Furthermore, from \eqref{Eqn:u_GS} and ${\mr E}\{\bf{a}^{\rm T}\bf{\pi}\}=0$, we have ${\rm E}\{\bf{u}^{\rm T}\bf{\pi}\}=0$. It can be seen that $\{u_n \pi_n\}$ are IID. Then from Lemma \ref{Pro:Chev_IID}, $\tfrac{1}{N}\bf{u}^{\rm T}\bf{\pi}\overset{\rm LLN}{\longrightarrow} 0$. This completes the proof.  
\end{IEEEproof} 

The conjecture below extends Lemma \ref{Pro:ESteinLemma}  to the PIIDG case.  

 \begin{conjecture}\label{Conj:ESteinLemma_PIIDG}
    Let $\bf{\pi}=\pi(\bf{a})$ where $\bf{a}$ is PIIDG and $\pi(\bf{a})$ separable-IID. \LLC{Let $\bf{u}$ be any PIIDG vector that is entry-wise jointly Gaussian with $\bf{a}$.} Let ${\rm E}\{\bf{a}\}= {\rm E}\{\bf{u}\}=\bf{0}$. Assume ${\rm E}\{\bf{a}^{\rm T}\bf{\pi}\}=0$. Then (i) $\tfrac{1}{N}\bf{a}^{\rm T}\bf{\pi}\overset{\rm LLN}{\longrightarrow} 0$ ; and (ii)  $\tfrac{1}{N}\bf{u}^{\rm T}\bf{\pi}\overset{\rm LLN}{\longrightarrow} 0$.
\end{conjecture}

\section{Proof of Theorem \ref{THE:IIDG}}\label{Sec:IIDG_proof} 
\LL{The asymptotically IID Gaussian (AIIDG) property of OAMP was conjectured in \cite{Ma2016} and rigorously proved in \cite{Takeuchi2017, Takeuchi2019}. A similar proof is given in \cite{Rangan2016} for VAMP, which is also applicable to OAMP due to the algorithmic equivalence between the two. The discussions below are inspired by these background works. In particular,  Proposition \ref{Pro:Haar} and Lemma \ref{Lem:Cond_Haar_IIDG} are inspired by \cite[Lemmas 3 and 5]{Rangan2016}, \cite[Lemmas 1 and 3]{Takeuchi2017} and \cite[Lemmas 1 and 3]{Takeuchi2019}. Also, the use of \eqref{Eqn:conds} follows the Bolthausen’s conditioning technique used in \cite{Rangan2016, Takeuchi2017, Takeuchi2019}.}

\LL{We will focus on half of an iteration in the symmetric model in Fig. \ref{Fig:model}, which is much simpler than tracking a full iteration as in \cite{Rangan2016, Takeuchi2017, Takeuchi2019}. This, together with several conjectures (i.e., Conjectures \ref{Conj:E_Var}, \ref{Conj:part_U_IIDG} and \ref{Conj:ESteinLemma_PIIDG}, which can be bridged by the rigorous treatments in \cite{Rangan2016, Takeuchi2017, Takeuchi2019}), results in conciseness in derivation. Our aim is to provide a clear insight into the mechanism of OAMP. Combining the discussions in Appendixes \ref{APP:Pro_PIIDG} and \ref{Sec:IIDG_proof}, we show that orthogonality suppresses the correlated error component in the iterative process, and hence ensures the correctness of SE.}

\subsection{Haar Distribution under a Linear Constraint}\label{APP:Haar_LC}
Fix $\bf{A}\in \bb{R}^{N\times m}$ and $\bf{B}\in \bb{R}^{N\times m}$. We first assume that $\bf{A}$ and $\bf{B}$ are fixed. Let $\bf{V}$ be a size $N\times N$ Haar distributed matrix. Denote by $\bf{\cal{U}}_{\bf{A}=\bf{VB}}$ the subset of $\bf{V}$ in the Haar ensemble meeting the following linear constraint $\bf{A}=\bf{VB}$.
The  following lemma is straightforward. 

\begin{lemma}\label{Pro:Haar_unifd}
Let $\bf{V}$ be Haar distributed independent of $\bf{A}$ and $\bf{B}$. The elements in $\bf{\mathcal{U}}_{\bf{A}=\bf{VB}}$ are uniformly distributed.
\end{lemma}

%$\bf{V}$ is said to be Haar constrained by $\bf{G}=\bf{VF}$ if $\bf{V}$ is uniformly distributed over $\bf{\mathcal{U}}_{\bf{G}=\bf{VF}}$, i.e., $p_V(\bf{V}) = p_V(\bf{V\Omega})$ provided that both $\bf{V}$ and $\bf{V\Omega}$ are in $\bf{\mathcal{U}}_{\bf{G}=\bf{VF}}$.
 
Let $\bf{U}_{\bf{A}}=[\bf{U}_{\bf{A}}^\parallel,\bf{U}_{\bf{A}}^\perp]$ and $\bf{U}_{\bf{B}}=[\bf{U}_{\bf{B}}^\parallel,\bf{U}_{\bf{B}}^\perp]$ be two orthogonal matrices with $\bf{U}_{\bf{A}}^\parallel\in\mr{Sp}(\bf{A})$, $\bf{U}_{\bf{A}}^\perp\in\mr{Sp}(\bf{A})^\perp$, $\bf{U}_{\bf{B}}^\parallel\in\mr{Sp}(\bf{B})$ and $\bf{U}_{\bf{B}}^\perp\in\mr{Sp}(\bf{B})^\perp$. Let
\BE\label{Eqn:U_parallel}
\bf{U}_{\bf{A}}^\parallel=\bf{V}\bf{U}_{\bf{B}}^\parallel.
\EE
From $\bf{A}=\bf{VB}$ and \eqref{Eqn:U_parallel}, $(\bf{U}_{\bf{A}}^\parallel)^{\rm T}\bf{VU}_{\bf{B}}^\parallel=\bf{I}$, $(\bf{U}_{\bf{A}}^\parallel)^{\rm T}\bf{VU}_{\bf{B}}^\perp =\bf{0}$ and $(\bf{U}_{\bf{A}}^\perp)^{\rm T}\bf{VU}_{\bf{B}}^\parallel=\bf{0}$, so
\BS\label{Eqn:V_tildeV}\BE\label{Eqn:UVW_a}
\bf{U}_{\bf{A}}^{\rm T}\bf{VU}_{\bf{B}}=\left[ \!\!\begin{array}{l}
(\bf{U}_{\bf{A}}^\parallel)^{\rm T}\vspace{0.1cm}\\
(\bf{U}_{\bf{A}}^\perp)^{\rm T}
\end{array} \!\!\right]\!\bf{V}\!\left[ \!\!{\begin{array}{*{20}{c}}
{\bf{U}_{\bf{B}}^\parallel,}&{\!\!\bf{U}_{\bf{B}}^\perp}
\end{array}} \!\!\right]=\left[ \!\!{\begin{array}{*{20}{c}}
{\bf{I}}&{\bf{0}}\\
{\bf{0}}&{\widetilde{\bf{V}}}
\end{array}} \!\!\right],
\EE
where
\BE\label{Eqn:UVW_b}
\widetilde{\bf{V}}\equiv(\bf{U}_{\bf{A}}^\perp)^{\rm T}\bf{VU}_{\bf{B}}^\perp.
\EE\ES
Left- and right multiplying \eqref{Eqn:UVW_a} by $\bf{U}_{\bf{A}}$ and $\bf{U}_{\bf{B}}^{\rm T}$  gives
\BS\label{Eqn:UVW}\begin{align}
 \!\!  \bf{V} \!\!=\!\! \left[ \!\!\! {\begin{array}{*{20}{c}}
{{\bf{U}_{\bf{A}}^\parallel },}&{{ \!\!\!\bf{U}_{\bf{A}}^ {\!\perp} }}
\end{array}}  \!\!\!\right]\!\!\left[  \!\!\!{\begin{array}{*{20}{c}}
\bf{I}&{\!\!\bf{0}}\\
{\bf{0}}&{{\!\!\widetilde{\bf{V}} }}
\end{array}} \! \!\!\!\right]\!\!\left[ \!\!\!\!\begin{array}{l}
(\bf{U}_{\bf{B}}^\parallel)^{\!\rm T}\vspace{0.1cm}\\
(\bf{U}_{\bf{B}}^\perp)^{\!\rm T}
\end{array} \!\!\!\!\right] \!\!= \bf{U}_{\bf{A}}^\parallel(\bf{U}_{\bf{B}}^\parallel)^{\!\rm T}\!\!+\!\bf{U}_{\bf{A}}^{\!\perp}\widetilde{\bf{V}} (\bf{U}_{\bf{B}}^{\!\perp})^{\!\rm T}\!.\label{Eqn:UVW_c}
\end{align}  
\ES
From \eqref{Eqn:V_tildeV}, if $\bf{V}$ is orthogonal, so is  $\widetilde{\bf{V}}$ and vice versa. Thus \eqref{Eqn:UVW_b} and \eqref{Eqn:UVW_c} establish a one-to-one affine mapping between $\bf{V}\in\bf{\mathcal{U}}_{\bf{A}=\bf{VB}}$ and $\widetilde{\bf{V}}\in\bf{\mathcal{U}}^{N-m}$, denoted as $\bf{V} \leftrightarrow\widetilde{\bf{V}}$. Fix a pair $\bf{V}\leftrightarrow\widetilde{\bf{V}}$. For any $\bf{\Omega}$ such that $\bf{V\Omega}\in\bf{\mathcal{U}}_{\bf{A}=\bf{VB}}$, we can always find a $\widetilde{\bf{\Omega}}$ such that $\widetilde{\bf{V}}\widetilde{\bf{\Omega}}\in\bf{\mathcal{U}}^{N-m}$, and vice versa. From Lemma \ref{Pro:Haar_unifd}, $\bf{V}$ is uniformly distributed over $\bf{\mathcal{U}}_{\bf{A}=\bf{VB}}$. Then for $\bf{\Omega}$ and $\widetilde{\bf{\Omega}}$ mentioned above, $p_V(\bf{V}) = p_V(\bf{V\Omega})$ and, due to the one-to-one affine mapping, 
\BE\label{Eqn:V_bot}
p_{\widetilde{V}}(\widetilde{\bf{V}}) = p_{\widetilde{V}}(\widetilde{\bf{V}}\widetilde{\bf{\Omega}}).
\EE
Eqn. \eqref{Eqn:V_bot} holds $\forall \widetilde{\bf{\Omega}}\in\bf{\mathcal{U}}^{N-m}$, so $\widetilde{\bf{V}}$ is Haar. This is summarized as follows.

%Now consider $\bf{Va}$ where $\bf{a}$ is a fixed vector. From Assumption \ref{Def:Haar_pro}, when $N-n\to\infty$, $\bf{e}\equiv\widetilde{\bf{V}}\bf{W}_\perp^{\rm T}\bf{a}\to$IIDG. Let $p_e({e})$ be the PDF of the entries of $\bf{e}$. We pad $n$ entries IID drawn from $p_e({e})$ to ${\bf{e}}$ to form a length-$N$ vector $\tilde{\bf{e}}$. Clearly, $\tilde{\bf{e}}$ is IIDG, so is $\bf{U}\tilde{\bf{e}}$. When $N-n\to\infty$, $\bf{Ua}=\bf{U}_\perp{\bf{e}}\to\bf{U}\tilde{\bf{e}}$, so $\bf{Va}$ is IIDG.

\begin{proposition}\label{Pro:Haar}
Let $\bf{A}$ and $\bf{B}$ be two $N\!\!\times\! m$ matrices of full column rank. Let $\bf{U}_{\bf{A}}^{\!\perp}$ and $\bf{U}_{\bf{B}}^{\!\perp}$ be respectively fixed orthonormal bases of $\mr{Sp}(\bf{A})$ and $\mr{Sp}(\bf{B})$. Assume that $\bf{V}$ is uniformly distributed over $\bf{\mathcal{U}}_{\bf{A}=\bf{VB}}$. Then $\widetilde{\bf{V}}\!\equiv\!(\bf{U}_{\bf{A}}^{\!\perp})^{\rm T}\bf{VU}_{\bf{B}}^\perp$ is Haar over $\bf{\mathcal{U}}^{N\!-\!m}$. 
\end{proposition}  

\LL{There is no loss of generality in the full-rank assumption in Proposition \ref{Pro:Haar}. Let the ranks of $\bm{A}_{N\times m}$ and $\bm{B}_{N\times m}$ be $m'$. When $m'<m$ (not full rank), we can obtain $\bm{A}'=\bm{VB}'$ with full-rank $\bm{A}'$ and $\bm{B}'$ by removing the linear correlated columns in $\bm{A}$ and $\bm{B}$. Proposition \ref{Pro:Haar} still applies to $\bm{A}'=\bm{VB}'$.}

\subsection{Random Part of a Constraint Haar Transform}
Let $\bf{V}\sim \bf{\cal{H}}^N$, $\bf{f}\in \bb{R}^N$ and
\BE
    \bf{g}=\bf{Vf}.
\EE
From \eqref{Eqn:UVW}, we  write $\bf{g}\!=\!\bf{g}_{\rm fixed} + \bf{g}_{\rm random}$ under $\bf{A}\!=\!\bf{VB}$ with 
 \begin{align}\label{Eqn:g_random}
    \bf{g}_{\rm fixed} = \bf{U}_{\bf{A}}^\parallel(\bf{U}_{\bf{B}}^\parallel)^{\rm T}\bf{f}, \qquad
    \bf{g}_{\rm random} = \bf{U}_{\bf{A}}^{\!\perp}\widetilde{\bf{V}} (\bf{U}_{\bf{B}}^{\!\perp})^{\rm T} \bf{f}.
\end{align} 
Recall from Proposition \ref{Pro:Haar} that $\widetilde{\bf{V}}\sim\bf{\cal{H}}^{N-m}$. Hence $\bf{g}_{\rm random}$ is the random part of $\bf{g}$. Let $N\to\infty$ and $m$ remain fixed. Then from Lemma \ref{Lem:PIIDG}, $\widetilde{\bf{V}} (\bf{U}_{\bf{B}}^{\!\perp})^{\rm T} \bf{f}$ is PIIDG with zero mean. Then from \LL{Conjecture \ref{Conj:part_U_IIDG} (see Appendix \ref{APP:orth_trans})} and \eqref{Eqn:g_random} (and since $\bf{U}_{\bf{A}}^{\!\perp}$ is of size $N\times (N-m)$), $\bf{g}_{\rm random}$ is PIIDG with zero mean. 

The above is for fixed $\bf{A}$ and $\bf{B}$. We now consider random $\bf{A}$ and $\bf{B}$. In this case, we define $\bf{\cal{U}}_{\bf{A}=\bf{VB}}$ using samples of $\bf{A}$ and $\bf{B}$. From Proposition \ref{Pro:Haar}, $\widetilde{\bf{V}}$  is not a function of $\bf{A}$ and $\bf{B}$ except the size of $\widetilde{\bf{V}}$. Let $N\to\infty$ and $m$ remain finite. Then from Property \ref{Lem:VT_Gau},  $\bf{g}_{\rm random}$ in \eqref{Eqn:g_random} is independent of $\bf{B}$. From \LL{Conjecture \ref{Conj:part_U_IIDG} (see Appendix \ref{APP:orth_trans})}, $\bf{g}_{\rm random}$ in \eqref{Eqn:g_random} is asymptotically independent of the columns of $\bf{U}_{\bf{A}}^{\!\perp}$ and so also independent of the columns of $\bf{A}$.  Hence $\bf{g}_{\rm random}$ is asymptotically independent of both $\bf{A}$ and $\bf{B}$. Furthermore, due to the one-to-one mapping between $\bf{V}\in\bf{\cal{U}}_{\bf{A}=\bf{VB}}$ and $\widetilde{\bf{V}}$  (see the discussions below \eqref{Eqn:UVW}), any $\bf{z}$ independent of $\bf{V}$ is also independent of $\widetilde{\bf{V}}$  and $\bf{g}_{\rm random}$. We summarize the above in a lemma below.

\begin{lemma}\label{Lem:Cond_Haar_IIDG}
    Let $\bf{V}$ be Haar distributed under the constraint $\bf{A}=\bf{VB}$. Let $N\to \infty$ and $m$ remains finite. Then $\bf{g}_{\rm random}$ is PIIDG with zero mean and asymptotically independent of $\bf{A}$ and $\bf{B}$. Furthermore, $\bf{g}_{\rm random}$ is asymptotically independent of any $\bf{z}$ if $\bf{z}$ is independent of $\bf{V}$. 
\end{lemma}

Lemma \ref{Lem:Cond_Haar_IIDG} is a generalization of Lemma \ref{Lem:PIIDG}. The latter is for $\bf{A}$ and $\bf{B}$ being empty sets, so $\bf{g}_{\rm fixed}$ vanishes and $\bf{Vf}=\bf{g}_{\rm random}$.  Note the subtle difference in the treatments of  $\bf{A}$, $\bf{B}$ and $\bf{z}$ in Lemma \ref{Lem:Cond_Haar_IIDG}. We cannot claim both $\bf{A}$ and $\bf{B}$ are independent of $\bf{V}$ since $\bf{A}=\bf{VB}$. For any finite $N$, $\bf{g}_{\rm random}$ is a function of $\bf{A}$ and $\bf{B}$, but the related influence becomes negligible when $N\to\infty$.   \vspace{-3mm}

\subsection{Proof of Theorem \ref{THE:IIDG}}\label{APP:IIDG_subsec}
The equations below are based on \eqref{Eqn:Ite_Proc_para}, \eqref{Eqn:errors}, \eqref{Eqn:Error_matrixc} and   \eqref{Eqn:G=VF}:\vspace{-0.4cm}
\BS\label{Eqn:conds}\begin{alignat}{2}
    \bf{A}_t &= \left[\bf{\mathcal{X}},  \bf{G}^{\mr{out}}_t, \bf{G}^{\mr{in}}_t\right],   &\quad \bf{B}_t& = \left[ \bf{x}, \bf{F}^{\mr{in}}_t, \bf{F}^{\mr{out}}_t\right],\label{Eqn:cond_a}\\
    \bf{A}_t &=\bf{VB}_t,&&\label{Eqn:cond_b}\\
    \bf{g}^{\mr{in}}_t &=\bf{V}\bf{f}^{\mr{out}}_t, &\quad \bf{f}^{\mr{in}}_t &=\bf{V}\bf{g}^{\mr{out}}_t.\label{Eqn:cond_c}
\end{alignat}\ES
Combining \eqref{Eqn:UVW}, \eqref{Eqn:cond_a} and \eqref{Eqn:cond_b}, we have
\begin{align} 
        \bf{V}  &= [\bf{u}^\parallel_{\bf{\mathcal{X}}}, \bf{U}_{\!\bf{G}_t^{\rm out}}^\parallel] [\bf{u}^\parallel_{\bf{x}}, \bf{U}_{\!\bf{F}_t^{\rm in}}^\parallel]^{\rm T} +\bf{U}_{\!\bf{G}_t^{\rm in}}^\parallel(\bf{U}_{\!\bf{F}_t^{\rm out}}^\parallel)^{\rm T} \nonumber\\
        & \quad+  \bf{U}^{\!\perp}_{\!\bf{A}_t}\widetilde{\bf{V}} (\bf{U}^\perp_{\!\bf{B}_t})^{\rm T}. \label{Eqn:V_Decomp}
\end{align} 
Substitute \eqref{Eqn:V_Decomp} into \eqref{Eqn:cond_c}:
\BS\label{Eqn:g_decom}\begin{align}
    \bf{g}^{\mr{in}}_t &=\bf{V}\bf{f}^{\mr{out}}_t\\
    &= \underbrace{[\bf{u}^\parallel_{\bf{\mathcal{X}}}, \bf{U}_{\!\bf{G}_t^{\rm out}}^\parallel] [\bf{u}^\parallel_{\bf{x}}, \bf{U}_{\!\bf{F}_t^{\rm in}}^\parallel]^{\rm T}\bf{f}^{\mr{out}}_t}_{\bf{\delta}_{t_1}} + \underbrace{\bf{U}_{\!\bf{G}_t^{\rm in}}^\parallel(\bf{U}_{\!\bf{F}_t^{\rm out}}^\parallel)^{\rm T}\bf{f}^{\mr{out}}_t}_{\bf{\delta}_{t_2}} \nonumber\\
    &\quad +  \underbrace{\bf{U}^{\!\perp}_{\!\bf{A}_t}\widetilde{\bf{V}} (\bf{U}^\perp_{\!\bf{B}_t})^{\rm T}\bf{f}^{\mr{out}}_t}_{\bf{\delta}_t^{\rm random}}.
\end{align}\ES
For OAMP under \eqref{Eqn:Orthogonality}, $ \tfrac{1}{N}[\bf{x}, \bf{F}^{\mr{in}}_t]^{\rm T}\bf{f}^{\mr{out}}_t \overset{\rm LLN}{\longrightarrow}\bf{0}$. Hence in \eqref{Eqn:g_decom} $ \bf{\delta}_{t_1} =\bf{o}(1) $ and \vspace{-2mm}
\BE\label{Eqn:delta_23}
    \bf{g}^{\mr{in}}_t \to \bf{\delta}_{t_2} + \bf{\delta}_t^{\rm random}.
\EE
Since $\bf{U}^\parallel_{\bf{G}_t^{\rm in}}\in {\rm Sp}(\bf{G}_t^{\rm in})$, we can find a $\bf{c}_t$ such that\vspace{-2mm}
\BE\label{Eqn:delta_2}
    \bf{\delta}_{t_2} = \bf{G}^{\mr{in}}_t\bf{c}_t. %\bf{G}^{\mr{in}}_t\big(\bf{F}^{\mr{out}}_t \big)^{\rm T} \bf{f}^{\mr{out}}_t=
\EE
From Lemma \ref{Lem:Cond_Haar_IIDG}, $\bf{\delta}_t^{\rm random}$ is PIIDG and independent of $\bf{A}_t, \bf{B}_t$ and $\bf{z}$. (Note: $\bf{\mathcal{X}}$ is a column in $\bf{A}_t$). Following \eqref{Eqn:delta_23} and \eqref{Eqn:delta_2}, the GS error $\bf{g}^{\mr{in}}_{t}$ is a linear combination of independent PIIDG vectors $\{\bf{\delta}_0^{\rm random}, \cdots,\bf{\delta}^{\rm random}_{t}\}$, which are zero mean and independent of $\bf{\mathcal{X}}$ and $\bf{z}$. Then, we have Lemma \ref{Lem:indep_PIIDG}.

\begin{lemma}\label{Lem:indep_PIIDG}
    We can find an upper triangular $\bf{C}_t$ such that
    \BE
        \bf{G}^{\mr{in}}_t \to [\bf{\delta}_0^{\rm random} \dots \bf{\delta}_{t-1}^{\rm random}]\bf{C}_t,
    \EE  
    where $\{\bf{\delta}_0^{\rm random}, \cdots,\bf{\delta}_{t-1}^{\rm random}\}$ are independent PIIDG  vectors with zero mean and are independent of $\bf{\mathcal{X}}$ and $\bf{z}$.
\end{lemma}

\LLC{Let $\bf{a}$ and $\bf{b}$ be two PIIDG  vectors and $\bf{c}=\alpha \bf{a} + \beta \bf{b}$ where $\alpha$ and $\beta$ are two constants. It can be shown that $\bf{c}$ is PIIDG.} Then following Lemma \ref{Lem:indep_PIIDG}, we obtain the claim (a) in Theorem \ref{THE:IIDG}. Due to the symmetry of the problem, we can prove claim (b) in a similar way. This completes the proof for Theorem \ref{THE:IIDG}.\vspace{-3mm}
 
%We now prove claim (a) using an induction on $t$. At $t=1$, claim (a) trivially holds with initialization  $\bf{g}_0^{\rm in} = \bf{0}$ that can be regarded as PIIDG with zero mean and zero variance. Such errors are independent of $\bf{\mathcal{X}}$ and $\bf{z}$, where $\bf{z}$ is an arbitrary variable independent of $\bf{V}$. 

% \LC{Now assume that, up to $t\ge 1$,  $\bf{G}^{\rm in}_t$ is CIIDG-{\color{blue}RJG} and independent of $\bf{\mathcal{X}}$ and $\bf{z}$. Then we have the following: 
% \begin{itemize}
%     \item From \eqref{Eqn:delta_2}, $[\bf{G}^{\rm in}_t, \bf{\delta}_{t_2}]$ is asymptotically CIIDG-{\color{blue}RJG} and independent of $\bf{\mathcal{X}}$ and $\bf{z}$. 
%     \item 
% \end{itemize}} 

\subsection{Discussions}
Here are some intuitions. Recall from \eqref{Eqn:alpha_SE} and \eqref{Eqn:Ite_Proc_para} that
\BE
    \alpha_t^{\gamma}\bf{\mathcal{X}}+{\bf{g}}^{\mr{out}}_{t} \!=\! \gamma_t\big(\alpha_{t-1}^{\phi}\bf{\mathcal{X}}+{\bf{g}}^{\mr{in}}_{t-1} \big),
\EE
where ${\bf{g}}^{\mr{out}}_{t}$ and ${\bf{g}}^{\mr{in}}_{t-1}$ are errors. Heuristically, in an iterative process, we should avoid correlation between ${\bf{g}}^{\mr{in}}_{t-1}$ and $\bf{\mathcal{X}}$,  otherwise the design of $\gamma_t $ becomes very complicated. Also, we should avoid correlation between ${\bf{g}}^{\mr{in}}_{t}$ and ${\bf{G}}^{\mr{out}}_{t}$. Such correlation may result in positive feedback in an iterative process and cause stability problem. %See the discussions in \ref{Sec:OAMP}. %since ${\bf{G}}^{\mr{out}}_{t}$ contains previous errors in the output of $\gamma_t $.

OAMP achieves such desirable correlation avoidance. To see this, rewrite \eqref{Eqn:g_decom} as
\BE
    \bf{g}^{\mr{in}}_t 
     = \underbrace{[\bf{u}^\parallel_{\bf{\mathcal{X}}}, \bf{U}_{\!\bf{G}_t^{\rm out}}^\parallel] [\bf{u}^\parallel_{\bf{x}}, \bf{U}_{\!\bf{F}_t^{\rm in}}^\parallel]^{\rm T}\bf{f}^{\mr{out}}_t}_{\bf{\delta}_{t_1}} + \bf{\delta}_{t_2} +\bf{\delta}_t^{\rm random}.
\EE
Clearly, $\bf{\delta}_{t_1}$ is potentially correlated with $\bf{\mathcal{X}}$ and ${\bf{G}}^{\mr{out}}_{t}$  since  $\bf{\delta}_{t_1}\in{\rm Sp}([\bf{\mathcal{X}}, {\bf{G}}^{\mr{out}}_{t}])$. In OAMP, $\bf{\delta}_{t_1}$ is statistically suppressed by a properly designed $\phi_t $, such that  $\tfrac{1}{N} [\bf{u}^\parallel_{\bf{x}}, \bf{U}_{\!\bf{F}_t^{\rm in}}^\parallel]^{\rm T}\bf{f}^{\mr{out}}_t \overset{\rm LLN}{\longrightarrow}\bf{0}$ as shown below \eqref{Eqn:g_decom}. Furthermore, we showed that $\bf{\delta}_{t_2}$ and $\bf{\delta}_t^{\rm random}$ will not cause the correlation problem above. 

In summary, the orthogonality of $\phi_t $ prevents the output of $\gamma_t $ from circulating back (partially or fully) to the input of $\gamma_{t+1} $. The similar can be said for $\phi_t $. Being orthogonal, the two local estimators help each other; each stops the input-output error circulation for the opposite one.

\section{Proof of Theorem \ref{THE:orth}}\label{APP:orth}
We  prove Theorem \ref{THE:orth} by induction.

 \subsubsection{Eqn. \texorpdfstring{\eqref{Eqn:Orthogonality}}{TEXT} holds for \texorpdfstring{$t=1$}{TEXT}}  
 
     \LL{Since $\gamma_{1}$ and $\phi_{1}$ are orthogonal estimators,} we have \eqref{Eqn:Orthogonalitya} for $t=1$. From the definition of the GS model in \eqref{Eqn:GP_model}, we have 
    \begin{align}\label{Eqn:orth_out}
        {\rm E}\{\bf{\mathcal{X}}^{\rm T}\bf{g}_1^{\rm out}\} =0, \qquad
        {\rm E}\{\bf{x}^{\rm T}\bf{f}_1^{\rm out}\}=0. 
    \end{align} 
    Furthermore, \LLC{$\bf{f}^{\mr{out}}_{1} =\phi_1(\bf{0})-\bf{x}$ is IID since  $\phi_1$ is separable-IID and $\bf{x}$ is IID.} Then, from \eqref{Eqn:orth_out} and Lemma \ref{Pro:Chev_IID}, we have
    \BE
       \tfrac{1}{N}\bf{x}^{\rm T} \bf{f}^{\mr{out}}_{1}  \overset{\rm LLN}{\longrightarrow}  {0}.
    \EE
   In addition, \LLC{$\bf{\mathcal{X}}$ and $\bf{g}_0^{\rm in}$ are independent PIIDG vectors.} Then $\bf{g}_1^{\rm out} = \gamma_1(\bf{0})-\bf{\mathcal{X}}$ is PIID since $\gamma_1$ is separable-IID. Then, from  \eqref{Eqn:orth_out} and Property \ref{Pro:PIID_LLN}, we have
    \BE
        \tfrac{1}{N}\bf{\mathcal{X}}^{\rm T}\bf{g}_1^{\rm out}    
        \overset{\rm LLN}{\longrightarrow}  {0}.
    \EE 
  Therefore, \eqref{Eqn:Orthogonalityb} holds for $t=1$.

 \subsubsection{Eqn. \texorpdfstring{\eqref{Eqn:Orthogonality}}{TEXT} holds for \texorpdfstring{$t-1\Rightarrow$}{TEXT} Eqn. \texorpdfstring{\eqref{Eqn:Orthogonality}}{TEXT} holds for \texorpdfstring{$t$}{TEXT}}
       
        Since \eqref{Eqn:Orthogonality} holds for $t-1$, from Theorem \ref{THE:IIDG} (see also the proof in Appendix \ref{Sec:IIDG_proof}), we have the following.
    \begin{enumerate}[(a)] 
     \item  $\left[\bf{g}_0^{\mr{in}},\dots,\bf{g}_{t-1}^{\mr{in}}\right]$ is \LLC{CPIIDG-RJG} and independent of  $\bf{\mathcal{X}}$;
    \item   $\left[\bf{f}_0^{\mr{in}},\dots,\bf{f}_{t-1}^{\mr{in}}\right]$ is \LLC{CPIIDG-RJG} and independent of $\bf{x}$. 
\end{enumerate} 
    Since $\gamma_{t}$ and $\phi_{t}$ are orthogonal estimators, we have
    \begin{align}\label{Eqn:orth_t}
        {\rm E}\{(\bf{g}_{t-1}^{\rm in})^{\rm T}\bf{g}_t^{\rm out}\} =0,\qquad
          {\rm E}\{(\bf{f}_{t-1}^{\rm in})^{\rm T}\bf{f}_t^{\rm out}\} =0.
    \end{align}  
     From  (a) and (b), \LLC{$\bf{g}^{\mr{in}}_{t-1}$ and $\bf{f}^{\mr{in}}_{t-1}$ are respectively PIIDG. Since $\gamma_{t}$ and $\phi_{t}$ are separable-IID, $\bf{g}^{\mr{out}}_{t} = \phi_t(\bf{g}^{\mr{in}}_{t-1} )-\bf{x}$ and $\bf{f}^{\mr{out}}_{t} = \gamma_t(\bf{f}^{\mr{in}}_{t-1} )-\bf{\mathcal{X}}$ are PIID.} Therefore, from \eqref{Eqn:orth_t} and Property \ref{Pro:PIID_LLN}, we have
   \begin{align}\label{Eqn:t_1_orth_t}
        \tfrac{1}{N}(\bf{g}_{t-1}^{\rm in})^{\rm T}\bf{g}_t^{\rm out}    \overset{\rm LLN}{\longrightarrow}  {0},  \qquad 
        \tfrac{1}{N}(\bf{f}_{t-1}^{\rm in})^{\rm T}\bf{f}_t^{\rm out}  \overset{\rm LLN}{\longrightarrow}  {0}.
    \end{align} 
    Then, from Conjecture \ref{Conj:ESteinLemma_PIIDG} and the joint Gaussianity in (a) and (b), the following hold for all $t'<t-1$.   
     \begin{align} \label{Eqn:pre_orth_t}
        \tfrac{1}{N}(\bf{g}_{t'}^{\rm in})^{\rm T}\bf{g}_t^{\rm out}    \overset{\rm LLN}{\longrightarrow}  {0},   \qquad
        \tfrac{1}{N}(\bf{f}_{t'}^{\rm in})^{\rm T}\bf{f}_t^{\rm out}  \overset{\rm LLN}{\longrightarrow}  {0}.
    \end{align} 
    Therefore, \eqref{Eqn:Orthogonalitya} holds for $t$. 
    
    From the definition of the GS model in \eqref{Eqn:GP_model}, we have 
    \begin{align}\label{Eqn:orth_GS_t}
        {\rm E}\{\bf{\mathcal{X}}^{\rm T}\bf{g}_t^{\rm out}\} =0, \qquad
        {\rm E}\{\bf{x}^{\rm T}\bf{f}_t^{\rm out}\} =0.
    \end{align} 
    Since \LLC{$\bf{x}$, $\bf{\mathcal{X}}$ $\bf{f}^{\mr{out}}_{t}$ and $\bf{g}_t^{\rm out}$ are PIID}, from  Property \ref{Pro:PIID_LLN}  and \eqref{Eqn:orth_GS_t}, we have  
      \begin{align}
        \tfrac{1}{N}\bf{x}^{\rm T} \bf{f}^{\mr{out}}_{t}   \overset{\rm LLN}{\longrightarrow}  {0}, \qquad
        \tfrac{1}{N} \bf{\mathcal{X}}^{\rm T}\bf{g}_t^{\rm out}    
         \overset{\rm LLN}{\longrightarrow}  {0}.
    \end{align} 
    Therefore, \eqref{Eqn:Orthogonalityb} holds for $t$.


\begin{thebibliography}{1}
\bibitem{Donoho2009}
D.~L. Donoho, A.~Maleki, and A.~Montanari, ``Message-passing algorithms for compressed sensing,''  in \textit{Proc. Nat. Acad. Sci.}, vol. 106, no.~45, Nov. 2009. 
\bibitem{Lodge1993}
J. Lodge, R. Young, P. Hoeher and J. Hagenauer, ``Separable MAP ``filters" for the decoding of product and concatenated codes,'' in \textit{IEEE International Conference on Communications (ICC)}, vol.~2, May 1993, pp. 1740--1745.
\bibitem{Moon1996EM}
T. K. Moon, ``The expectation-maximization algorithm,'' \textit{IEEE Signal Process. Mag.}, vol. 13, no. 6, pp. 47-60, Nov. 1996.

% \bibitem{Naffouri}
% T. Y. Al-Naffouri, ``An EM-based forward-backward Kalman filter for the estimation of time-variant channels in OFDM,'' \textit{IEEE Trans. Signal
%   Process.}, vol. 55, no. 7, pp. 3924-3930, July 2007.
\bibitem{Loeliger2009EM}
J. Dauwels, A. Eckford, S. Korl, H. A. Loeliger, ``Expectation maximization as message passing-part I: Principles and Gaussian messages,'' \textit{arXiv preprint arXiv:0910.2832}, 2009. 
\bibitem{TurboCode}
C.~Berrou, A.~Glavieux, and P.~Thitimajshima, ``Near shannon limit
  error-correcting coding and decoding: Turbo-codes,'' in \textit{IEEE
  International Conference on Communications (ICC)}, vol.~2, May 1993, pp.
  1064--1070.
\bibitem{Richardson2001}
T.~Richardson and R.~Urbanke, ``The capacity of low-density parity-check codes
  under message-passing decoding,'' \textit{{IEEE} Trans. Inf. Theory}, vol.~47,
  no.~2, pp. 599--618, Feb. 2001.

% \bibitem{Brink2001}
% S.~ten Brink, ``Convergence behavior of iteratively decoded parallel
%   concatenated codes,'' \textit{{IEEE} Trans. Commun.}, vol.~49, no.~10, pp.
%   1727--1737, Oct 2001.
% \bibitem{Brink2004}
% S.~ten Brink, G. Kramer, and A. Ashikhmin, ``Design of low-density parity-check codes for modulation and detection,'' \textit{IEEE Trans. Commun.}, vol. 52, no. 4, pp. 670-678, April 2004.
\bibitem{Wang1999}
X.~Wang and H.~V. Poor, ``Iterative (turbo) soft interference cancellation and decoding for coded \protect{CDMA},'' \textit{{IEEE} Trans. Commun.}, vol.~47, no.~7, pp. 1046--1061, Jul 1999.
\bibitem{Douillard1995}
 C. Douillard et al., ``Iterative correction of intersymbol interference: Turbo-equalization,'' \textit{Eur. Trans. Telecommun.}, vol. 6, no. 5, pp. 507-511, Sep./Oct. 1995.
\bibitem{Tuchler2002}
M. T$\ddot{\mr{u}}$chler, R. Koetter, and A. C. Singer, ``Turbo equalization: Principles
and new results,'' \textit{IEEE Trans. Commun.}, vol. 50, no. 5, pp. 754-767, May 2002.
\bibitem{Loeliger2007}
H. A. Loeliger, J. Dauwels, J. Hu, S. Korl, L. Ping, and F. R. Kschischang, ``The factor graph approach to model-based signal processing,'' \textit{Proc. IEEE}, vol. 95, no. 6, pp. 1295-1322, June 2007.
% \bibitem{CSXiao2011}
% J. Tao, J. Wu, Y. R. Zheng and C. Xiao, ``Enhanced MIMO LMMSE turbo equalization: Algorithm, simulations, and undersea experimental results,'' \textit{IEEE Trans. Signal Process.}, vol. 59, no. 8, pp. 3813-3823, Aug. 2011.
\bibitem{Yuan2014}
X. Yuan, L. Ping, C. Xu and A. Kavcic, ``Achievable rates of MIMO systems with linear precoding and iterative LMMSE detector,'' \textit{IEEE Trans. Inf. Theory}, vol. 60, no.11, pp. 7073-7089, Oct. 2014.
\bibitem{LiuLei2019TSP}
L.~{Liu}, Y.~{Chi}, C.~{Yuen}, Y.~L. {Guan}, and Y.~{Li}, ``Capacity-achieving
  {MIMO-NOMA}: Iterative LMMSE detection,'' \textit{IEEE Trans. Signal
  Process.}, vol.~67, no.~7, pp. 1758--1773, April 2019.
\bibitem{Minka2001}
T.~P. Minka, ``Expectation propagation for approximate \LL{Bayesian} inference,'' in
  \textit{Proceedings of the Seventeenth conference on Uncertainty in artificial
  intelligence}, 2001, pp. 362--369.
\bibitem{Cakmak2018}
B.~{\c{C}}akmak and M.~Opper, ``Expectation propagation for approximate
  inference: Free probability framework,'' \textit{arXiv preprint
  arXiv:1801.05411}, 2018.     
\LL{\bibitem{Bolthausen2014}
  Bolthausen, E., ``An iterative construction of solutions of the TAP equations for the Sherrington-Kirkpatrick model'', \textit{Communications in Mathematical Physics}, vol. 325, no. 1, pp. 333–366, 2014.}
% \bibitem{MengVTC2015}
% X. Meng, S. Wu, L. Kuang, Z. Ni and J. Lu, ``Expectation propagation based iterative multi-user detection for MIMO-IDMA systems,'' \textit{2014 IEEE 79th VTC (VTC Spring)}, Seoul, 2014, pp. 1-5.
\bibitem{Takeuchi2017}
K.~Takeuchi, ``Rigorous dynamics of expectation-propagation-based signal recovery from unitarily invariant measurements,'' \textit{{IEEE} Trans. Inf. Theory}, vol. 66, no. 1, 368 - 386, Oct. 2019.  
\bibitem{Ma2016}
J.~Ma and L.~Ping, ``Orthogonal {AMP},'' \textit{IEEE Access}, vol.~5, pp. 2020--2033, 2017, preprint arXiv:1602.06509, 2016.
% \bibitem{MaSPL2015a}
% J. Ma, X. Yuan and L. Ping, ``Turbo compressed sensing with partial DFT sensing matrix,'' \textit{IEEE Signal Process. Lett.}, vol. 22, no. 2, pp. 158-161, Feb. 2015.
% \bibitem{Ma_SPL2015b}
% J. Ma, X. Yuan and L. Ping, ``On the performance of Turbo signal recovery with partial DFT sensing matrices,'' \textit{IEEE Signal Process. Lett.}, vol. 22, no. 10, pp. 1580-1584, Oct. 2015.
 \bibitem{Bayati2011}
M.~Bayati and A.~Montanari, ``The dynamics of message passing on dense graphs,
  with applications to compressed sensing,''  \textit{{IEEE} Trans. Inf. Theory},
  vol.~57, no.~2, pp. 764--785, Feb. 2011.
  \bibitem{Rangan2016}
S.~Rangan, P.~Schniter, and A.~Fletcher, ``Vector approximate message passing,''  \textit{{IEEE} Trans. Inf. Theory},  vol. 65, no. 10, pp. 6664-6684, Oct. 2019. 
\bibitem{Dudeja2022}
R. Dudeja, Y. M. Lu, and S. Sen, ``Universality of approximate message passing with semi-random matrices," \textit{arXiv preprint arXiv:2204.04281}, 2022.
\bibitem{LiuAMP2019}
L. Liu, C. Liang, J. Ma, and L. Ping, ``Capacity optimality of AMP in coded systems,'' \textit{{IEEE} Trans. Inf. Theory}, vol. 67, no. 7, 4929-4445, July 2021.
\bibitem{MaLiu2018}
J. Ma, L. Liu, Y. Xiao, and L. Ping, ``On orthogonal AMP in coded linear vector systems,'' \textit{IEEE Trans. Wireless Commun.}, vol. 18, no. 12, 6558-6572, Oct. 2019.  
%\bibitem{kabashima2003cdma}
%Y.~Kabashima, ``A cdma multiuser detection algorithm on the basis of belief propagation,'' \textit{Journal of Physics A: Mathematical and General}, vol.~36, no.~43, p. 11111, 2003.

\bibitem{Takeuchi2019}
K.~Takeuchi, ``A unified framework of state evolution for message-passing algorithms,'' \textit{IEEE International Symposium on Information Theory (ISIT)}, 2019, pp. 151-155.
\bibitem{He2018AI}
H. He, C. Wen, S. Jin and G. Y. Li, ``A model-driven deep learning network for MIMO detection,'' \textit{2018 IEEE Global Conference on Signal and Information Processing (GlobalSIP)}, Anaheim, CA, USA, 2018, pp. 584-588.
\bibitem{Zhang2019AI}
J. Zhang, C. Wen, S. Jin and G. Y. Li, ``Artificial intelligence-aided receiver for a CP-free OFDM system: Design, simulation, and experimental test,'' \textit{IEEE Access}, vol. 7, pp. 58901-58914, 2019.
% \bibitem{Ito2019}
% D. Ito, S. Takabe and T. Wadayama, ``Trainable ISTA for sparse signal recovery,'' \textit{IEEE Trans. Signal Process.}, vol. 67, no. 12, pp. 3113-3125, June, 2019.
%\bibitem{Takabe2018}
%S. Takabe, M. Imanishi, T. Wadayama, K. Hayashi, ``Deep learning-aided projected gradient detector for massive overloaded MIMO channels," \textit{arXiv preprint arXiv:1806.10827v2}, 2018.
\bibitem{Takabe2019}
S. Takabe, M. Imanishi, T. Wadayama, R. Hayakawa and K. Hayashi, ``Trainable projected gradient detector for massive overloaded MIMO channels: Data-driven tuning approach,'' \textit{IEEE Access}, vol. 7, pp. 93326-93338, 2019.

\bibitem{Lei_TSP_2_2019}
Y. Cheng, L. Liu, S. Liang, J. H. Manton and L. Ping, ``Orthogonal AMP for problems with multiple measurement vectors and/or multiple transforms,''  under preparation.
 \bibitem{Fletcher2016}
A. K. Fletcher, M. Saharee, S. Rangan, and P. Schniter, ``Expectation consistent approximate inference: Generalizations and convergence.'' \textit{arXiv preprint arXiv:1602.07795v2}, 2016. 
\LL{\bibitem{Yiyao_integral}
  Y. Cheng, L. Liu, L. Ping, ``An integral-based approach to orthogonal AMP,'' \textit{IEEE Signal Process. Lett.}, vol. 28, 194-198, Dec. 2020.
  \bibitem{OAMP_ISIT22}
L. Liu, S. Liang and L. Ping, ``Capacity optimality of OAMP in coded large unitarily invariant systems," \textit{IEEE ISIT}, pp. 1384-1389, July 2022.
 \bibitem{OAMP_TCOM}
 L. Liu, S. Liang, and L. Ping, ``Capacity optimality of OAMP: Beyond IID sensing matrices and Gaussian signaling," \textit{arXiv preprint: arXiv:2108.08503}, Aug. 2021.}
\bibitem{Senel2018}
K. Senel and E. G. Larsson, ``Grant-free massive MTC-enabled massive MIMO: A compressive sensing approach,'' \textit{IEEE Trans. Commun.}, vol. 66, no. 12, pp. 6164-6175, Dec. 2018.
\bibitem{ChenAMP2019}
Z. Chen, F. Sohrabi and W. Yu, ``Multi-cell sparse activity detection for massive random access: Massive MIMO versus cooperative MIMO,'' \textit{IEEE Trans. Commun.}, vol. 18, no. 8, pp. 4060-4074, Aug. 2019.
\bibitem{LiangAMP2020}
S. Liang, C. Liang, J. Ma and L. Ping, ``Compressed coding, AMP-based decoding, and analog spatial coupling,'' \textit{IEEE Trans. Commun.}, vol. 68, no. 12, pp. 7362-7375, Dec. 2020.
\bibitem{SunAMP2019}
Z. Sun, Z. Wei, L. Yang, J. Yuan, X. Cheng and L. Wan, ``Exploiting transmission control for joint user identification and channel estimation in massive connectivity,'' \textit{IEEE Trans. Commun.}, vol. 67, no. 9, pp. 6311-6326, Sept. 2019.
\bibitem{ZhuAMP2021}
W. Zhu, M. Tao, X. Yuan and Y. Guan, ``Deep-learned approximate message passing for asynchronous massive connectivity,'' \textit{IEEE Trans. Commun.}, vol. 20, no. 8, pp. 5434-5448, Aug. 2021.
\bibitem{Ruan2022}
C. Ruan, Z. Zhang, H. Jiang, J. Dang, L. Wu and H. Zhang, ``Approximate message passing for channel estimation in reconfigurable intelligent surface aided MIMO multi-user systems,'' \textit{IEEE Trans. Commun.}, 2022. 
\bibitem{KhaniANSD2020}
M. Khani, M. Alizadeh, J. Hoydis and P. Fleming, ``Adaptive neural signal detection for massive MIMO,'' \textit{IEEE Trans. Wireless Commun.}, vol. 19, no. 8, pp. 5635-5648, Aug. 2020.
\bibitem{IC-SRC2021}
W. Li, L. Liu and B. M. Kurkoski, ``Irregularly clipped sparse regression codes," \textit{IEEE Information Theory Workshop (ITW)}, 2021, pp. 1-6.
\bibitem{HWang2019}
H. Wang, W. -T. Shih, C. -K. Wen and S. Jin, ``Reliable OFDM receiver with ultra-low resolution ADC," \textit{IEEE Trans. Commun.}, vol. 67, no. 5, pp. 3566-3579, May 2019.
\bibitem{XZhou2022}
X. Zhou, J. Zhang, C. -W. Syu, C. -K. Wen, J. Zhang and S. Jin, ``Model-driven deep learning-based MIMO-OFDM detector: Design, simulation, and experimental results," \textit{IEEE Trans. Commun.}, 2022.
\bibitem{Yiyao_ofdm}
Y. Cheng, M. A. Van Wyk and L. Ping, ``Orthogonal AMP detection techniques for massive access over OFDM," \textit{IEEE Commun. Lett.}, vol. 25, no. 10, pp. 3384-3388, Oct. 2021.
\bibitem{ZhangOAMP2017}
S. Zhang, C. -K. Wen, K. Takeuchi and S. Jin, ``Orthogonal approximate message passing for GFDM detection,'' \textit{IEEE 18th International Workshop on Signal Processing Advances in Wireless Communications (SPAWC)}, 2017, pp. 1-5.  
\bibitem{SLiu2022}
S. Liu, H. Zhang and Q. Zou, ``Decentralized channel estimation for the uplink of grant-free massive machine-type communications," \textit{IEEE Trans. Commun.}, vol. 70, no. 2, pp. 967-979, Feb. 2022.  
\bibitem{Yiyao_mmv}
Y Cheng, L. Liu and L. Ping, ``Orthogonal AMP for massive access in channels with spatial and temporal correlations" \textit{IEEE J. Sel. Areas Commun.}, vol. 39, no. 3, 726-740, March 2021.
\bibitem{LiOTFS2022}
S. Li, W. Yuan, Z. Wei and J. Yuan, ``Cross domain iterative detection for orthogonal time frequency space modulation,'' \textit{IEEE Trans. Wireless Commun.}, vol. 21, no. 4, pp. 2227-2242, April 2022.

%  \bibitem{LeiMAMP} 
% L. Liu, S. Huang, and B. M. Kurkoski, ``Memory AMP,'' \textit{{IEEE} Trans. Inf. Theory}, 2022, early access.


 \bibitem{Arfken1985_polar}
G. Arfken, \textit{Spherical Polar Coordinates}, §2.5 in Mathematical Methods for Physicists, 3rd ed. Orlando, FL: Academic Press, 1985.
\bibitem{Bronshtein2004}
I. N. Bronshtein, K. A. Semendyayev, G. Musiol, and H. Muehlig, \textit{Handbook of Mathematics}, 4th ed. New York: Springer-Verlag, 2004.
% \bibitem{Kim2011}
% J. Kim, W. Chang, B. Jung, D. Baron, J. C. Ye, ``Belief propagation for joint sparse recovery," \textit{arXiv preprint arXiv:1102.3289}, 2011.
% \bibitem{Chen2006MMV}
% J. Chen and X. Huo, ``Theoretical results on sparse representations of multiple measurement vectors,” \textit{IEEE Trans. on Signal Process.}, vol. 54, no. 12, pp. 4634–4643, 2006.
% \bibitem{Yuwei2018}
% L. Liu and W. Yu, ``Massive connectivity with massive MIMO---part I: Device activity detection and channel estimation,'' \textit{IEEE Trans. Signal Process.}, vol. 66, no. 11, pp. 2933-2946, June 2018.

%\bibitem{Kim2011}
%J. Kim, W. Chang, B. Jung, D. Baron, J. C. Ye, ``Belief propagation for joint sparse recovery," \textit{arXiv preprint arXiv:1102.3289}, 2011.
%\bibitem{Fletcherdeepnet2017}
%A. K. Fletcher, S. Rangan, ``Inference in deep networks in high dimensions,'' \textit{arXiv preprint arXiv:1706.06549}, 2017.



\bibitem{Tulino2004}
A. M. Tulino and S. Verdu, ``Random matrix theory and wireless communications''. \textit{Now Publishers Inc.}, 2004, vol. 1.
%\bibitem{Reffy2019}
%J. R$\acute{\mathrm{e}}$ffy, \textit{Asymptotics of random unitaries}, PhD Thesis, BUTE Institute of Mathematics, 2005. 

% \bibitem{Arfken1974}
% G. Arfken, ``Discrete orthogonality--discrete fourier transform," \textit{Mathematical Methods for Physicists}, vol. 3, pp. 787-792, 1985.
\bibitem{Ahmed1974}
N. Ahmed, T. Natarajan and K. R. Rao, ``Discrete cosine transform," \textit{IEEE Transactions on Computers}, vol. C-23, no. 1, pp. 90-93, Jan. 1974
\bibitem{Yarlagadda1997}
R. K. R. Yarlagadda and J. E. Hershey, \textit{Hadamard matrix analysis and synthesis: With applications to communications and signal/image processing.} Boston, MA: Kluwer, 1997.
%  \bibitem{David2005}
%  Tse David and P. Viswanath, \textit{Fundamentals of wireless communication.} Cambridge university press, 2005.
% \bibitem{Blum2020}
% A. Blum, J. Hopcroft and R. Kannan, \textit{Foundations of data science}. New York, NY, USA: Cambridge Univ. Press, 2020.

\bibitem{Meckes2014}
E. Meckes, \textit{Concentration of measure and the compact classical matrix groups}, 2014.

\bibitem{Schmidt1908}
E. Schmidt, ``$\mr{\ddot{U}}$ber die aufl$\mr{\ddot{o}}$sung linearer gleichungen mit unendlich vielen unbekannten,'' \textit{Rend. Circ. Mat. Palermo (1884-1940)}, vol. 25, no. 1, 53-77, 1908.
\bibitem{TakeuchiCG2017}
K. Takeuchi and C.-K. Wen, “Rigorous dynamics of expectation-propagation signal detection via the conjugate gradient method,” \textit{IEEE 18th International Workshop on Signal Processing Advances in Wireless Communications (SPAWC)}, 2017, pp. 1–5. 
%\bibitem{Price1958}
%R. Price, ``A useful theorem for nonlinear devices having Gaussian inputs", \textit{IRE Trans. Inf. Theory}, vol. IT-4, pp. 69-72, Jun. 1958.
%\bibitem{Bussgang1952}{Ermolova2004}
%J. J. Bussgang, ``Cross-correlation function of amplitude-distorted Gaussian signals", \textit{Res. Lab. Elec., Mas. Inst. Technol.}, Cambridge MA, Tech. Rep. 216, Mar. 1952.
%\bibitem{Ermolova2004}
%N. Y. Ermolova and S. -G. Haggman, ``An extension of Bussgang's theory to complex-valued signals," \textit{the 6th Nordic Signal Processing Symposium}, Espoo, Finland, 2004, pp. 45-48.

% \bibitem{OMAP_III}
% L. Liu, Y. Cheng, S. Liang, J. H. Manton and L. Ping, ``Some notes on OAMP,''  \LL{\textit{arXiv preprint arXiv: ...}}, 2021.  


\bibitem{Stein1972}
C.~Stein, ``A bound for the error in the normal approximation to the
  distribution,'' in \textit{Proc. 6th Berkeley Symp. Math. Statist. Probab.},
  1972. 
% \bibitem{Campese2015}  
% S. Campese, ``Fourth moment theorems for complex Gaussian approximation,'' 2015, arXiv:1511.00547. %[Online]. Available: https://arxiv.org/abs/1511.00547 
\bibitem{Replica_Guo2005}
Dongning Guo and S. Verdu, ``Randomly spread CDMA: Asymptotics via statistical physics,'' \textit{IEEE Trans. Inf. Theory}, vol. 51, no. 6, pp. 1983-2010, June 2005. 
\LL{\bibitem{Sun2015}
P. Sun, C. Zhang, Z. Wang, C. N. Manch\'{o}n and B. H. Fleury, ``Iterative receiver design for ISI channels using combined belief- and expectation-propagation," \textit{IEEE Signal Process. Lett.}, vol. 22, no. 10, pp. 1733-1737, Oct. 2015.
\bibitem{Donoho2006}
D. L. Donoho, ``Compressed sensing," \textit{IEEE Trans. Inf. Theory}, vol. 52,
no. 4, pp. 1289–1306, Apr. 2006.
\bibitem{Vila2015}
J. Vila, P. Schniter, S. Rangan, F. Krzakala, and L. Zdeborová, ``Adaptive
damping and mean removal for the generalized approximate message
passing algorithm," \textit{in ICASSP 2015}, pp. 2021–2025.
\bibitem{Ahn2019}
J. Ahn, B. Shim and K. B. Lee, ``EP-Based Joint Active User Detection and Channel Estimation for Massive Machine-Type Communications," \textit{IEEE Trans. Commun.}, vol. 67, no. 7, pp. 5178-5189, July 2019.
\bibitem{Donoho1995}
D. L. Donoho, ``De-noising by soft-thresholding," \textit{IEEE Trans. Inf. Theory}, vol. 41, no. 3, pp. 613-627, May 1995.}

%\bibitem{Urbanke2000}
%T. Richardson and R. Urbanke, ``The capacity of low-density parity check codes under message-passing decoding,'' \textit{IEEE Trans. Inform. Theory}, vol. 47, pp. 599-618, Feb. 2000.


%\bibitem{RBose2008}
%R. Bose. \textit{Information theory, coding and cryptography}, Tata McGraw-Hill Education, 2008.
% \bibitem{DTse2005}
% D. Tse and P. Viswanath, \textit{Fundamentals of Wireless Communication}. Cambridge, U.K.: Cambridge Univ. Press, 2005.
% \bibitem{Ribeiro2010}
% A. Ribeiro, ``Ergodic stochastic optimization algorithms for wireless communication and networking,'' \textit{IEEE Trans. Signal Process.}, vol. 58, no. 12, pp. 6369-6386, Dec. 2010.
%\bibitem{Cover1991}
%T. M. Cover and J. A. Thomas, \textit{Elements of Information Theory}. New
%York: Wiley, 1991.


%\bibitem{berthier2017state}
%R.~Berthier, A.~Montanari, and P.-M. Nguyen, ``State evolution for approximate
%  message passing with non-separable functions,'' \textit{arXiv preprint
%  arXiv:1708.03950}, 2017.
%\bibitem{Stein1981}
%Charles M. Stein, ``Estimation of the mean of a multivariate normal distribution,''
%\textit{The Annals of Statistics}, 1981, Vol. 9, No. 6, 1135-1151.

% \bibitem{He2018AI}
% H. He, C. Wen, S. Jin and G. Y. Li, ``A model-driven deep learning network for MIMO detection,'' \textit{2018 IEEE Global Conference on Signal and Information Processing (GlobalSIP)}, Anaheim, CA, USA, 2018, pp. 584-588.
% \bibitem{Zhang2019AI}
% J. Zhang, C. Wen, S. Jin and G. Y. Li, ``Artificial intelligence-aided receiver for a CP-free OFDM system: Design, simulation, and experimental test,'' \textit{IEEE Access}, vol. 7, pp. 58901-58914, 2019.
% \bibitem{Ito2019}
% D. Ito, S. Takabe and T. Wadayama, ``Trainable ISTA for sparse signal recovery,'' \textit{IEEE Trans. Signal Process.}, vol. 67, no. 12, pp. 3113-3125, June, 2019.
%\bibitem{Takabe2018}
%S. Takabe, M. Imanishi, T. Wadayama, K. Hayashi, ``Deep learning-aided projected gradient detector for massive overloaded MIMO channels," \textit{arXiv preprint arXiv:1806.10827v2}, 2018.
% \bibitem{Takabe2019}
% S. Takabe, M. Imanishi, T. Wadayama, R. Hayakawa and K. Hayashi, ``Trainable projected gradient detector for massive overloaded MIMO channels: Data-driven tuning approach,'' \textit{IEEE Access}, vol. 7, pp. 93326-93338, 2019.
\bibitem{Lu2021}
Y. M. Lu, ``Householder dice: A matrix-free algorithm for simulating dynamics on Gaussian and random orthogonal ensembles," \textit{IEEE Trans.  Inf. Theory}, vol. 67, no. 12, pp. 8264-8272, Dec. 2021. 
\bibitem{Dekking2005}
F. M. Dekking \textit{et al.}, \textit{A Modern Introduction to Probability and Statistics}. London, U.K.: Springer-Verlag, 2005. 
\bibitem{Stein1981}
C. M. Stein, ``Estimation of the mean of a multivariate normal distribution,'' \textit{The Annals of Statistics}, 1981, Vol. 9, No. 6, 1135-1151.
\bibitem{Williams2001}
D. Williams, \textit{Probability with martingales}, Cambridge Univ. Press, 2001.

 

 



% \bibitem{Hannak2018}
% G. Hannak, A. Perelli, N. Goertz, G. Matz, and M. E. Davies, ``Performance analysis of approximate message passing for distributed compressed sensing," \textit{IEEE J. Select. Topics Signal Process.}, arXiv: 1212.0489v1, Jun. 2018.  
% \bibitem{Fletcher2018}
% A. K. Fletcher, S. Rangan, S. Sarkar, and P. Schniter, ``Plug-in estimation in high-dimensional linear inverse problems: A rigorous analysis,'' \textit{in Proc. Neural Inf. Process. Syst. Conf.}, 2018, pp. 7440-7449.
%\bibitem{Takeuchi2020}
%K. Takeuchi, ``Convolutional Approximate Message-Passing," \textit{arXiv preprint arXiv:2002.08632}, Feb. 2020.
%\bibitem{Schniter2016}
%P. Schniter, S. Rangan, and A. K. Fletcher, ``Vector AMP for the generalized linear model,'' \textit{Proc. Asilomar Conf. on Signals, Systems, and Computers (Pacific Grove, CA)}, Nov. 2016.
%\bibitem{Rangan2010}
%S. Rangan, ``Generalized approximate message passing for estimation with random linear mixing," \textit{arXiv preprint arXiv:1010.5141}, 2010.
% \bibitem{Meng2015}
% X. Meng, S. Wu, L. Kuang and J. Lu, ``An expectation propagation perspective on approximate message passing," \textit{IEEE Signal Process. Lett.}, vol. 22, no. 8, pp. 1194-1197, Aug. 2015.
% \bibitem{Meng2018SPL}
%  X. Meng, S. Wu and J. Zhu, ``A unified Bayesian inference framework for generalized linear models," \textit{IEEE Signal Process. Lett.}, vol. 25, no. 3, pp. 398-402, 2018.
%\bibitem{LeiTVT2019} 
%L. Liu, Y. Li, C. Huang, C. Yuen and Y. L. Guan, ``A New Insight Into GAMP and AMP," \textit{IEEE Trans. Vehi. Tech.}, vol. 68, no. 8, pp. 8264-8269, Aug. 2019.
%  \bibitem{Viterbi1967}
% A. Viterbi, ``Error bounds for convolutional codes and an asymptotically optimum decoding algorithm," \textit{IEEE Trans. Inf. Theory}, vol. 13, no. 2, pp. 260-269, April 1967.
% \bibitem{Forney_Viterbi1973}
% G. D. Forney, ``The viterbi algorithm," \textit{Proceedings of the IEEE}, vol. 61, no. 3, pp. 268-278, March 1973.
% \bibitem{BCJR1974}
% L. Bahl, J. Cocke, F. Jelinek and J. Raviv, ``Optimal decoding of linear codes for minimizing symbol error rate (Corresp.)," \textit{IEEE Trans. Inf. Theory}, vol. 20, no. 2, pp. 284-287, March 1974.


%\bibitem{Loeliger2004}
%H.-A. Loeliger, ``An introduction to factor graphs,,'' \textit{IEEE Signal Process.
%Mag.}, vol. 21, no. 1, pp. 28-41, Jan. 2004.
 
%\bibitem{Fletcherdeepnet2017}
%A. K. Fletcher, S. Rangan, ``Inference in deep networks in high dimensions,'' \textit{arXiv preprint arXiv:1706.06549}, 2017.
%\bibitem{Pandit2019}
%P. Pandit, M. Sahraee, S. Rangan, A. K. Fletcher, ``Asymptotics of map inference in deep networks,'' \textit{arXiv preprint arXiv:1903.01293}, 2019
%\bibitem{HHe2018}
%H. He, C. Wen, and S. Jin, ``Bayesian optimal data detector for hybrid mmwave MIMO-OFDM systems with low-resolution ADCs,'' \textit{IEEE J. Select. Topics Signal Process.}, vol. 12, no. 3, pp. 469-483, Jun 2018.
%\bibitem{Schniter2018}
%J. Mo, P. Schniter, and R. W. Heath, ``Channel estimation in broadband millimeter wave MIMO systems with few-bit ADCs,'' \textit{IEEE J. Select. Topics Signal Process.}, vol. 66, no. 5, pp. 1141-1154, Mar 2018.
%\bibitem{Liang2019}
%S. Liang, J. Tong, and L. Ping, ``On iterative compensation of clipping distortion in OFDM systems,'' \textit{IEEE Wireless Commun. Lett.} vol.8, no.2, pp. 436-439, April 2019.
%\bibitem{JTan2015}
%J. Tan, Y. Ma and D. Baron, ``Compressive imaging via approximate message passing with image denoising," \textit{IEEE Trans. Signal Process.}, vol. 63, no. 8, pp. 2085-2092, 2015.
%\bibitem{Davies2015}
% C. Guo and M. E. Davies, ``Near optimal compressed sensing without priors: Parametric SURE approximate message passing," \textit{IEEE Trans. Signal Process.}, vol. 63, no. 8, pp. 2130-2141, 2015.
%\bibitem{Yuan2017}
%Z. Xue, J. Ma and X. Yuan, ``Denoising-based turbo compressed sensing,'' \textit{IEEE Access}, vol. 5, pp. 7193-7204, 2017.
%\bibitem{Dabov2006}
%K. Dabov, A. Foi, V. Katkovnik, and K. Egiazarian, ``Image denoising with block-matching and 3D filtering,'' \textit{Proc. SPIE Electronic Imaging}, no. 6064A-30, San Jose, California, USA, Jan. 2006.
%\bibitem{Meckes2019}
%E. S. Meckes, \textit{The random matrix theory of the classical compact groups}, Cambridge, U.K.: Cambridge Univ. Press, 2019.
%\bibitem{Yanai2011}
%H. Yanai, K. Takeuchi and Y. Takane, \textit{Projection matrices, generalized inverse matrices, and singular value decomposition}, Springer 2011.
%\bibitem{Rush2017}{Barbier2017}
%C. Rush, A. Greig, and R. Venkataramanan, ``Capacity-achieving sparse
%superposition codes via approximate message passing decoding,'' \textit{IEEE
%Trans. Inf. Theory}, vol. 63, no. 3, pp. 1476-1500, Mar. 2017.
%\bibitem{Barbier2017}
%J. Barbier and F. Krzakala, ``Approximate message-passing decoder and
%capacity-achieving sparse superposition codes,'' \textit{IEEE
%Trans. Inf. Theory}, vol. 63, no. 8, pp. 4894-4927, Aug. 2017.
%\bibitem{Ma2015SPL}
%J. Ma, X. Yuan and L. Ping, ``On the Performance of Turbo Signal Recovery with Partial DFT Sensing Matrices,'' \textit{IEEE Signal Processing Letters}, vol. 22, no. 10, pp. 1580-1584, Oct. 2015.
%\bibitem{Liang_cl2017}
%S. Liang, J. Ma, and L. Ping, ``Clipping can improve the performance of spatially coupled sparse superposition codes,'' \textit{IEEE Commun. Lett.}, vol. 21, no. 12, pp. 966-969, Dec. 2017.
%\bibitem{Opper2005}
%M. Opper and O. Winther, ``Expectation consistent approximate inference," \textit{Journal of Machine Learning Research}, vol. 6, no. Dec, pp. 2177-2204, 2005.
%\bibitem{Jeon2018arxiv}
%C. Jeon, R. Ghods, A. Maleki, and C. Studer, ``Optimal data detection in large MIMO,''
%\bibitem{Schniter2016learning}
%P. Schniter, S. Rangan, A. K. Fletcher, and M. Borgerding, ``Vector AMP and its connections to deep learning,'' \textit{IEEE Info. Thy. Workshop}, Cambridge, UK, Sept. 2016.
%\bibitem{Micciancio2001}
%D. Micciancio, ``The hardness of the closest vector problem with
%preprocessing,''  \textit{IEEE Trans. Inf. Theory}, vol. 47, no. 3, pp. 1212-1215, Mar. 2001.
%\bibitem{verdu1984_1}
%S. Verd\'{u}, ``Optimum multi-user signal detection,''  Ph.D. dissertation, Department of Electrical and Computer Engineering, University of Illinois at Urbana-Champaign, Urbana, IL, Aug. 1984.
%\bibitem{Tuchler2002}
%M.~Tuchler, R.~Koetter, and A.~C. Singer, ``Turbo equalization: {P}rinciples
%  and new results,'' \textit{{IEEE} Trans. Commun.}, vol.~50, no.~5, pp.
%  754--767, May 2002.
%\bibitem{Foucher2001}
%S. Foucher, G. B. Benie and J. M. Boucher, ``Multiscale MAP filtering of SAR images,'' in \textit{IEEE Trans. Image Process.}, vol. 10, no. 1, pp. 49-60, Jan. 2001. 
%\bibitem{Ping2009SCM}
%L. Ping, J. Tong, X. Yuan, and Q. Guo, ``Superposition coded modulation and iterative linear MMSE detection,'' \textit{IEEE J. Select. Areas Commun.}, vol. 27, no.6, pp. 995-1004, Aug. 2009.
%\bibitem{Liang2019clip}
%S. Liang, Jun Tong, and Li Ping, ``On iterative compensation of clipping distortion in OFDM systems,''  \textit{IEEE Wireless Commun. Lett.}, vol.8, no.2, pp.436-439, April 2019.
%\bibitem{Guill2018}
%C. E. Gonz$\acute{\mathrm{a}}$lez-Guill$\acute{\mathrm{e}}$n, C. Palazuelos and I. Villanueva, ``Euclidean distance between Haar orthogonal and Gaussian matrices,'' \textit{Journal of Theoretical Probability}, Vol. 31, Iss. 1, pp 93-118, March 2018.
%\bibitem{XYuan2018}
%A. Liu, L. Lian, V. K. N. Lau and X. Yuan, ``Downlink channel estimation in multiuser massive MIMO with hidden markovian sparsity," \textit{IEEE Trans. Signal Process.}, vol. 66, no. 18, pp. 4796-4810, 2018.
%\bibitem{Berthier2017}
%R. Berthier, A. Montanari, P. M. Nguyen, ``State evolution for approximate message passing with non-separable functions," \textit{arXiv preprint arXiv:1708.03950}, 2017.
%\bibitem{YMa2016}
%Y. Ma, J. Zhu and D. Baron, ``Approximate message passing algorithm with universal denoising and Gaussian mixture learning," \textit{IEEE Trans. Signal Process.}, vol. 64, no. 21, pp. 5611-5622, 2016.
%\bibitem{Kim2011}
%J. Kim, W. Chang, B. Jung, D. Baron, J. C. Ye, ``Belief propagation for joint sparse recovery," \textit{arXiv preprint arXiv:1102.3289}, 2011.
%\bibitem{Yuwei2018}
%L. Liu and W. Yu, ``Massive connectivity with massive MIMO-part I: Device activity detection and channel estimation,'' \textit{IEEE Trans. Signal Process.}, vol. 66, no. 11, pp. 2933-2946, 1 June1, 2018.
%\bibitem{yuwei20180}
%Z. Chen, F. Sohrabi and W. Yu, "Sparse activity detection for massive connectivity," \textit{IEEE Trans. Signal Process.}, vol. 66, no. 7, pp. 1890-1904, April, 2018.
%\bibitem{Hlawatsch2013}
%C. Novak, G. Matz and F. Hlawatsch, "IDMA for the Multiuser MIMO-OFDM Uplink: A Factor Graph Framework for Joint Data Detection and Channel Estimation," \textit{IEEE Trans. Signal
%  Process.}, vol. 61, no. 16, pp. 4051-4066, Aug.15, 2013.
%\bibitem{Huang2019}
%C. Huang, L. Liu, C. Yuen and S. Sun, ``Iterative Channel Estimation Using LSE and Sparse Message Passing for MmWave MIMO Systems,'' \textit{IEEE Trans. Signal
%  Process.}, vol. 67, no. 1, pp. 245-259, 1 Jan.1, 2019.
%\bibitem{LiuLei2016TWC}
%L.~{Liu}, C.~{Yuen}, Y.~L. {Guan}, Y.~{Li}, and Y.~{Su}, ``Convergence analysis
%  and assurance for gaussian message passing iterative detector in massive
%  {MU-MIMO} systems,'' \textit{IEEE Trans. Wireless Commun.},
%  vol.~15, no.~9, pp. 6487--6501, Sep. 2016.
%\bibitem{LiuLei2019TWC}
%L.~{Liu}, C.~{Yuen}, Y.~L. {Guan}, Y.~{Li}, and C.~{Huang}, ``Gaussian message
%  passing for overloaded massive {MIMO-NOMA},'' \textit{IEEE Trans. Wireless Commun.}, vol.~18, no.~1, pp. 210--226, Jan 2019.

\end{thebibliography}
\end{document}